%% file: 1682.tex
\documentclass[runningheads]{llncs}
\usepackage{graphicx}
\usepackage{amsmath,amssymb} 
\usepackage{color}
\usepackage{multirow}

\usepackage{hyperref}
\usepackage{url}
\usepackage{makecell}
\usepackage{pdfpages}

\usepackage[utf8]{inputenc}

\usepackage{multirow}
\usepackage{multicol}
\usepackage[utf8]{inputenc} 
\usepackage[T1]{fontenc}    
\usepackage{booktabs} 
\usepackage{balance} 
\usepackage{amssymb}
\usepackage{subcaption}
\usepackage{array}
\usepackage{multirow}
\usepackage[linesnumbered,ruled]{algorithm2e}
\usepackage{algpseudocode}
\usepackage{amsmath,amssymb,mathtools}
\usepackage{cuted}
\usepackage{afterpage}
\usepackage{bbm}
\usepackage{color}
\usepackage{algorithm2e}
\usepackage{mathtools}

\newcommand{\V}[1]{\mathbf{#1}}

\newcommand{\redd}[1]{\textcolor{red}{#1}}

\SetCommentSty{mycommfont}

\DeclareMathOperator*{\argmax}{argmax}

\newcommand{\ignorethis}[1]{}

\usepackage{subcaption}
\usepackage{array}
\usepackage{multirow}
\usepackage[linesnumbered,ruled]{algorithm2e}
\usepackage{algpseudocode}
\usepackage{amsmath,amssymb,mathtools}
\usepackage{cuted}
\usepackage{amsthm}
\usepackage{afterpage}
\usepackage{color}
\usepackage{rotating} 
\usepackage{hhline}

\let\oldnl\nl
\newcommand{\nonl}{\renewcommand{\nl}{\let\nl\oldnl}}

\newif\ifsubmit

\submitfalse
\ifsubmit
\newcommand{\bo}[1]{}
\newcommand{\chaowei}[1]{}
\newcommand{\dawn}[1]{}
\newcommand{\mingyan}[1]{}
\newcommand{\wh}[1]{}
\newcommand{\george}[1]{}
\else
\newcommand{\bo}[1]{\textcolor{blue}{Bo: #1}}
\newcommand{\dawn}[1]{\textcolor{red}{Dawn: #1}}
\newcommand{\chaowei}[1]{\textcolor{cyan}{Chaowei: #1}}
\newcommand{\mingyan}[1]{\textcolor{purple}{Mingyan: #1}}
\newcommand{\wh}[1]{\textcolor{blue}{[wh: #1]}}
\newcommand{\george}[1]{\textcolor{cyan}{George: #1}}
\fi

\begin{document}

\input{def.tex}
\pagestyle{headings}
\mainmatter
\def\ECCV18SubNumber{1682}  

\title{Characterizing Adversarial Examples Based on Spatial Consistency Information for Semantic Segmentation} 

\titlerunning{Characterizing Adversarial Examples Based on Spatial Consistency}
\authorrunning{Xiao et al.}
\author{%
   \small
   Chaowei Xiao\inst{1}
   \and
   Ruizhi Deng\inst{2}
   \and
   Bo Li\inst{3,4} \and
   Fisher Yu\inst{4} \and
   Mingyan Liu\inst{1} \and
   Dawn Song\inst{4} 
}

\institute{
   \small
   $^1$University of Michigan \quad
   $^2$Simon Fraser University \quad
  $^3$UIUC \quad
   $^4$UC Berkeley \quad\\
}

\maketitle

\begin{abstract}
Deep Neural Networks (DNNs) have been widely applied in various recognition tasks. However, recently DNNs have been shown to be vulnerable against adversarial examples, which can mislead DNNs to make arbitrary incorrect predictions.
While adversarial examples are well studied in classification tasks, other learning problems may have different properties. For instance, semantic segmentation requires additional components such as dilated convolutions and multiscale processing.
In this paper, we aim to characterize adversarial examples based on spatial context information in semantic segmentation. We observe that spatial consistency information can be potentially leveraged to detect adversarial examples robustly even when a strong adaptive attacker has access to the model and detection strategies. We also show that adversarial examples based on attacks considered within the paper barely transfer among models, even though transferability is common in classification. Our observations shed new light on developing adversarial attacks and defenses to better understand the vulnerabilities of DNNs.

\keywords{Semantic segmentation, adversarial example, spatial consistency}
\end{abstract}

\section{Introduction}
Deep Neural Networks (DNNs) have been shown to be highly expressive and have achieved state-of-the-art performance on a wide range of tasks, such as speech recognition \cite{hinton2012deep}, image classification \cite{krizhevsky2012imagenet}, natural language understanding~\cite{zeng2014relation}, and robotics~\cite{noda2014multimodal}. However, recent studies have found that DNNs are vulnerable to \emph{adversarial examples}~\cite{szegedy2013intriguing,goodfellow2014explaining,nguyen2015deep,xiao2018spatially,ijcai2018-543,tong2017hardening,chen2017zoo,chen2017ead,chen2018attacking}. 
Such examples are intentionally perturbed inputs with small magnitude adversarial perturbation added, which can induce the network to make arbitrary incorrect predictions at test time, even when the examples are generated against different models \cite{liu2016delving,carlini2017towards,papernot2016transferability,217571}. The fact that the adversarial perturbation required to fool a model is often small and (in the case of images) imperceptible to human observers makes detecting such examples very challenging.
This undesirable property of deep networks has become a major security concern in real-world applications of DNNs, such as self-driving cars and identity recognition systems~\cite{evtimov2017robust,sharif2016accessorize}.
Furthermore, both white-box and black-box attacks have been performed against DNNs successfully when an attacker is given full or zero knowledge about the target systems~\cite{bhagoji2017exploring,goodfellow2014explaining,ijcai2018-543}. 
Among black-box attacks, transferability is widely used for generating attacks against real-world systems which do not allow white-box access. Transferability refers to the property of adversarial examples in classification tasks where one adversarial example generated against a local model can mislead another unseen model without any modification~\cite{papernot2016transferability}. 

Given these intriguing properties of adversarial examples, various analyses for understanding adversarial examples have been proposed~\cite{ma2018characterizing,madry2017towards,weng2018evaluating,weng2018towards}, and potential defense/detection techniques have also been discussed mainly for the image classification problem~\cite{das2017keeping,hosseini2017blocking,madry2017towards}. For instance, image pre-processing~\cite{dziugaite2016study}, adding another type of random noise to the inputs~\cite{xie2017mitigating}, and adversarial retraining~\cite{goodfellow2014explaining} have been proposed for defending/detecting adversarial examples when classifying images. However, researchers~\cite{carlini2017adversarial,he2017adversarial} have shown that these defense or detection methods are easily attacked again by attackers with or even without knowledge of the defender's strategy. 
Such observations bring up concerns about safety problems within diverse machine learning based systems. 

\begin{figure}[thb]
    \centering
     \centering
     \includegraphics[width=0.6\textwidth]{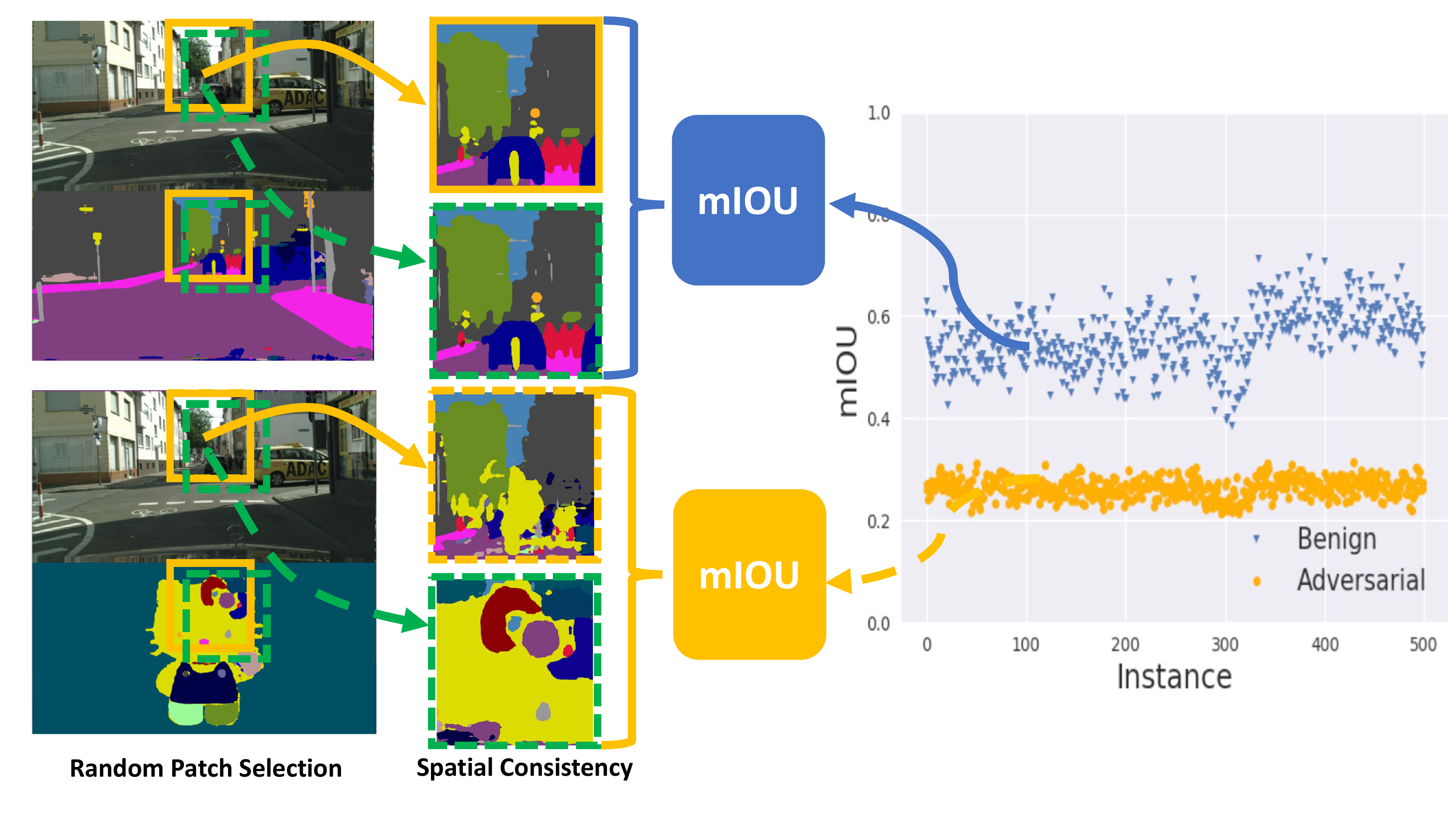}
    \caption{Spatial consistency analysis for adversarial and benign instances in semantic segmentation. }
    \label{fig:arch}
\end{figure}

In order to better understand adversarial examples against different tasks, in this paper we aim to analyze adversarial examples in the semantic segmentation task instead of classification. We hypothesize that adversarial examples in different tasks may contain unique properties that provide in-depth understanding for such examples and encourage potential defensive mechanisms.
Different from image classification, in semantic segmentation, each pixel will be given a prediction label which is based on its surrounding information~\cite{cui2013localized}. Such spatial context information plays a more important role for segmentation algorithms, such as~\cite{Yu2016,zhao2017pyramid,lin2016efficient,krahenbuhl2011efficient}.
Whether adversarial perturbation would break such spatial context is unknown to the community. In this paper we propose and conduct image spatial consistency analysis, which randomly selects overlapping patches from a given image and checks how consistent the segmentation results are for the overlapping regions. Our pipeline of spatial consistency analysis for adversarial/benign instances is shown in Figure ~\ref{fig:arch}. 
We find that in segmentation task, adversarial perturbation can be weakened for separately selected patches, and therefore adversarial and benign images will show very different behaviors in terms of the spatial consistency information. Moreover, since such spatial consistency is highly random, it is hard for adversaries to take such constraints into account when performing adaptive attacks. This renders the system less brittle even facing the sophisticated adversaries, who have full knowledge about the model as well as the detection/defense method applied.. 

We use image scale transformation to perform detection of adversarial examples as a baseline, which has been used for detection in classification tasks~\cite{tabacof2016exploring}.
We show that by randomly scaling the images, adversarial perturbation can be destroyed and therefore adversarial examples can be detected. However, when the attacker knows the detection strategy (adaptive attacker), even without the exact knowledge about the scaling rate, attacker can still perform adaptive attacks against the detection mechanism, which is similar with the findings in classification tasks~\cite{carlini2017adversarial}.
On the other hand, we show that by incorporating spatial consistency check, existing semantic segmentation networks can detect adversarial examples (average AUC 100\%), which are generated by the state-of-the-art attacks considered in this paper, regardless of whether the adversary knows the detection method. 
Here, we allow the adversaries to have full access to the model and any detection method applied to analyze the robustness of the model against adaptive attacks.
We additionally analyze the defense in a black-box setting, which is more practical in real-world systems.

In this paper, our goal is to further understand adversarial attacks by conducting spatial consistency analysis in the semantic segmentation task, and we make the following contributions:
\begin{enumerate}
    \item We propose the spatial consistency analysis for benign/adversarial images and conduct large scale experiments on two state-of-the-art attack strategies against both DRN and DLA segmentation models with diverse adversarial targets on different dataset, including Cityscapes and real-world autonomous driving video dataset.
    \item We are the first to analyze spatial information for adversarial examples in segmentation models. We show that spatial consistency information can be potentially leveraged to distinguish  adversarial examples. We also show that spatial consistency check mechanism induce a high degree of randomness and therefore is robust against adaptive adversaries. 
    We evaluate image scaling and spatial consistency, and show that spatial consistency outperform standard scaling based method.
    \item In addition, we empirically show that adversarial examples generated by
    the attack methods considered in our studies barely transfer among models, even when these models are of the same architecture with different initialization, different from the transferability phenomena in classification tasks. 
\end{enumerate}

\section{Related work}

\paragraph{Semantic Segmentation} has received long lasting attention in the computer vision community~\cite{leung2001representing}. Recent advances in deep learning~\cite{krizhevsky2012imagenet} also show that deep convolutional networks can achieve much better results than traditional methods~\cite{long2015fully}. Yu et al.~\cite{Yu2016} proposed using dilated convolutions to build high-resolution feature maps for semantic segmentation. They can improve the performance significantly compared to upsampling approaches~\cite{long2015fully,ronneberger2015u,badrinarayanan2017segnet}. Most of the recent state-of-the-art approaches are based on dilated convolutions~\cite{Yu2017,zhao2017pyramid,wu2016wider} and residual networks~\cite{he2016deep}. Therefore, in this work, we choose dilated residual networks (DRN)~\cite{Yu2017} and deep layer aggregation (DLA)~\cite{yu2017deep} as our target models for attacking and defense. 

\paragraph{Adversarial Examples for Semantic Segmentation}
have been studied recently in addition to adversarial examples in image classification. 
Xie et al.\ proposed a gradient based algorithm to attack pixels within the whole image iteratively until most of the pixels have been misclassified into the target class~\cite{xie2017adversarial}, which is called dense adversary generation (DAG). Later an optimization based attack algorithm has been studied by introducing a surrogate loss function called Houdini in the objective function~\cite{cisse2017houdini}. 
The Houdini loss function is made up of two parts. The first part represents the stochastic margin between the score of actual and predicted targets, which reflects the confidence of model prediction. The second part is the task loss, which is independent with the model and corresponds to the actual task. The task loss enables Houdini algorithm to generate adversarial examples in different tasks, including image segmentation, human pose estimation, and speech recognition. 

Various detection and defense methods have also been studied against adversarial examples in image classification. For instance, adversarial training~\cite{goodfellow2014explaining} and its variations~\cite{tramer2017ensemble,madry2017towards} have been proposed and demonstrated to be effective in classification task, which is hard to adapt for the segmentation task. Currently no defense or detection methods have been studied in image segmentation.

\begin{figure}[tbh]
    \centering
    \begin{minipage}{.45\textwidth}
     \begin{subfigure}{\textwidth}
     \centering
     \includegraphics[width=\textwidth]{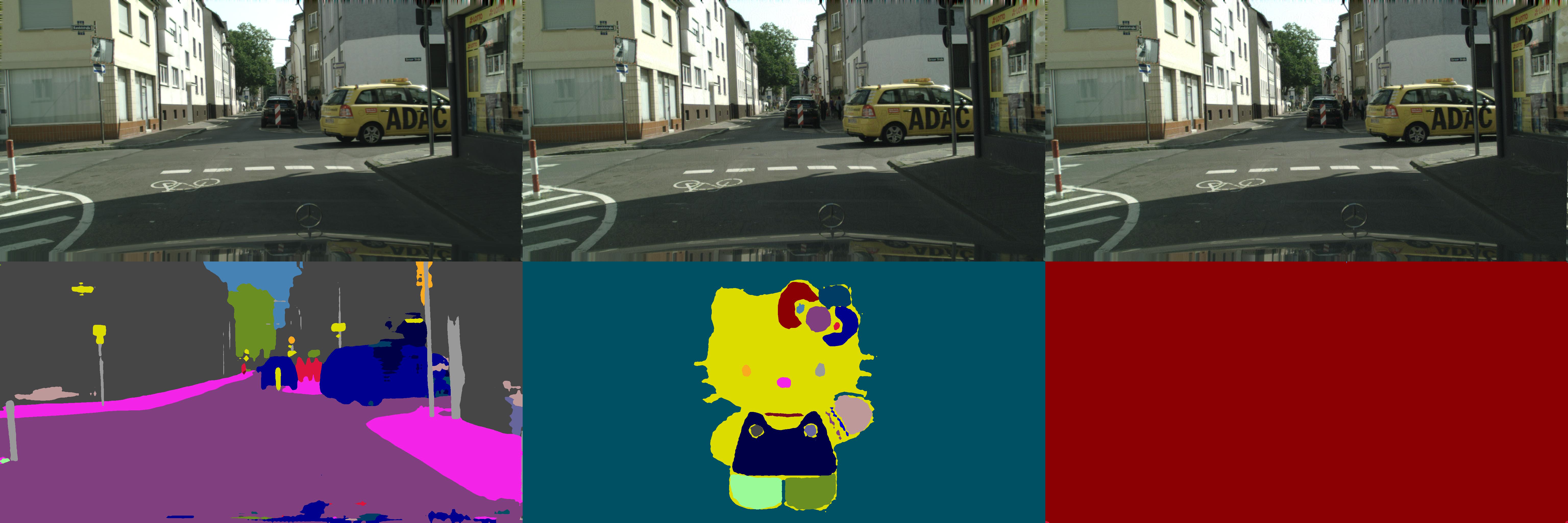}
     \caption{Cityscapes }
     \label{fig:attention-a}
     \end{subfigure}
    \end{minipage}
    \begin{minipage}{.40\textwidth}
     \begin{subfigure}{\textwidth}
     \centering
     \includegraphics[width=\textwidth]{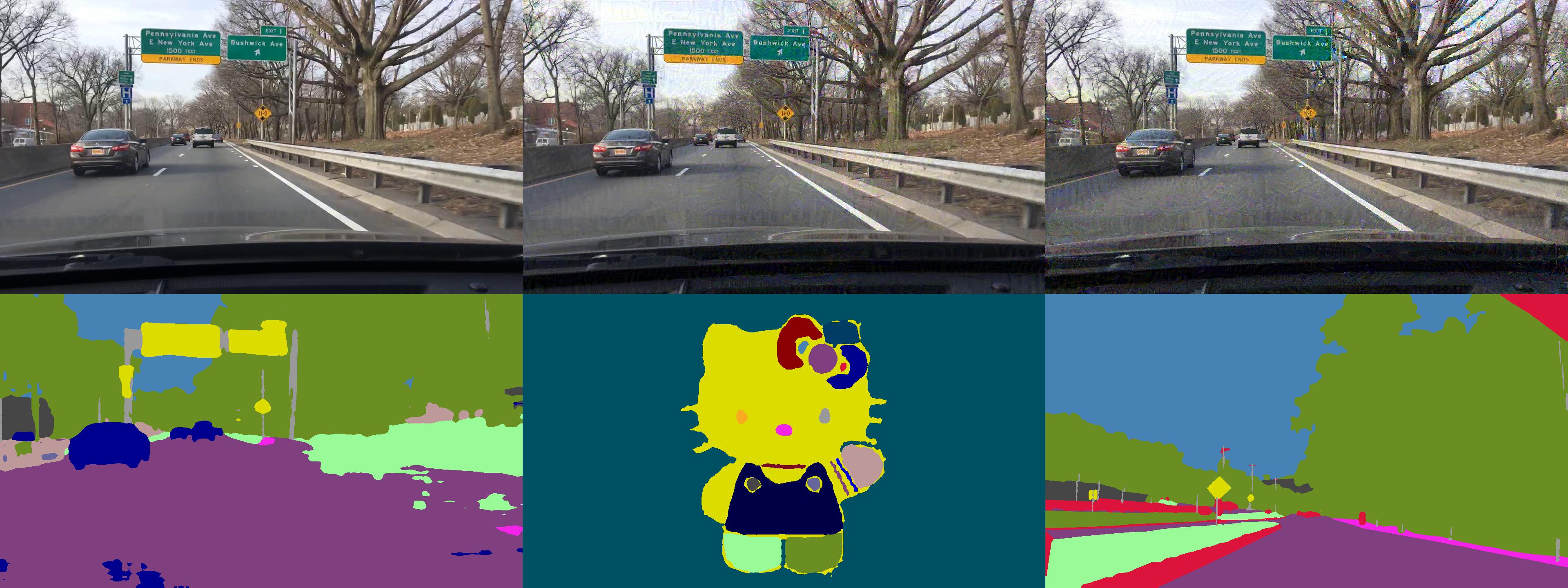}
     \caption{BDD }
     \label{fig:attention-a}
     \end{subfigure}
    \end{minipage}
    \caption{Samples of benign and adversarial examples generated by Houdini on Cityscapes~\cite{cordts2016cityscapes} (targeting on Kitty/Pure) and BDD100K~\cite{yu2018bdd100k} (targeting on Kitty/Scene). We select DRN as our target model here. Within each subfigure, the first column shows benign images and corresponding segmentation results, and the second and third columns show adversarial examples with different adversarial targets. 
    }
    \label{fig:example}
\end{figure}

\section{\Spatialconsis}
In this section, we will explore the effects that spatial context information has on benign and adversarial examples in segmentation models. We conduct different experiments based on various models and datasets, and due to the space limitation, we will use a small set of examples to demonstrate our discoveries and relegate other examples to the supplementary materials.
Figure~\ref{fig:example} shows the benign and adversarial examples targeting diverse adversarial targets: ``Hello Kitty'' (Kitty) and random pure color (Pure) on Cityscapes; and ``Hello Kitty'' (Kitty) and a real scene without any cars (Scene) on BDD video dataset, respectively. In the rest of the paper, we will use the format ``attack method | target'' to label each adversarial example. Here we consider both DAG~\cite{xie2017adversarial} and Houdini~\cite{cisse2017houdini} attack methods.
\begin{figure}[tbh]
\centering
\begin{minipage}{.39\textwidth}
 \begin{subfigure}{\textwidth}
 \centering
 \includegraphics[width=\textwidth]{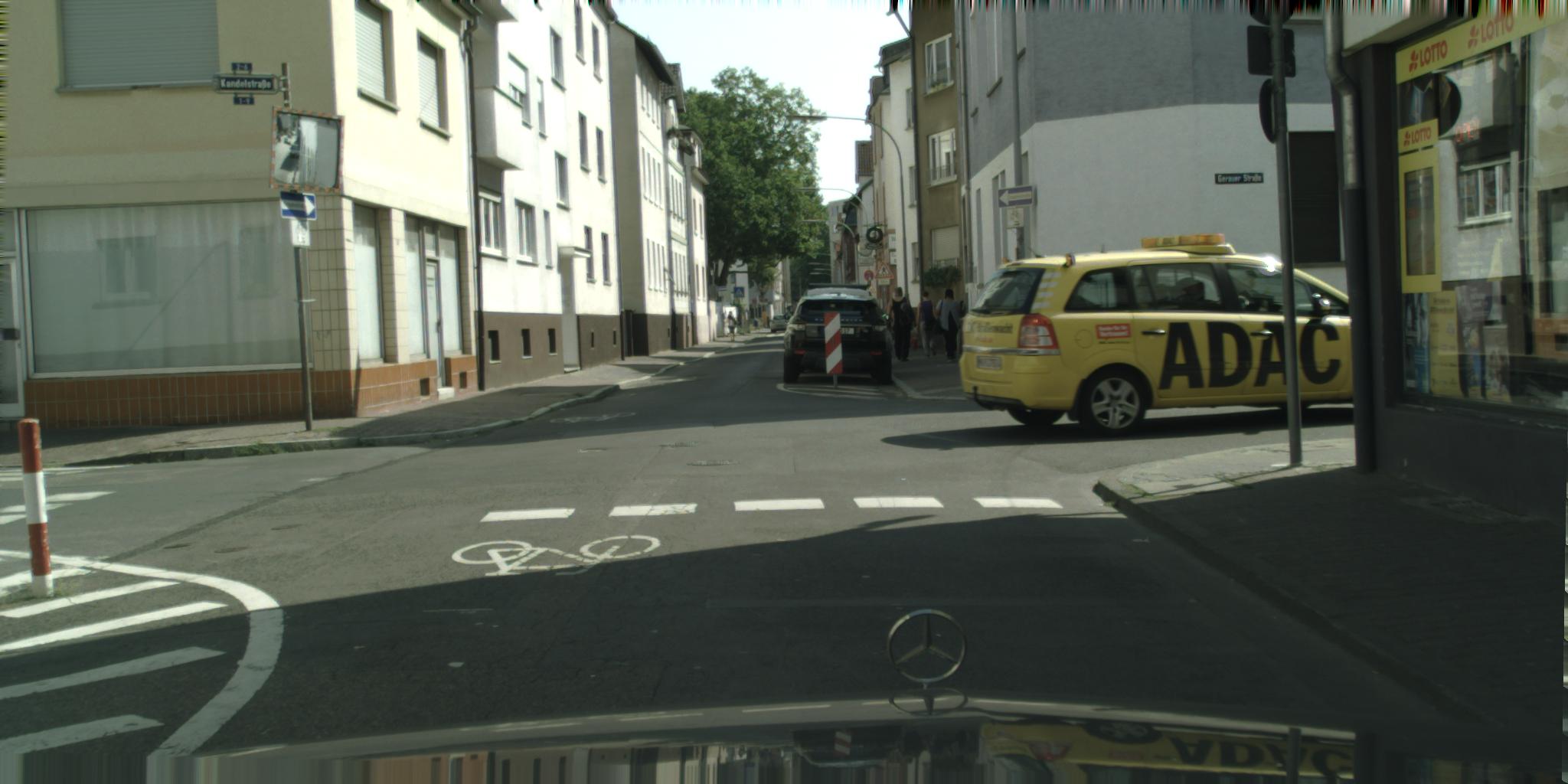}
 \caption{ Benign example }
 \label{fig:attention-a}
 \end{subfigure}
\end{minipage}
\begin{minipage}{.45\textwidth}
 \begin{subfigure}{\textwidth}
 \centering
 \includegraphics[width=\textwidth]{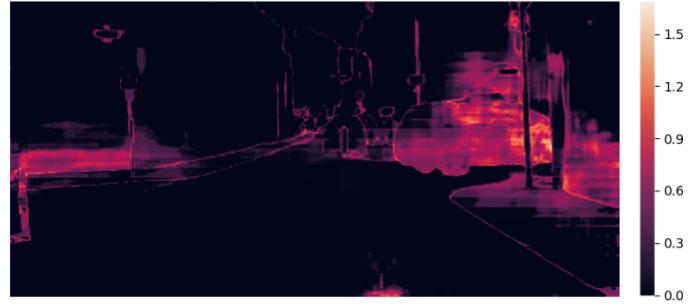}
 \caption{Heatmap of benign image}
 \label{fig:attention-a}
 \end{subfigure}
\end{minipage}
\begin{minipage}{.24\textwidth}
 \begin{subfigure}{\textwidth}
 \centering
 \includegraphics[width=\textwidth]{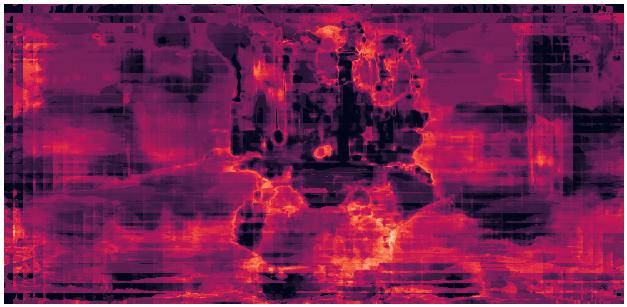}
 \caption{DAG | Kitty  }
 \label{fig:attention-a}
 \end{subfigure}
\end{minipage}
\begin{minipage}{.24\textwidth}
 \begin{subfigure}{\textwidth}
 \centering
 \includegraphics[width=\textwidth]{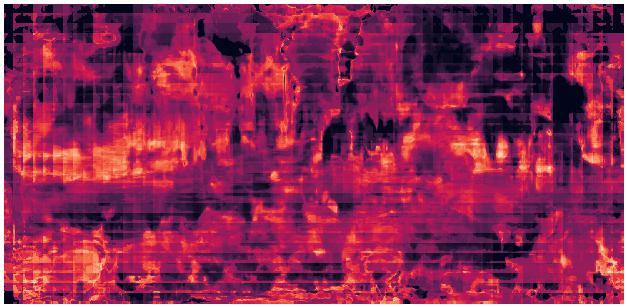}
 \caption{DAG | Pure }
 \label{fig:attention-a}
 \end{subfigure}
\end{minipage}
\begin{minipage}{.24\textwidth}
 \begin{subfigure}{\textwidth}
 \centering
 \includegraphics[width=\textwidth]{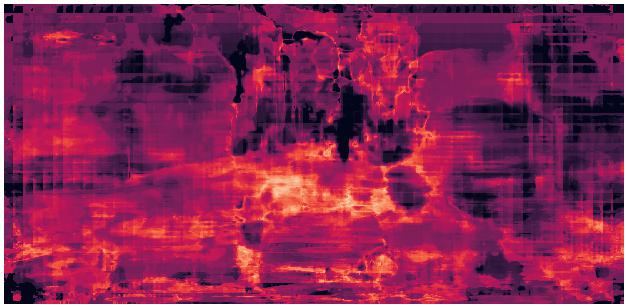}
 \caption{Houdini | Kitty }
 \label{fig:attention-a}
 \end{subfigure}
\end{minipage}
\begin{minipage}{.24\textwidth}
 \begin{subfigure}{\textwidth}
 \centering
 \includegraphics[width=\textwidth]{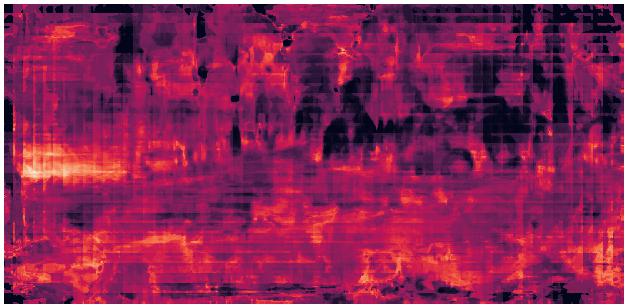} 
 \caption{Houdini |  Pure}
 \label{fig:attention-a}
 \end{subfigure}
\end{minipage}
\caption{Heatmap of per-pixel self-entropy
on Cityscapes dataset against DRN model. 
(a) and (b) show a benign image and its corresponding per-pixel self-entropy heatmap.
(c)-(f) show the heatmaps of the adversarial examples generated by DAG and Houdini attacks targeting ``Hello Kitty'' (Kitty) and random pure color (Pure). }
\label{fig:drn_entropy}
\end{figure}

\begin{figure}[h]
    \centering
     \begin{subfigure}{.99\textwidth}
     \centering
     \includegraphics[width=\textwidth]{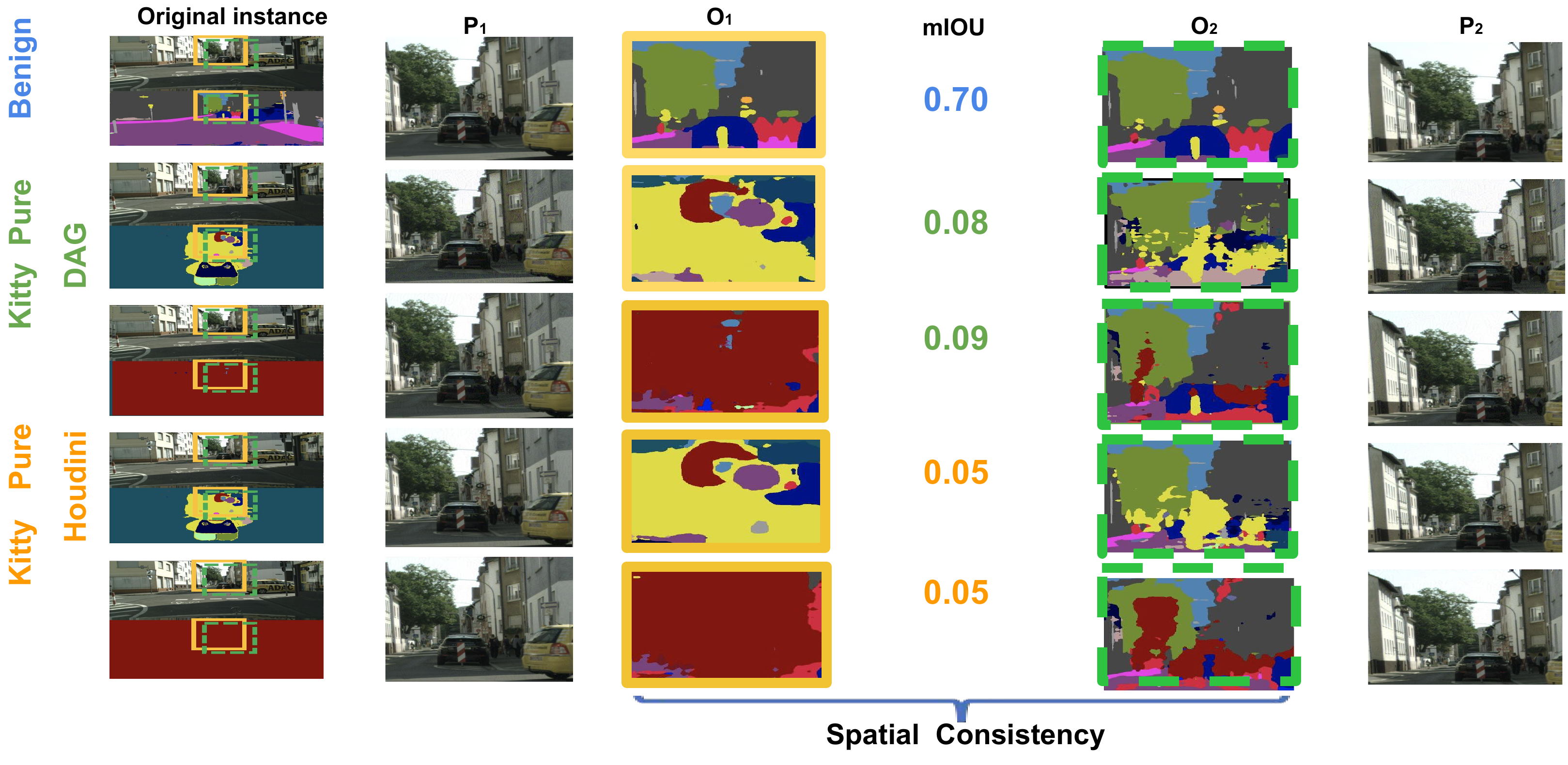}
     \end{subfigure}
    \label{fig:scc}
    \caption{Examples of \spatialconsis on adversarial examples generated by DAG and Houdini attacks targeting on Kitty and Pure. First column shows the original image and corresponding segmentation results. Column $P_1$ and $P_2$ show two randomly selected patches, while column $O_1$ and $O_2$ represent the segmentation results of the overlapping regions from these two patches, respectively. The mIOU between $O_1$ and $O_2$ are reported. It is clear that the segmentation results of the overlapping regions from two random patches are very different for adversarial images (low mIOU), but relatively consistent for benign instance (high mIOU).
    }
\end{figure}

\subsection{Spatial Context Analysis}
To quantitatively analyze the contribution of spatial context information to the segmentation task, we first evaluate the entropy of prediction based on different spatial context.  
For each pixel $m$ within an image, we randomly select $K$ patches $\{P_1, P_2, ..., P_K\}$ which contain $m$. 
Afterwards, within each patch $P_i$, the pixel $m$ will be assigned with a confidence vector based on Softmax prediction, so pixel $m$ will correspond to $K$ vectors in total. We discretize each vector to a one-hot vector and sum up these $K$ one-hot vectors to obtain vector $\mathcal{V}_{m}$. Each component $\mathcal{V}_{m}[j]$ of the vector represents the number of times pixel $m$ is predicted to be class $j$. We then normalize $\mathcal{V}_{m}$ by dividing $K$. 
Finally, for each pixel $m$, we calculate its self-entropy
\[\mathcal{H}(m) = -\sum_{j} \mathcal{V}_{m}[j] \log \mathcal{V}_{m}[j] \]
and therefore calculate the self entropy for each vector. 
We utilize such entropy information of each pixel to convey the consistency of different surrounding patches and plot this information in the heatmaps in Figure~\ref{fig:drn_entropy}.
It is clear that for benign instances, the boundaries of original objects have higher entropy, indicating that these are places harder to predict and can gain more information by considering different surrounding spatial context information.

\subsection{Patch Based Spatial Consistency}
The fact that surrounding spatial context information shows different spatial consistency behaviors for benign and adversarial examples motivates us to perform the spatial consistency check hoping to potentially tell these two data distributions apart. 

First, we introduce how to generate overlapping spatial contexts by selecting random patches and then validate the spatial consistency information.
Let $s$ be the patch size and $w,h$ be the width and height of an image $\mathbf{X}$. 
We define the first and second patch based on the coordinates of their top-left and bottom-right vertices $(u_1, u_2, u_3, u_4 ), (v_1, v_2, v_3, v_4)$, where 
Let $(d_{u_{1},v_{1}}, d_{u_2, v_2})$ be displacement between the top-left coordinate of the first and second patch: $d_{u_1, v_1}=v_1 - u_1, d_{u_2, v_2} = v_2 - u_2$.
To guarantee that there is enough overlap, we require $(d_{u_1, v_1}$ and $d_{u_2, v_2})$ to be in the range $(b_{\V{low}}, b_{\V{upper}})$. 
Here we randomly select the two patches, aiming to capture diverse enough surrounding spatial context, including information both near and far from the target pixel. 
The {\bf patch selection algorithm (getOverlapPatches)} is shown in supplementary materials. 

Next we show how to apply the \spatialconsis to a given input and therefore recognize adversarial examples. The detailed algorithm is shown in Algorithm~\ref{algo:consistency}. Here $K$ denotes the number of overlapping regions for which we will check the spatial consistency. We use the mean Intersection Over Union (mIOU) between the overlapping regions $O_1$, $O_2$ from two patches $P_1$, $P_2$ 
to measure their spatial consistency. The mIOU is defined as $\frac{1}{n_\textit{cls}}\sum_{i} n_{ii} / (\sum_{j}n_{ij} + \sum_{j} n_{ji} - n_{ii})$, where $n_{ij}$ denotes the number of pixels predicted to be class $i$ in $O_1$ and class $j$ in $O_2$, and $n_\textit{cls}$ is the number of the unique classes appearing in both $O_1$ and $O_2$.  
$\V{getmIOU}$ is a function that computes the mIOU given patches $P_1$, $ P_2$ along with their overlapping regions $O_1$ and $O_2$ shown in supplementary materials.

\begin{algorithm}[h]
\SetAlgoLined
\SetKwInput{Input}{input}
\SetKwInput{Output}{output}
\SetKwInput{Init}{Initialization}
\SetKwInput{Blank}{}
\SetKwInput{Ret}{Return}

\tabcolsep=0pt
\nonl \begin{tabular}{@{}ll}
    \Input{}&Input image $\mathbf{X}$\;\\
 &number of overlapping regions $K$\;\\
&patch size $s$\; \\
&segmentation model $f$\;\\
&bound $b_{\V{low}}, b_{\V{upper}}$\;\\
    \Output{}&Spatial consistency threshold $c$\;\\
\end{tabular}
\BlankLine
 $\V{Initialization : }$ $\V{cs} \gets $[], $w \gets x.width, h \gets x.height$\;
 \For{$k\leftarrow 0$ \KwTo $K$}{
 $(u_1,u_2, u_3, u_4), (v_1, v_2, v_3, v_4) \gets \V{getOverlapPatches}(s, w, h, b_{\V{low}}, b_{\V{upper}})$\;
 $P_1 =X[u1: u3, u2:u4], P_2=X[v_1:v_3, v_3:v_4]$\; 
  \tcc*[h]{get prediction result of two random patches from $f$}\;
 $\textit{pred}^1 \gets \argmax_{c} f_{c}(P_1), \textit{pred}^2 \gets \argmax_{c} f_{c}(P_2)$\;
  \tcc*[h]{get prediction of the overlap area between two patches}\;
 $p_1 \gets \{\textit{pred}^1_{i,j} | \forall (i,j) \in \textit{pred}^1, i>v_1-u_1, j > v_2 - u_2\} $\;
 $p_2 \gets \{\textit{pred}^2_{i,j} | \forall (i,j) \in \textit{pred}^2, i<s-(v_1-u_1), j<s-(v_2 - u_2)\}$\;
    \tcc*[h]{get consistency value (mIOU) from two patches}\;
 $\V{cs} \overset{+}{\leftarrow} \V{getmIOU}(p1,p2)$\;
}
$c \gets \V{Mean(cs)}$\;
\Ret{c}
 \caption{Spatial Consistency Check Algorithm}
 \label{algo:consistency}
\end{algorithm}

\section{Scale Consistency Analysis}
We have discussed how spatial consistency can be utilized to potentially characterize adversarial examples in segmentation task. In this section, we will discuss another baseline method: image scale transformation, which is another natural factor considered in semantic segmentation~\cite{johnson2011unsupervised,long2015fully}.
Here we focus on image blur operation by applying Gaussian blur to given images~\cite{chan1998total}, which is studied for detecting adversarial examples in image classification~\cite{tabacof2016exploring}.
Similarly, we will analyze the effects of image scaling on benign/adversarial samples.
Since spatial context information is important for segmentation task, scaling or performing segmentation on small patches may damage the global information and therefore affect the final prediction. Here we aim to provide quantitative results to understand and explore how image scale transformation would affect adversarial perturbation. 

\subsection{Scale Consistency Property}
\begin{figure}[thb]
    \centering
    \begin{minipage}{.9\textwidth}
     \begin{subfigure}{\textwidth}
     \centering
     \includegraphics[width=\textwidth]{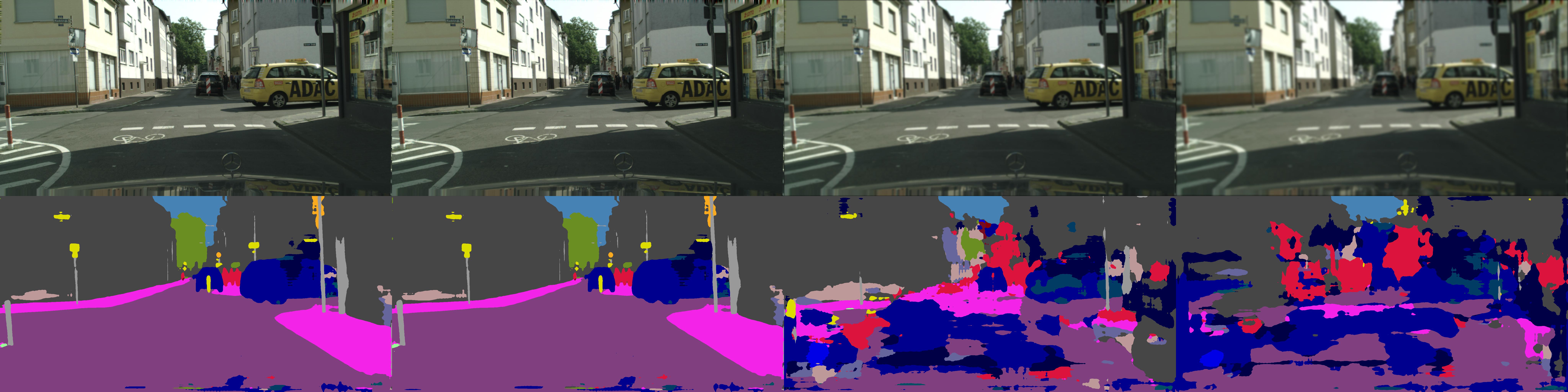}
     \caption{Benign example}
     \end{subfigure}
    \end{minipage}
    \begin{minipage}{.44\textwidth}
     \begin{subfigure}{\textwidth}
     \centering
     \includegraphics[width=\textwidth]{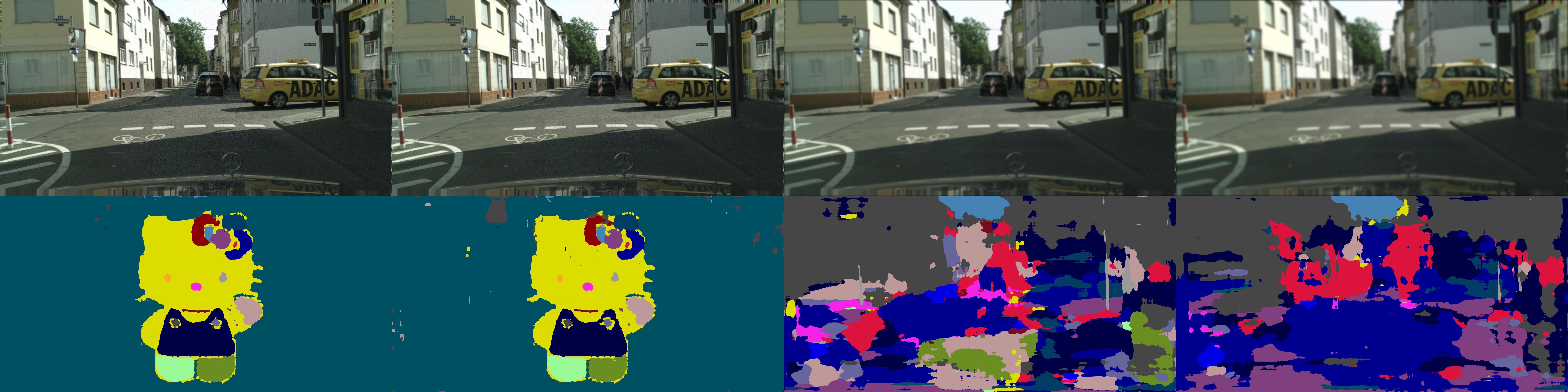}
     \caption{DAG | Kitty}
     \end{subfigure}
    \end{minipage}
    \begin{minipage}{.44\textwidth}
     \begin{subfigure}{\textwidth}
     \centering
     \includegraphics[width=\textwidth]{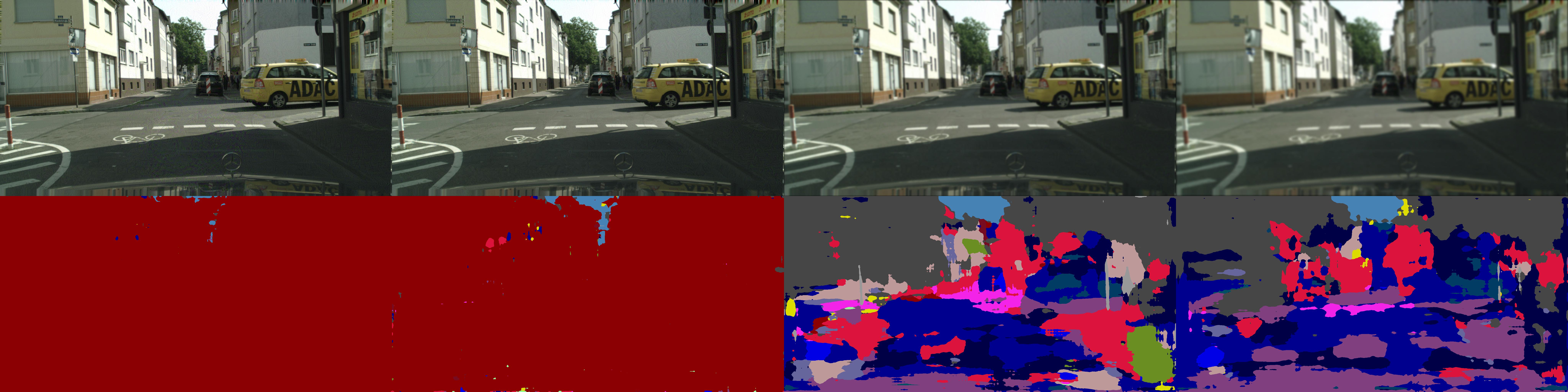}
     \caption{DAG | Pure}
     \end{subfigure}
    \end{minipage}
    \begin{minipage}{.44\textwidth}
     \begin{subfigure}{\textwidth}
     \centering
     \includegraphics[width=\textwidth]{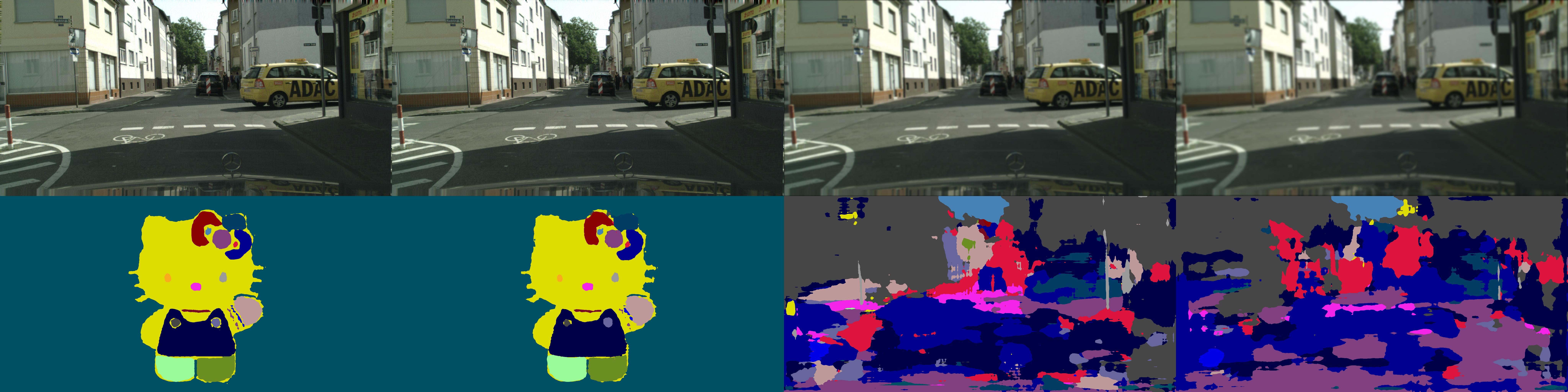}
     \caption{Houdini | Kitty}
     \end{subfigure}
    \end{minipage}
    \begin{minipage}{.44\textwidth}
     \begin{subfigure}{\textwidth}
     \centering
     \includegraphics[width=\textwidth]{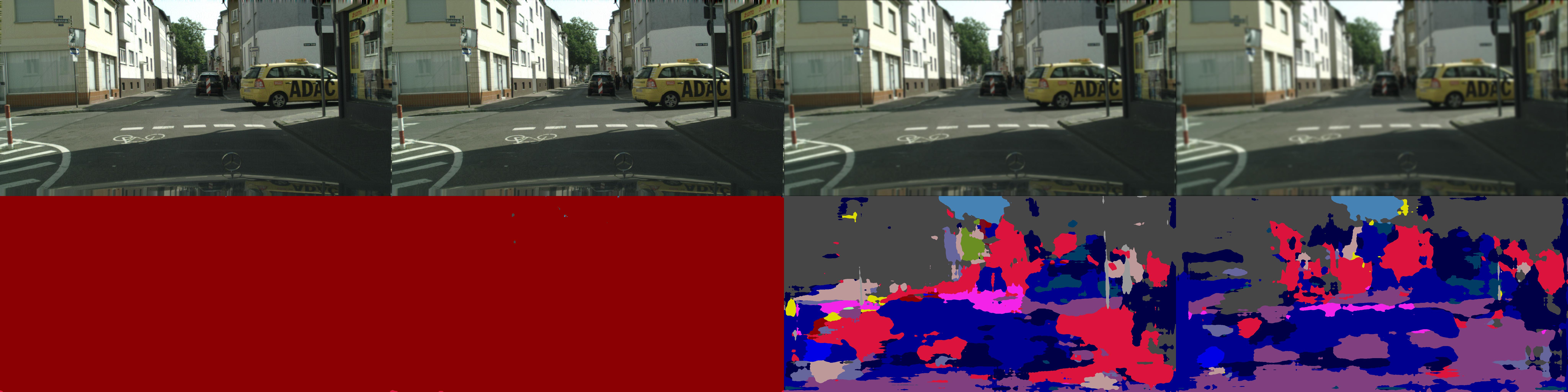}
     \caption{Houdini | Pure}
     \end{subfigure}
    \end{minipage}
    \caption{Examples of images and corresponding segmentation results before/after image scaling on Cityscapes against DRN model.
    For each subfigure, the first column shows benign/adversarial image, while the later columns represent images after scaling by applying Gaussian kernel with std as 0.5, 3, and 5, respectively. 
    (a) shows benign images before/after image scaling and the corresponding segmentation results; (b)-(e) present similar results for adversarial images generated by DAG and Houdini attacks targeting on Kitty and Pure.
    }
    \label{fig:blur}
\end{figure}

Scale theory is commonly applied in image segmentation task~\cite{saha2000scale}, and therefore we train scale resilient models to obtain robust ones, which we perform attacks against. On these scale resilient models, we first analyze how image scaling affect segmentation results for benign/adversarial samples. We applied the DAG~\cite{xie2017adversarial} and Houdili~\cite{cisse2017houdini} attacks against the DRN and DLA models with different adversarial targets. The images and corresponding segmentation results before and after scaling are shown in Figure~\ref{fig:blur}. We apply Gaussian kernel with different standard deviations (std) to scale both benign and adversarial instances. 
It is clear that when we apply Gaussian blurring with higher std (3 and 5), adversarial perturbation is harmed and the segmentation results are not longer adversarial targets for scale transformed adversarial examples as shown in Figure~\ref{fig:blur} (a)-(e).

\section{Experimental Results}
In this section, we conduct comprehensive large scale experiments to evaluate the image spatial and scale consistency information for benign and adversarial examples generated by different attack methods. We will also show that the spatial consistency based detection method is robust against sophisticated adversaries with knowledge about defenders, while scale transformation method is not.

\subsection{Implementation Details}

\paragraph{Datasets.} We apply both Cityscapes~\cite{cordts2016cityscapes} and BDD100K~\cite{yu2018bdd100k} in our evaluation. We show results on the validation set of both datasets, which contains 500 high resolution images with a combined 19 categories of segmentation labels.
These two datasets are both outdoor datasets containing instance-level annotations, which would raise real-wold safety concerns if they were attacked. 
Comparing with other datasets such as Pascal VOC~\cite{Everingham15} and CamVid~\cite{BrostowSFC:ECCV08}, these two dataset are more challenging due to the relatively high resolution and diverse scenes within each image. 

\paragraph{Semantic Segmentation Models.}
We apply Dilated residual networks (DRN)~\cite{Yu2017} and Deep Layer Aggregation (DLA)~\cite{yu2017deep} as our target models.
More specifically, we select DRN-D-22 and DLA-34. For both models, we use 512 crop size and 2 random scale during training to obtain scale resilient models for both the BDD and Cityscapes datasets. The mIOU of these two models on pristine training data are shown in Table~\ref{tbl:city}.  More result on different models can be found in supplementary materials.

\paragraph{Adversarial Examples}
We generate adversarial examples 
based on two state-of-the-art attack methods: DAG~\cite{xie2017adversarial} and Houdini~\cite{cisse2017houdini} using our own implementation of the methods. 
We select a complex image, Hello Kitty (Kitty), with different background colors and a random pure color (Pure) as our targets on Cityscapes dataset. Furthermore, in order to increase the diversity, we also select a real-world driving scene (Scene) without any cars from the BDD training dataset as another malicious target on BDD. 
Such attacks potentially show that every image taken in the real world can be attacked to the same scene without any car showing on the road, which raises great security concerns for future autonomous driving systems. Furthermore, we also add three additional adversarial targets, including ``ECCV 2018'', ``Remapping'', and ``Color strip'' in supplementary materials to increase the diversity of adversarial targets.

We generate 500 adversarial examples for Cityscapes and BDD100K datasets against both DRN and DLA segmentation models targeting on various malicious targets  (More results can be found in supplementary materials).   
\subsection{Spatial Consistency Analysis}
To evaluate the spatial consistency analysis quantitatively for segmentation task, we leverage it to build up a simple detector to demonstrate its property.
Here we perform patch based spatial consistency analysis, and we select patch size and region bound as $s=512$, $b_{low}=32, b_{upper} = 64$. We select the number of overlapping regions as $K \in \{1, 5, 10, 50\}$. 
Here we first select some benign instances, and calculate the normalize mIOU of overlapping regions from two random patches. We record the lower bound of theses mIOU as the threshold of the detection method.
Note that when reporting detection rate in the rest of the paper, we will use the threshold learned from a set of benign training data; while we also report Area Under Curve (AUC) of Receiver Operating Characteristic Curve (ROC) curve of a detection method to evaluate its overall performance.
Therefore, given an image, for each overlapping region of two random patches, we will calculate the normalize mIOU and compare with the threshold calculated before. If it is larger, the image is recognized as benign; vice versa. This process is illustrated in Algorithm~\ref{algo:consistency}. 
We report the detection results in terms of AUC in Table~\ref{tbl:city} for adversarial examples generated in various settings as mentioned above. 
We observed that such simple detection method based on spatial consistency information can achieve AUC as nearly 100\% for adversarial examples that we studied here.  
In addition, we also select $s$ with a random number between 384 to 512 (too small patch size will affect the segmentation accuracy even on benign instances, so we tend not to choose small patches on the purpose of control variable) and show the result in supplementary materials.
We observe that random patch sizes achieve similar detection result. 

\begin{table}[htb]
    \centering
    \begin{small}\resizebox{\linewidth}{!}{%
    
    \begin{tabular}{l|c|c|c|c|c|c|c|c|c|c|cc}
        \toprule
    \multicolumn{2}{c|}{\multirow{3}{*}{Method}} &
    \multicolumn{1}{c|}{\multirow{3}{*}{Model}} &
    \multirow{3}{*}{ mIOU} & 
    \multicolumn{4}{c|}{\shortstack{Detection} }& \multicolumn{4}{c}{\shortstack{Detection Adap }}\\ 
   \multicolumn{2}{c|}{} & & & \multicolumn{2}{c}{DAG} & \multicolumn{2}{c|}{Houdini} & \multicolumn{2}{c}{DAG} & 
    \multicolumn{2}{c}{Houdini} \\ 
    \multicolumn{2}{c|}{} & & &Pure&Kitty&Pure&Kitty&Pure&Kitty&Pure&Kitty\\ \midrule
   \multirow{6}{*}{\shortstack{Scale\\{(std)}}} & 0.5 & \multirow{3}{*}{\shortstack{DRN\\(16.4M)}} & \multirow{3}{*}{66.7} &  100\% & 95\% &  100\% & 99\% & 100\% & \redd{67\%} & 100\% & \redd{78\%}\\ 
     & 3.0 &  &  &  100\% & 100\% &  100\% & 100\% & 100\% & \redd{0\%} & 97\% & \redd{0\%}\\ 
   &5.0 &  &  &  100\% & 100\% &  100\% & 100\% & 100\% & \redd{0\%} & \redd{71\%} & \redd{0\%}\\ \cmidrule{2-12} 
    
    &0.5 & \multirow{3}{*}{\shortstack{DLA\\(18.1M)}} & \multirow{3}{*}{74.5} &  100\% & 98\% &  100\% & 100\% & 100\% & \redd{75\%} & 100\% & 81\%\\ 
    &3.0 &  &  &  100\% & 100\% &  100\% & 100\% & 100\% & \redd{24\%} & 100\% & \redd{34\%}\\ 
    &5.0 &  &  &  100\% & 100\% &  100\% & 100\% & 97\% & \redd{0\%} & 95\% & \redd{0\%}\\ 
    
    \midrule
    
   \multirow{8}{*}{\shortstack{Spatial\\(K)}}&  1 & \multirow{4}{*}{\shortstack{DRN\\(16.4M)} }& \multirow{4}{*}{66.7} &  91\% & 91\% &  94\% & 92\% & 98\% & 94\% & 92\% & 94\%\\ 
   &  5 &  &  & 100\% & 100\% & 100\% & 100\% & 100\% & 100\% &100\% & 100\%\\ 
   &  10 &  &  &  100\% & 100\% &  100\% & 100\% & 100\% & 100\% &100\% &100\%\\ 
   & 50 &  &  & 100\% & 100\% &  100\% & 100\% & 100\% & 100\% & 100\% & 100\%\\ \cmidrule{2-12} 
    
    & 1 & \multirow{4}{*}{\shortstack{DLA\\(18.1M)}} & \multirow{4}{*}{74.5} &  96\% & 98\% &  97\% & 97\% & 99\% & 99\% & 100\% & 100\%\\ 
   &  5 &  &  & 100\% & 100\% & 100\% & 100\% & 100\% & 100\% &100\% & 100\%\\ 
    & 10 &  &  & 100\% & 100\% &  100\% & 100\% & 100\% & 100\% &100\% &100\%\\ 
   &  50 &  &  &  100\% & 100\% &  100\% & 100\% & 100\% & 100\% & 100\% & 100\%\\

        \bottomrule
    \end{tabular} }
\caption{ {\small Detection results (AUC) of image spatial (Spatial) and scale consistency (Scale) based methods on Cityscapes dataset.
   The number in parentheses of the Model shows the number of parameters for the target mode, and mIOU shows the performance of segmentation model on pristine data. We color all the AUC less than 80\% with red.}}
    \label{tbl:city}
    
    \end{small}
\end{table}

\subsection{Image Scale Analysis}
As a baseline, we also utilize image scale information to perform as a simple detection method and compare it with the \spatialconsis.
We apply Gaussian kernel to perform the image scaling based detection, and select $\V{std}_{detect}\in \{0.5, 3,5\}$ as the standard deviation of Gaussian kernel. We compute the normalize mIOU between the original and scalled images. Similarly, the detection results of corresponding AUC are shown in Table~\ref{tbl:city}. 
It is demonstrated that detection method based on image scale information can achieve similarly high AUC compared with \spatialconsis. 

\subsection{Adaptive Attack Evaluation}
\begin{figure}[tbh]
    \centering
    \begin{minipage}{.35\textwidth}
     \begin{subfigure}{\textwidth}
     \centering
     \includegraphics[width=\textwidth]{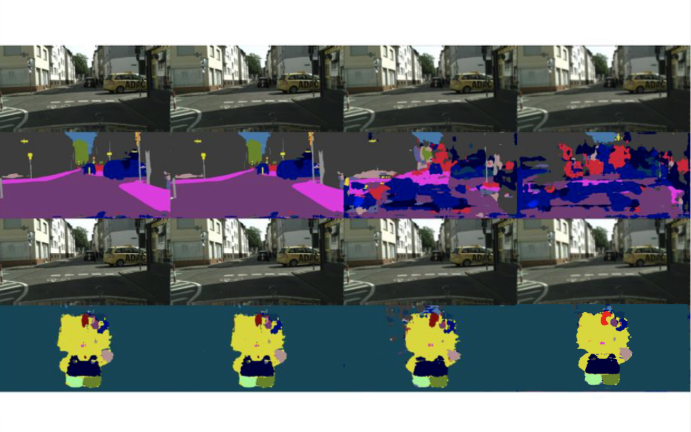}
     \caption{Image scaling}
     \end{subfigure}
    \end{minipage}
     \begin{minipage}{.31\textwidth}
     \begin{subfigure}{\textwidth}
     \centering
     \includegraphics[width=\textwidth]{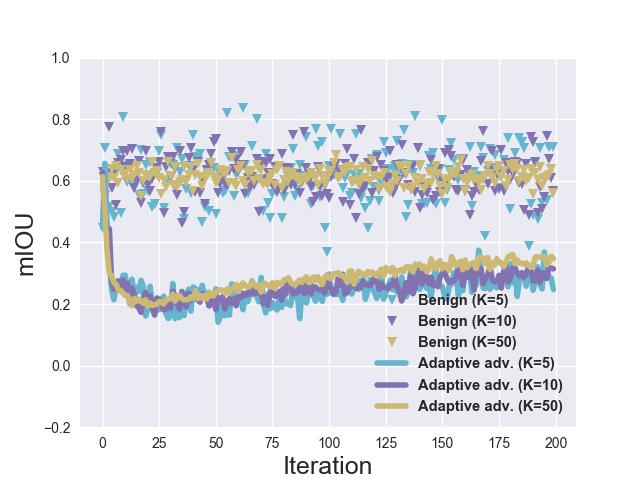}
     \caption{Convergence analysis}
     \label{fig:adap_iter}
     \end{subfigure}
    \end{minipage}
     \begin{minipage}{.31\textwidth}
     \begin{subfigure}{\textwidth}
     \centering
     \includegraphics[width=\textwidth]{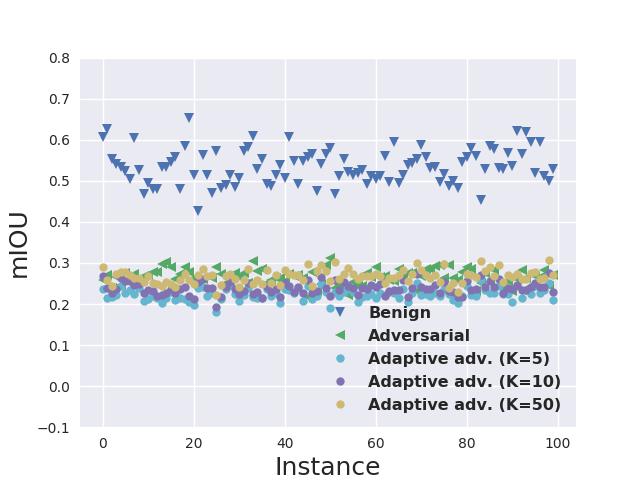}
     \caption{spatial consistency}
     \label{fig:adap_miou}
     \end{subfigure}
    \end{minipage}

    \caption{Performance of adaptive attack. (a) shows adversarial image and corresponding segmentation result for adaptive attack against image scaling. The first two rows show benign images and the corresponding segmentation results; the last two rows show the adaptive adversarial images and corresponding segmentation results under different std of Gaussian kernel (0.5, 3, 5 for column 2-4).  (b) and (c) show the performance of adaptive attack against spatial consistency based method with different $K$. (b) presents mIOU of overlapping regions for benign and adversarial images during along different iterations. (c) shows mIOU for overlapping regions of benign and adversarial instances at iteration 200. }
    \label{fig:adaptive}
\end{figure}
Regarding the above detection analysis, it is important to evaluate \emph{adaptive attacks}, where adversaries have knowledge of the detection strategy. 

As Carlini \& Wagner suggest~\cite{carlini2017adversarial}, we conduct attacks with full access to the detection model to evaluate the adaptive adversary based on Kerckhoffs principle \cite{shannon1949communication}.
To perform adaptive attack against the image scaling detection mechanism, instead of attacking the original model, we 
add another convolutional layer after the input layer of the target model similarly with~\cite{carlini2017adversarial}. We select std $\in \{0.5, 3, 5\}$ to apply adaptive attack, which is the same with the detection model.
To guarantee that the attack methods will converge, when performing the adaptive attacks, we select 0.06 for the upper bound for adversarial perturbation, in terms of $L_2$ distance (pixel values are in range [0,1]), since larger than that the perturbation is already very visible.
The detection results against such adaptive attacks are shown in Table~\ref{tbl:city} on Cityscapes (We omit the results on BDD to supplementary materials). 
Results on adaptive attack show that the image scale based detection method is easily to be attacked (AUC of detection drops dramatically), which draws similar conclusions as in classification task~\cite{carlini2017adversarial}.
We show the qualitative results in Figure~\ref{fig:adaptive} (a), and it is obvious that even under large std of Gaussian kernel, the adversarial example can still be fooled into the malicious target (Kitty).

\begin{figure}[tbh]
\centering
\begin{minipage}{.45\textwidth}
 \begin{subfigure}{\textwidth}
 \centering
 \includegraphics[width=\textwidth]{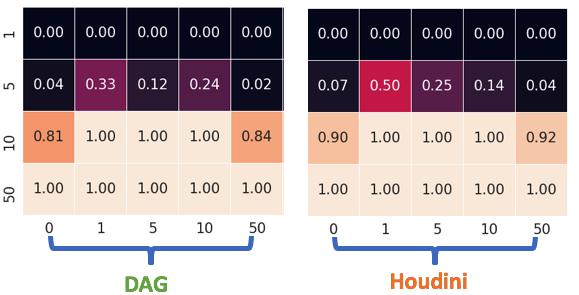}
 \caption{Kitty }
 \label{fig:attention-a}
 \end{subfigure}
\end{minipage}
\begin{minipage}{.51\textwidth}
 \begin{subfigure}{\textwidth}
 \centering
 \includegraphics[width=\textwidth]{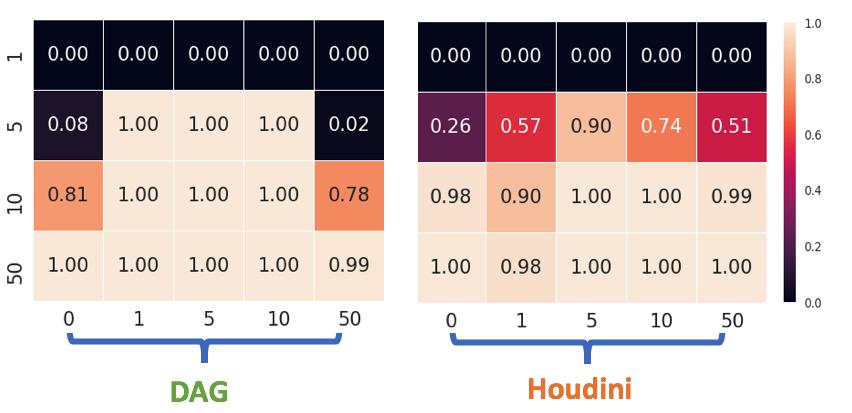}
 \caption{Pure }
 \end{subfigure}
\end{minipage}
\caption{Detection performance of \spatialconsis against adaptive attack with different $K$ on Cityscapes with DRN model. X-axis indicates the number of patches selected to perform the adaptive attack (0 means regular attack). Y-axis indicates the number of overlapping regions selected for during detection.}
\label{fig:adapconfusion}
\end{figure}

Next, we will apply adaptive attack against the \spatialconsis. Due to the randomness of the approach, we propose to develop a strong adaptive adversary that we can think of by randomly select $K$ patches (the same value of $K$ used by defender). Then the adversary will try to attack both the whole image and the selected $K$ patches to the corresponding part of malicious target. The detailed attack algorithm is shown in the supplementry materials. 
The corresponding detection results of the \spatialconsis against such adaptive attacks on Cityscapes are shown in Table~\ref{tbl:city}. 
It is interesting to see that even against such strong adaptive attacks, the \spatialconsis can still achieve nearly 100\% detection results. We hypothesize that it is because of the high dimension randomness induced by the \spatialconsis since the search space for patches and the overlapping regions is pretty high. 
Figure~\ref{fig:adaptive} (b) analyzes the convergence of such adaptive attack against \spatialconsis.
From figure~\ref{fig:adaptive} (b) and (c), we can see that with different $K$, the selected overlapping regions still remain inconsistent with high probability.  

Since the \spatialconsis can induce large randomness, 
we generate a confusion matrix of detection results for adversaries and detection method choosing various $K$ as shown in Figure~\ref{fig:adapconfusion}.
It is clear that for different malicious targets and attack methods, choosing $K=50$ is already sufficient to detect sophisticated attacks. 
In addition, based on our empirical observation, attacking with higher $K$ increases the computation complexity of adversaries dramatically. 

\begin{figure}[htb]
    \centering
    \begin{minipage}{0.4\textwidth}
     \begin{subfigure}{\textwidth}
     \centering
     \includegraphics[width=\textwidth]{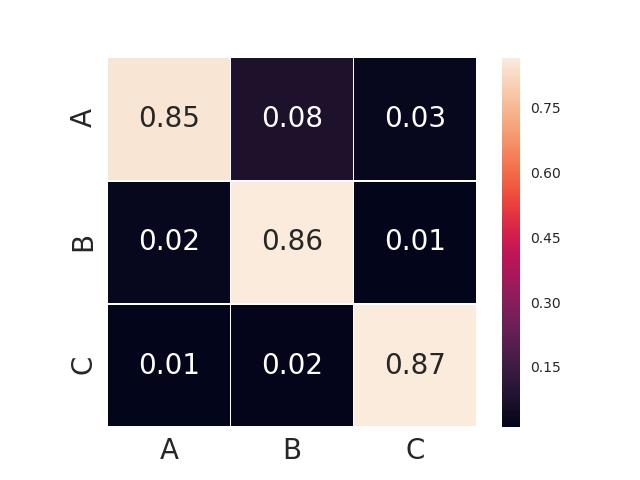}
     \caption{DAG}
     \label{fig:attention-a}
     \end{subfigure}
    \end{minipage}
    \begin{minipage}{0.4\textwidth}
     \begin{subfigure}{\textwidth}
     \centering
     \includegraphics[width=\textwidth]{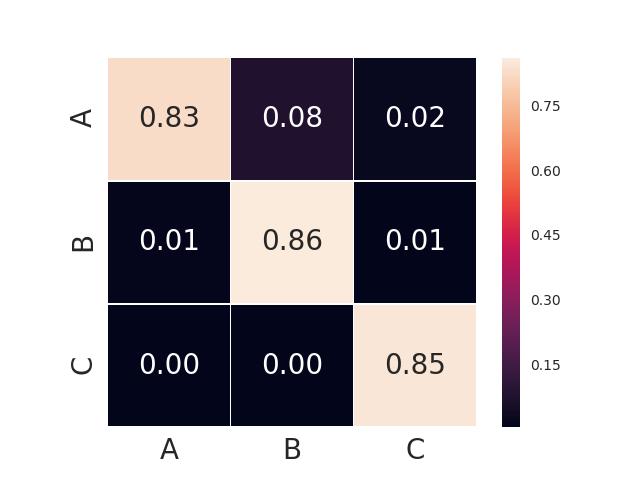}
     \caption{Houdini }
     \label{fig:attention-a}
     \end{subfigure}
    \end{minipage}
    \caption{Transferability analysis: cell $(i,j)$ shows the normalized mIoU value or pixel-wise attack success rate of adversarial examples generated against model $j$ and evaluate on model $i$. Model A,B,C are DRN (DRN-D-22) with different initialization. We select ``Hello Kitty'' as target}
    \label{fig:transfer}
\end{figure}

\subsection{Transferability Analysis}
Given the common properties of adversarial examples for both classifier and segmentation tasks, next we will analyze whether transferability of adversarial examples exists in segmentation models considering they are particularly sensitive to spatial and scale information.
\emph{Transferability} is demonstrated to be one of the most interesting properties of adversarial examples in classification task, where adversarial examples generated against one model is able to mislead the other model, even if the two models are of different architectures. Given this property, transferability has become
the foundation of a lot of black-box attacks in classification task. Here we aim to analyze whether adversarial examples in segmentation task still retain high transferability. 
First, we train three DRN models with the same architecture (DRN-D-22) but different initialization and generate adversarial images with the same target.

Each adversarial image has at least 96\% pixel-wise attack success rate against the original model.
We evaluate both the DAG and Houdini attacks and 
evaluate the transferability using normalized mIoU excluding pixels with the same label for the ground truth adversarial target. 
We show the transferability evaluation among different models in the confusion matrices in Figure~\ref{fig:transfer}\footnote{Since the prediction of certain classes presents low IoU value due to imperfect segmentation, we eliminate K classes with the lowest IoU values to avoid side effects. In our experiments, we set K to be 13. }. We observe that the transferability rarely appears in the segmentation task. More results on different network architectures and data sets are in the supplementary materials.

As comparison with classification task, for each network architecture we train a classifier on it and evaluate the transferability results as shown in supplementary materials. As a control experiments, we observe that classifiers with the same architecture still have high transferability aligned with existing findings, which shows that the low transferability is indeed due to the natural of segmentation instead of certain network architectures.

This observation here is quite interesting, which indicates that black-box attacks against segmentation models may be more challenging. Furthermore, the reason for such low transferability in segmentation is possibly because adversarial perturbation added to one image could have focused on a certain region, while such spatial context information is captured differently among different models. We plan to analyze the actual reason for low transferability in segmentation in the future work.

\section{Conclusions}
Adversarial examples have been heavily studied recently, pointing out vulnerabilities of deep neural networks and raising a lot of security concerns. However, most of such studies are focusing on image classification problems, and in this paper we aim to explore the spatial context information used in semantic segmentation task to better understand adversarial examples in segmentation scenarios. We propose to apply spatial consistency information analysis to recognize adversarial examples in segmentation, which has not been considered in either image classification or segmentation as a potential detection mechanism.
We show that such spatial consistency information is different for adversarial and benign instances and can be potentially leveraged to detect adversarial examples even when facing strong adaptive attackers. These observations open a wide door for future research to explore diverse properties of adversarial examples under various scenarios and develop new attacks to understand the vulnerabilities of DNNs.

\subsubsection*{Acknowledgments}
We thank Warren He, George Philipp, Ziwei Liu, Zhirong Wu, Shizhan Zhu and Xiaoxiao Li for their valuable discussions
on this work. This work was supported in part by Berkeley DeepDrive, Compute Canada, NSERC and National Science Foundation under grants CNS-1422211, CNS-1616575, CNS-1739517, JD Grapevine plan, and by the DHS via contract number FA8750-18-2-0011.  
\clearpage
\bibliographystyle{splncs04}
\bibliography{egbib}
\clearpage
\input{appendix_cr.tex}
\end{document}

%% file: def.tex
\newcommand{\Spatialconsis}{Spatial Consistency Based Method\xspace}
\newcommand{\spatialconsis}{spatial consistency based method\xspace}

%% file: appendix_cr.tex
\appendix
{
\setlength\parindent{0pt}
\textbf{\Large Appendix}
}
\section{Adversarial Examples For Cityscapes and BDD Datasets Against DRN and DLA models}
\begin{figure}[h!]
    \centering
    \begin{minipage}{.45\textwidth}
     \begin{subfigure}{\textwidth}
     \centering
     \includegraphics[width=\textwidth]{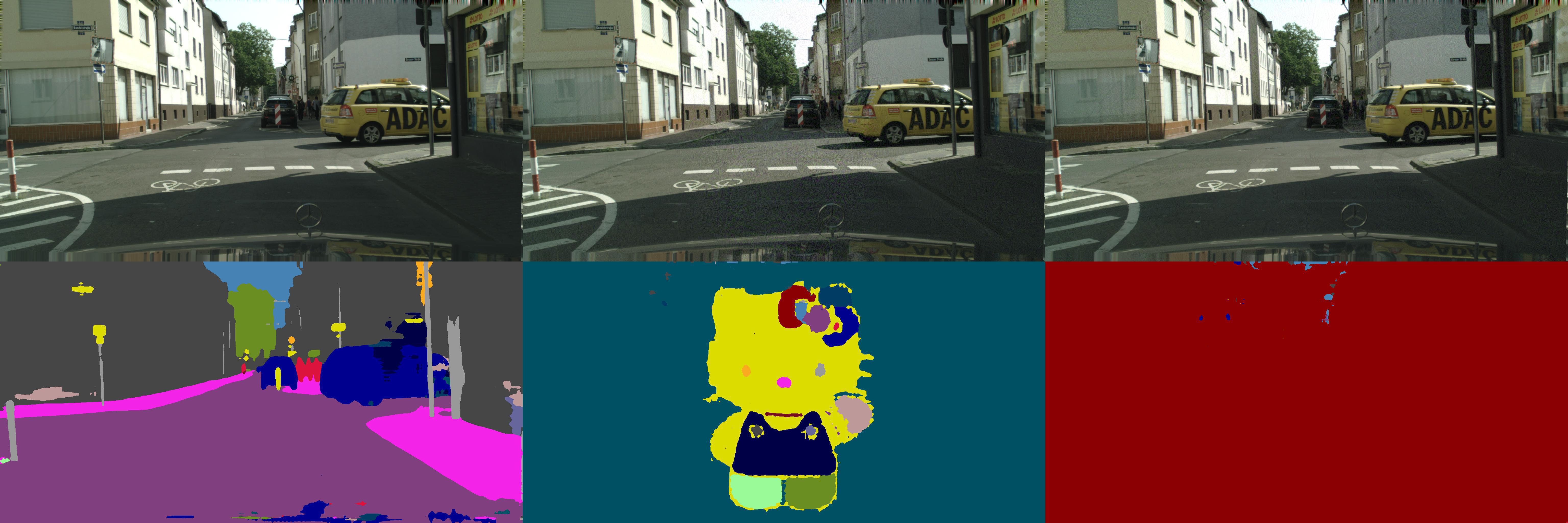}
     \caption{DAG | DRN | Cityscapes }
     \label{fig:attention-a}
     \end{subfigure}
    \end{minipage}
    \begin{minipage}{.40\textwidth}
     \begin{subfigure}{\textwidth}
     \centering
     \includegraphics[width=\textwidth]{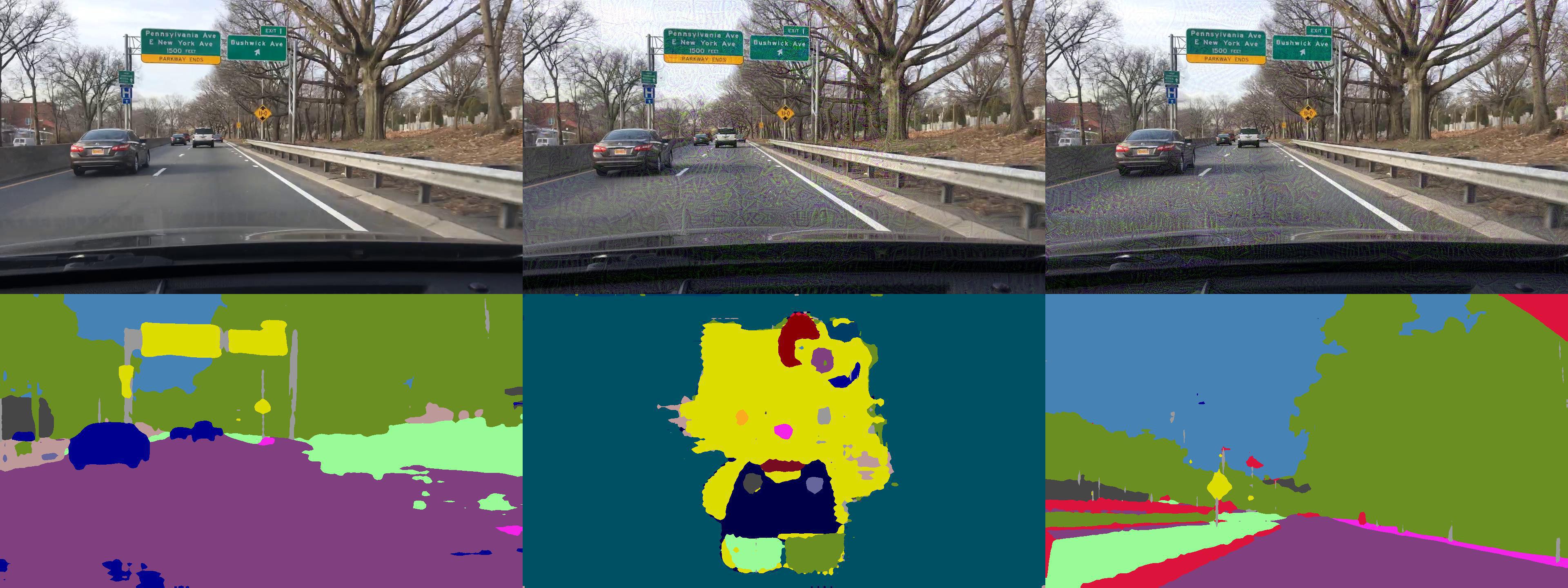}
     \caption{DAG | DRN | BDD }
     \label{fig:attention-a}
     \end{subfigure}
    \end{minipage}
    \begin{minipage}{.45\textwidth}
     \begin{subfigure}{\textwidth}
     \centering
     \includegraphics[width=\textwidth]{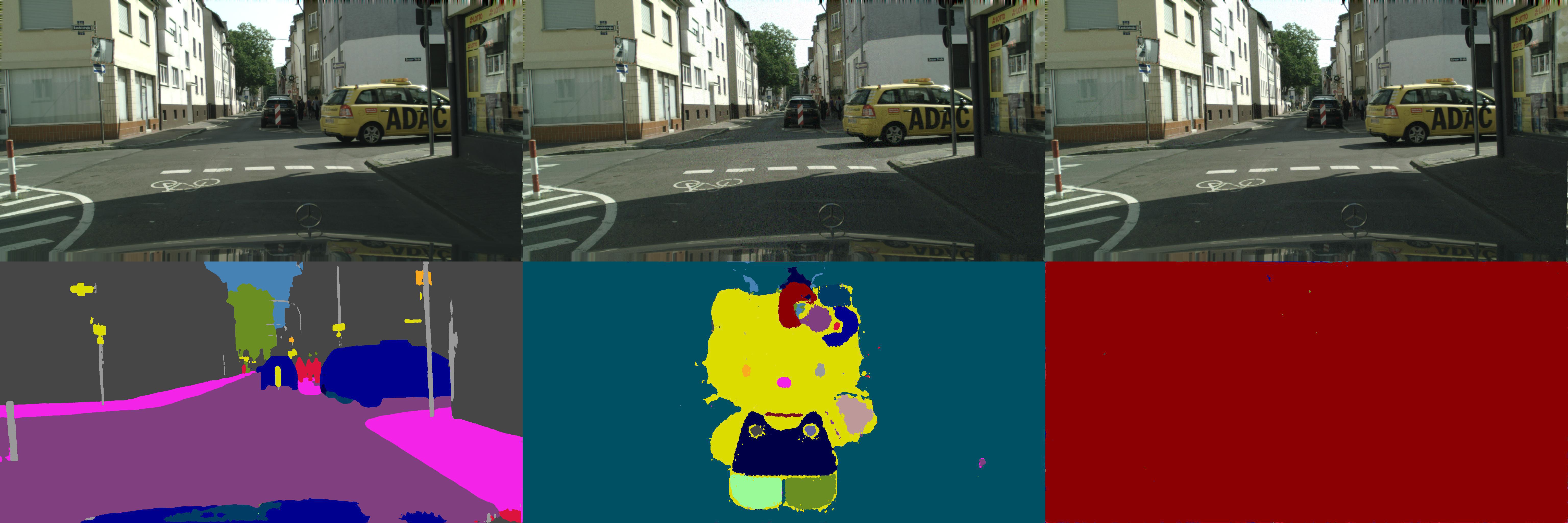}
     \caption{DAG | DLA | Cityscapes }
     \label{fig:attention-a}
     \end{subfigure}
    \end{minipage}
    \begin{minipage}{.40\textwidth}
     \begin{subfigure}{\textwidth}
     \centering
     \includegraphics[width=\textwidth]{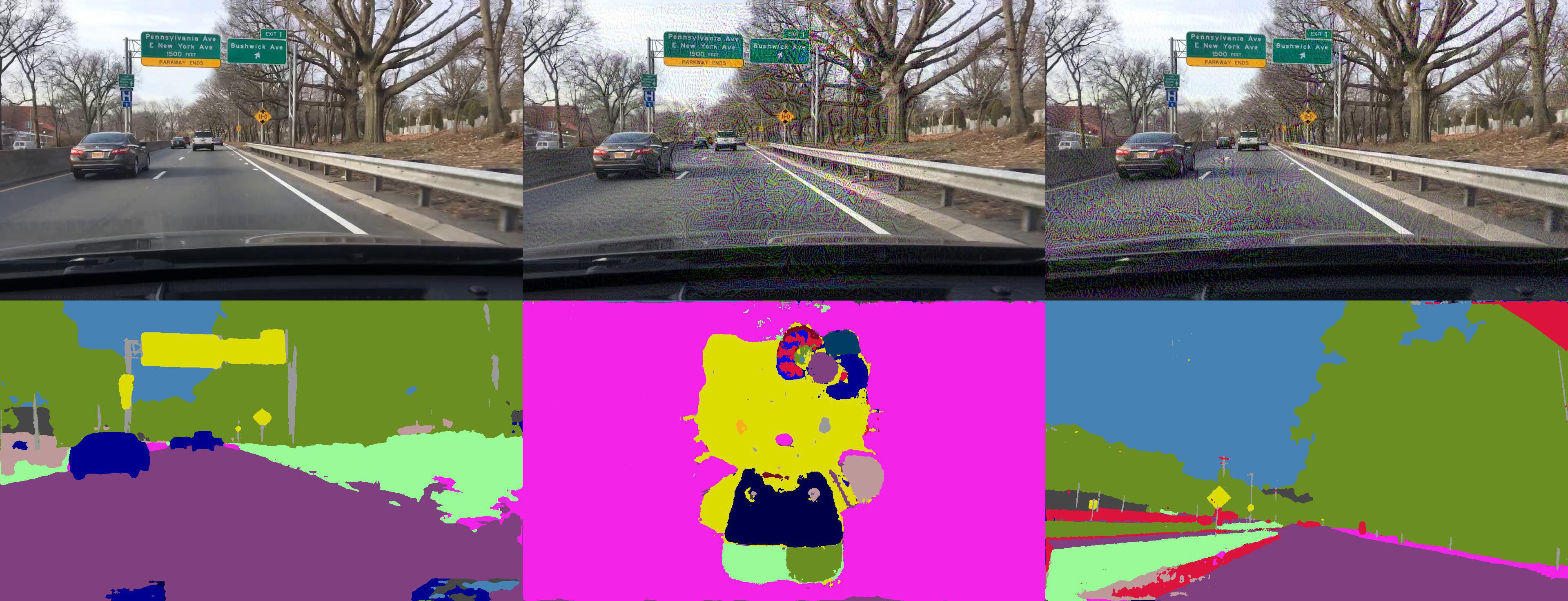}
     \caption{DAG | DLA | BDD }
     \label{fig:attention-a}
     \end{subfigure}
    \end{minipage}
    \begin{minipage}{.45\textwidth}
     \begin{subfigure}{\textwidth}
     \centering
     \includegraphics[width=\textwidth]{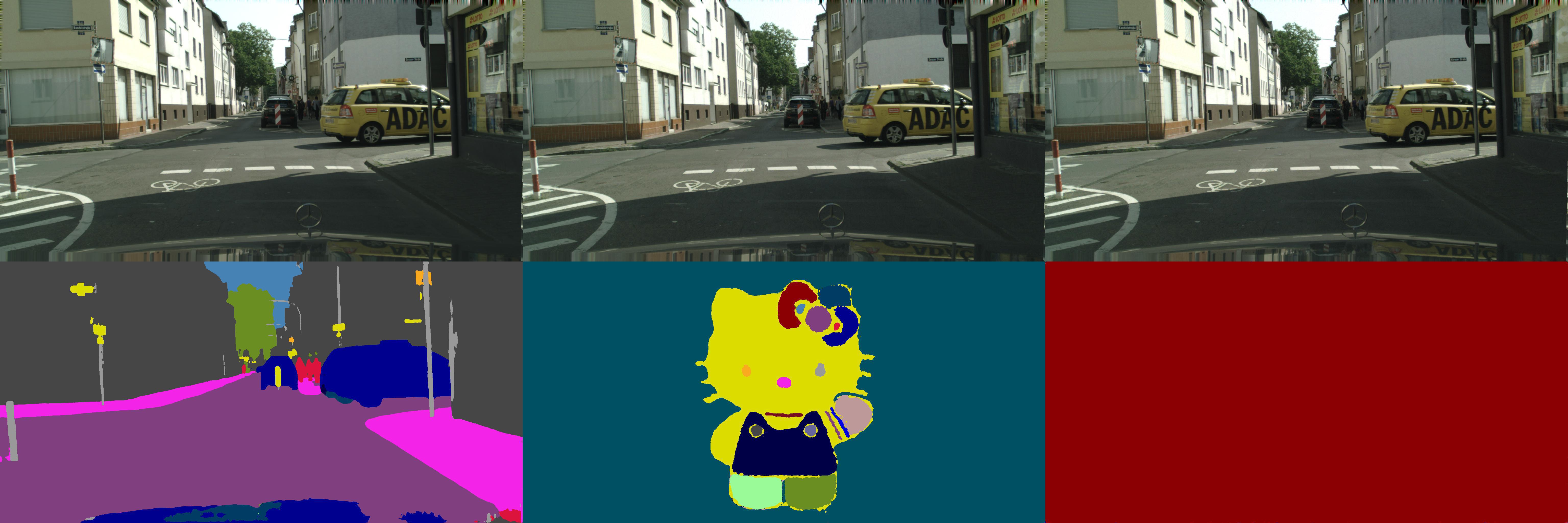}
     \caption{Houdni | DLA | CityScapes }
     \label{fig:attention-a}
     \end{subfigure}
    \end{minipage}
    \begin{minipage}{.40\textwidth}
     \begin{subfigure}{\textwidth}
     \centering
     \includegraphics[width=\textwidth]{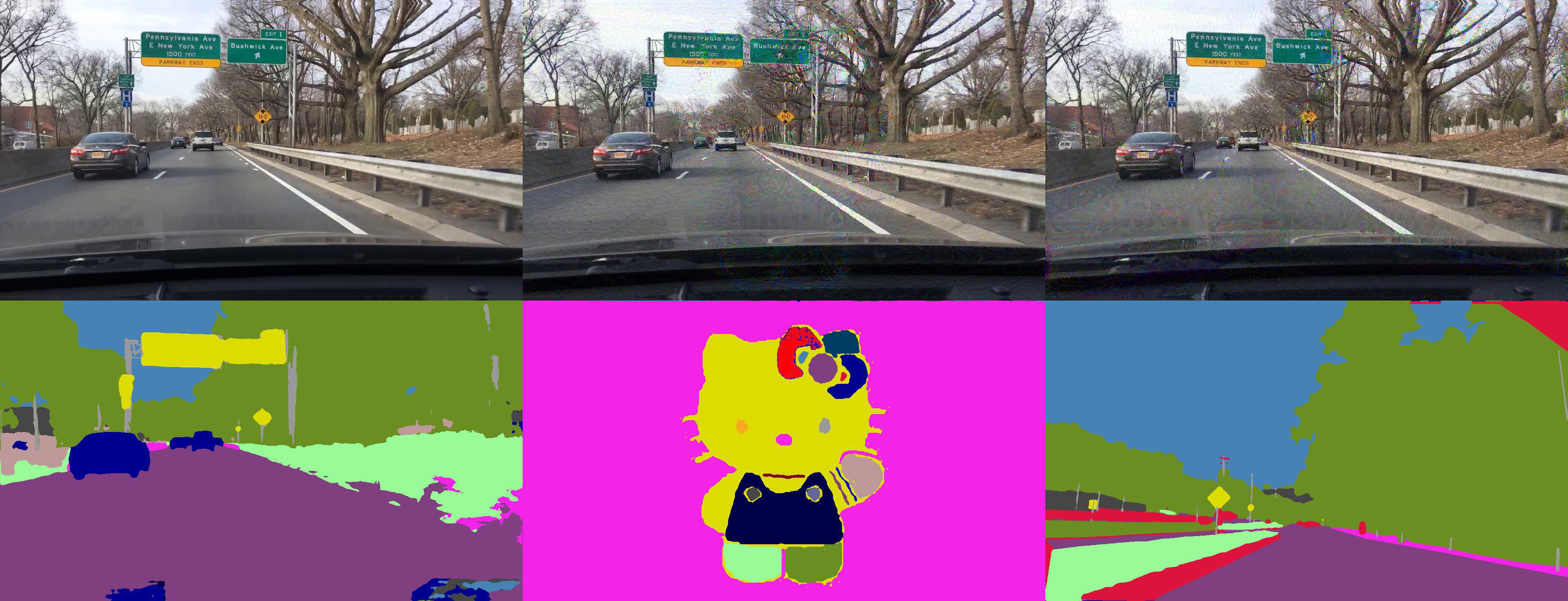}
     \caption{Houdini | DLA | BDD }
     \label{fig:attention-a}
     \end{subfigure}
    \end{minipage}
    \caption{Samples of benign and adversarial examples. We use the format ``attack method |attack model | dataset'' to label the settings of each adversarial examples. Within each subfigure, the first column shows benign images and corresponding segmentation results, the second and third columns show adversarial examples with different adversarial targets (targeting on Kitty/Pure in (a),(c), (d) and on Kitty and Scene in (b),(d),(f)).
    }
    \label{fig:all_examples}
\end{figure}
Figure~\ref{fig:all_examples} shows the benign and adversarial examples targeting at diverse adversarial targets: ``Hello Kitty'' (Kitty) and random pure color (Pure) on Cityscapes~\cite{cordts2016cityscapes}; and ``Hello Kitty'' (kitty) and a real scene without any cars (Scene) on BDD~\cite{xu2017end} dataset against DRN~\cite{Yu2017} and DLA~\cite{yu2017deep} segmentation models. In order to increase the diversity of our target set, we also apply different colors for the background of ``Hello Kitty'' on BDD dataset against DLA model. 
\begin{figure}[t]
    \centering
     \centering
     \includegraphics[width=\textwidth]{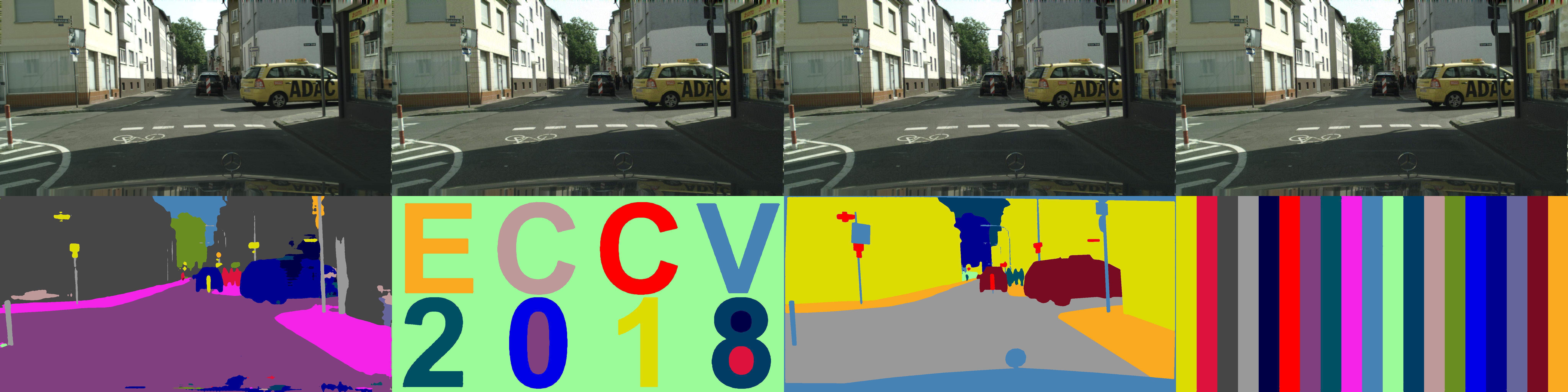}
    \caption{
    Attack results of additional targets on Cityscapes. The first column shows benign instance, while 2-4 columns show adversarial examples with target ``ECCV 2018'', ``Remapping'', and ``Color strip'', respectively.
    }
    \label{fig:new-target}
\end{figure}

Figure~\ref{fig:new-target} shows the additional adversarial targets, including ``ECCV 2018'', ``Remapping'', and ``Color strip''.
Here remapping means we generate an adversarial target by shifting the numerical label of each class in the ground truth by a constant offset. This way, we can guarantee that each target has no overlap with the ground truth mask.
For ``Color strip", we divide the target into 19 strips evenly, each of which is filled with a class label, aiming to mitigate possible bias for different classes.


\section{Spatial Consistency Based Method}
\begin{algorithm}[h!]
\SetAlgoLined
\SetKwInOut{Input}{input}\SetKwInOut{Output}{output}
\SetKwInput{Ret}{Return}
\tabcolsep=0pt
\nonl \begin{tabular}{@{}ll}
    \Input{} & patch size $s$ \\
     &image width $w$\\
    &image height $h$ \\
    &bound $b_{\V{low}}, b_{\V{upper}}$\\
    \Output{}& Two random patches $P_1$ and $P_2$: $(u_1, u_2, u_3, u_4), (v_1, v_2, v_3, v_4)$\\
\end{tabular}

\nl Generate two random integer numbers $I_1, I_2$, where $0<I_1<w-s-b_{\V{upper}}, 0<I_2<h-s-b_{\V{upper}}$. \;
 Generate two random integer numbers $I_3, I_4$, where $b_{\V{low}} <I_3, I_4 < b_{\V{upper}}$. \;
 \Ret{$( I_1, I_2,I_1+s, I_2+s ), (I_1 + I_3, I_2 + I_4,I_1 + I_3 + s, I_2+I_4+s)$ }
 \caption{Patch Selection Algorithm ($\V{getOverlapPatches}$)}
\label{algo:patchselection}
\end{algorithm}

\begin{table}[h]
    \centering
    \begin{small}
    \begin{tabular}{l|c|c|c|c|c|c|c|c|c|c}
        \toprule
    \multirow{3}{*}{Method} &
    \multirow{3}{*}{Model} &
    \multirow{3}{*}{ mIOU} &
    \multicolumn{4}{c|}{\shortstack{Detection} }& \multicolumn{4}{c}{\shortstack{Detection Adap }}\\ \cline{4-11}
    & & &  \multicolumn{2}{c}{DAG} & \multicolumn{2}{|c|}{Houdini} & \multicolumn{2}{c|}{DAG} & 
    \multicolumn{2}{c}{Houdini} \\ \cline{4-11}
    & & &Scene&Kitty&Scene&Kitty&Scene&Kitty&Scene&Kitty\\ \hline
    Scale (std=0.5) & \multirow{3}{*}{\shortstack{DRN\\(16.4M)}} & \multirow{3}{*}{54.5} &  96\% & 100\% &  99\% & 100\% & \redd{69\%} & 89\% & \redd{46\%} & 91\%\\ \cline{4-11}
    Scale (std=3.0) &  &  & 100\% & 100\% &  100\% & 100\% & \redd{31}\% & 89\% & \redd{1\%} & \redd{48\%}\\ \cline{4-11}
    Scale  (std=5.0) &  &  & 100\% & 100\% &  100\% & 100\% & \redd{8\%} & 84\% & \redd{0\%} & \redd{36\%}\\ \hline
    
    Scale (std=0.5) & \multirow{3}{*}{\shortstack{DLA\\(18.1M)}} & \multirow{3}{*}{46.29} & 96\% & 88\% &  99\% & 99\% & 89\% & 90\% & 80\% & \redd{58\%}\\ \cline{4-11}
    Scale (std=3.0) &  &  & 100\% & 100\% &  100\% & 100\% & \redd{66\%} & 88\% & \redd{11\%} & \redd{26\%}\\ \cline{4-11}
    Scale (std=5.0) &  &  & 98\% & 100\% &  99\% & 100\% & \redd{32\%} & \redd{78\%} & \redd{2\%} & \redd{12\%}\\ \hline \hline
    
    Spatial (K=1) & \multirow{4}{*}{\shortstack{DRN\\(16.4M)} }& \multirow{4}{*}{54.5} &  98\% & 100\% &  99\% & 99\% & 89\% & 99\% & 89\% & 99\%\\ \cline{4-11}
    Spatial (K=5) &  &   & 100\% & 100\% & 100\% & 100\% & 100\% & 100\% & 100\% & 100\%\\ \cline{4-11}
    Spatial (K=10) &  &   & 100\% & 100\% &  100\% & 100\% & 100\% & 100\% &100\% &100\%\\ \cline{4-11}
    Spatial (K=50) &  &  & 100\% & 100\% &  100\% & 100\% & 99\% & 100\% & 99\% & 100\%\\ \hline
    
    Spatial (K=1) & \multirow{4}{*}{\shortstack{DLA\\(18.1M)}} & \multirow{4}{*}{46.29} &   98\% & 99\% &  95\% & 95\% & 96\% & 99\% & 98\% & 95\%\\ \cline{4-11}
    Spatial (K=5) &  &  &  100\% & 100\% & 98\% & 98\% & 99\% & 100\% &99\% & 96\%\\ \cline{4-11}
    Spatial (K=10) &  &  &  100\% & 100\% &  99\% & 99\% & 99\% & 100\% &99\% &96\%\\ \cline{4-11}
    Spatial (K=50) &  &  &  100\% & 100\% &  99\% & 99\% & 100\% & 100\% & 99\% & 93\%\\ \hline

        \bottomrule
    \end{tabular}
    \caption{ Detection results (AUC) of image spatial (Spatial) and scale consistency (Scale) based methods on BDD dataset.
   The number in parentheses of the ``Model'' shows the number of parameters for the target mode, and ``mIOU'' shows the performance of segmentation model on pristine data. We color all the AUC less than 80\% with red.} 
    \label{tbl:bdd} 
    
    \end{small}
\end{table}
\begin{table}[h]
    \centering 
    \begin{small}
    \resizebox{\linewidth}{!}{%
    \begin{tabular}{l|c|c|c|c|c|c|c|c|c|c}
        \toprule
    \multirow{3}{*}{ \shortstack{Method \\(Spatial)}} &
    \multirow{3}{*}{Model} &
    \multirow{3}{*}{ mIOU} & 
    \multicolumn{4}{c|}{\shortstack{Detection} }& \multicolumn{4}{c}{\shortstack{Detection Adap }}\\ \cline{4-11}
    & & & \multicolumn{2}{c}{DAG} & \multicolumn{2}{|c|}{Houdini} & \multicolumn{2}{c|}{DAG} & 
    \multicolumn{2}{c}{Houdini} \\ \cline{4-11}
    & & &Pure&Kitty&Pure&Kitty&Pure&Kitty&Pure&Kitty\\ \hline
    
     K=1 & \multirow{4}{*}{\shortstack{DRN\\(16.4M)} }& \multirow{4}{*}{66.7} &  $91\pm0.1$\% & $88\pm0.1$\% &$91\pm 0.3$\% & $90\pm0.1$\% & $97\pm0$\% & $92\pm0.1 $\% & $90\pm2.5$\% & $93\pm0.1$\%\\ \cline{4-11}
    K=5 &  &  &$99\pm0.1$\% & $99\pm0.1$\% & $100\pm0.0$\% & $99\pm0.1$\% &$100\pm0$\% & $100\pm0$\% & $100\pm0$\% & $100\pm0$\%\\ \cline{4-11}
    K=10 &  &  &  $100\pm0$\% & $100\pm0$\% &  $100\pm0$\% & $100\pm0$\% & $100\pm0$\% & $100\pm0$\% & $100\pm0$\% &$100\pm0$\%\\ \cline{4-11}
    K=50 &  &  & $100\pm0$\% & $100\pm0$\% & $100\pm0$\% & $100\pm0$\% & $100\pm0$\% & $100\pm0$\% & $100\pm0$\% & $100\pm0$\%\\ \hline

    K=1 & \multirow{4}{*}{\shortstack{DLA\\(18.1M)}} & \multirow{4}{*}{74.5} &  $96\pm0.1$\% & $98\pm0.1$\% &  $96\pm0.1$\% & $96\pm0.1$\% & $99\pm0.3$\% & $99\pm0.1$\% & $98\pm0.4$\% & $99\pm0.1$\%\\ \cline{4-11}
    K=5 &  &  & $100\pm0$\% & $100\pm0$\% & $100\pm0$\% & $100\pm0$\% & $100\pm0$\% & $100\pm0$\% &$100\pm0$\% & $100\pm0$\%\\ \cline{4-11}
    K=10 &  &  &  $100\pm0$\% & $100\pm0$\% &  $100\pm0$\% & $100\pm0$\% & $100\pm0$\% & $100\pm0$\% & $100\pm0$\% &$100\pm0$\%\\ \cline{4-11}
    K=50 &  &  & $100\pm0$\% & $100\pm0$\% & $100\pm0$\% & $100\pm0$\% & $100\pm0$\% & $100\pm0$\% & $100\pm0$\% & $100\pm0$\%\\ \hline 
        \bottomrule
    \end{tabular} }
    \caption{ Detection results (AUC) of image spatial (Spatial) based method with random patch size on Cityscapes dataset.
   }
    \label{tbl:city-random}

    \end{small}
\end{table}
\begin{table}[h]
    \centering 
    \begin{small}
    
\resizebox{\linewidth}{!}{%
    \begin{tabular}{l|c|c|c|c|c|c|c|c|c|c}
        \toprule
    \multirow{3}{*}{ \shortstack{Method \\(Spatial)}} &
    \multirow{3}{*}{Model} &
    \multirow{3}{*}{ mIOU} & 
    \multicolumn{4}{c|}{\shortstack{Detection} }& \multicolumn{4}{c}{\shortstack{Detection Adap }}\\ \cline{4-11}
    & & & \multicolumn{2}{c}{DAG} & \multicolumn{2}{|c|}{Houdini} & \multicolumn{2}{c|}{DAG} & 
    \multicolumn{2}{c}{Houdini} \\ \cline{4-11}
    & & &Pure&Kitty&Pure&Kitty&Pure&Kitty&Pure&Kitty\\ \hline
    
     K=1 & \multirow{4}{*}{\shortstack{DRN\\(16.4M)} }& \multirow{4}{*}{54.5} &  $91\pm0.1$\% & $88\pm0.1$\% &$91\pm 0.3$\% & $90\pm0.1$\% & $97\pm0$\% & $92\pm0.1 $\% & $90\pm2.5$\% & $93\pm0.1$\%\\ \cline{4-11}
    K=5 &  &  &$99\pm0.1$\% & $99\pm0.1$\% & $100\pm0.0$\% & $99\pm0.1$\% &$100\pm0$\% & $100\pm0$\% & $100\pm0$\% & $100\pm0$\%\\ \cline{4-11}
    K=10 &  &  &  $100\pm0$\% & $100\pm0$\% &  $100\pm0$\% & $100\pm0$\% & $100\pm0$\% & $100\pm0$\% & $100\pm0$\% &$100\pm0$\%\\ \cline{4-11}
    K=50 &  &  & $100\pm0$\% & $100\pm0$\% & $100\pm0$\% & $100\pm0$\% & $100\pm0$\% & $100\pm0$\% & $100\pm0$\% & $100\pm0$\%\\ \hline

    K=1 & \multirow{4}{*}{\shortstack{DLA\\(18.1M)}} & \multirow{4}{*}{46.29} &  $98\pm0$\% & $96\pm0.1$\% &  $95\pm0.1$\% & $92\pm0$\% & $98\pm0.1$\% & $94\pm0$\% & $97\pm0.1$\% & $90\pm0$\%\\ \cline{4-11}
    K=5 &  &  & $100\pm0$\% & $99\pm0$\% & $99\pm0$\% & $98\pm0$\% & $99\pm0$\% & $98\pm0$\% &$99\pm0$\% & $90\pm0$\%\\ \cline{4-11}
    K=10 &  &  &  $100\pm0$\% & $100\pm0$\% &  $100\pm0$\% & $98\pm0$\% & $100\pm0$\% & $99\pm0$\% & $100\pm0$\% &$96\pm0$\%\\ \cline{4-11}
    K=50 &  &  & $100\pm0$\% & $100\pm0$\% & $100\pm0$\% & $99\pm0$\% & $100\pm0$\% & $100\pm0$\% & $100\pm0$\% & $100\pm0$\%\\ \hline 
        \bottomrule
    \end{tabular} }
    \caption{ Detection results (AUC) of image spatial (Spatial) based method with random patch size on BDD dataset.
   }
    \label{tbl:bdd-random}
    \end{small}
\end{table}
\begin{table}[h]
    \centering 
    \begin{small}
        \noindent
\resizebox{\linewidth}{!}{%
    \begin{tabular}{l|c|c|c|c|c|c|c|c|c|c|c|c|c|c}
        \toprule
    \multirow{3}{*}{ \shortstack{Method\\(Spatial)}} &
    \multirow{3}{*}{Model} &
    \multirow{3}{*}{ mIOU} & 
    \multicolumn{6}{c|}{\shortstack{Detection} }& \multicolumn{6}{c}{\shortstack{Detection Adap }}\\ \cline{4-15}
    & & & \multicolumn{3}{c}{DAG} & \multicolumn{3}{|c|}{Houdini} & \multicolumn{3}{c|}{DAG} & 
    \multicolumn{3}{c}{Houdini} \\ \cline{4-15}
    & & &ECCV&Remap&Strip&ECCV&Remap&Strip&ECCV&Remap&Strip&ECCV&Remap&Strip\\ \hline
    
     K=1 & \multirow{4}{*}{\shortstack{DRN\\(16.4M)} }& \multirow{4}{*}{66.7} &  93\% & 91\% &  91\% & 91\% & 91\% & 91\% & 90\% & 92\% &  89\% & 90\% & 92\% & 90\%\\ \cline{4-15}
    K=5 &  &  & 99\% & 100\% & 99\% &  99\% & 100\% & 99\% & 100\% & 99\% & 100\% & 100\% & 100\%& 99\%\\ \cline{4-15}
    K=10 &  &  &  100\% & 100\% &  100\% & 100\% & 100\% & 100\% & 100\% &100\%&100\%&100\%&100\%&99\%\\ \cline{4-15}
    K=50 &  &  & 100\% & 100\% & 100\% & 100\% & 100\% & 100\% & 100\% & 100\%& 100\%& 100\%& 100\%& 100\%\\ \hline

     K=1 & \multirow{4}{*}{\shortstack{DLA\\(18.1M)} }& \multirow{4}{*}{74.5} &  96\% & 99\% &  97\% & 95\% & 97\% & 96\% & 99\% & 99\% &  98\% & 99\% & 99\% & 98\%\\ \cline{4-15}
    K=5 &  &  & 100\% & 100\% & 100\% &  100\% & 100\% & 100\% & 100\% & 100\% & 100\% & 100\% & 100\%& 100\%\\ \cline{4-15}
    K=10 &  &  &  100\% & 100\% &  100\% & 100\% & 100\% & 100\% & 100\% &100\%&100\%&100\%&100\%&99\%\\ \cline{4-15}
    K=50 &  &  & 100\% & 100\% & 100\% & 100\% & 100\% & 100\% & 100\% & 100\%& 100\%& 100\%& 100\%& 100\%\\ \hline 
   
        \bottomrule
    \end{tabular}
    }
    \caption{ Detection results (AUC) of  spatial consistency (Spatial) based method on Cityscapes dataset for additional targets.
   }
    \label{tbl:city-additional}
    
    \end{small}
\end{table}
\begin{table}[h]
    \centering 
    \begin{small}
        \noindent
\resizebox{\linewidth}{!}{%
    \begin{tabular}{l|c|c|c|c|c|c|c|c|c|c|c|c|c|c}
        \toprule
    \multirow{3}{*}{ \shortstack{Method\\(Spatial)}} &
    \multirow{3}{*}{Model} &
    \multirow{3}{*}{ mIOU} & 
    \multicolumn{6}{c|}{\shortstack{Detection} }& \multicolumn{6}{c}{\shortstack{Detection Adap }}\\ \cline{4-15}
    & & & \multicolumn{3}{c}{DAG} & \multicolumn{3}{|c|}{Houdini} & \multicolumn{3}{c|}{DAG} & 
    \multicolumn{3}{c}{Houdini} \\ \cline{4-15}
    & & &ECCV&Remap&Strip&ECCV&Remap&Strip&ECCV&Remap&Strip&ECCV&Remap&Strip\\ \hline
    
     K=1 & \multirow{4}{*}{\shortstack{DRN\\(16.4M)} }& \multirow{4}{*}{54.5} &  99\% & 99\% &  99\% & 99\% & 99\% & 99\% & 99\% & 99\% &  99\% & 99\% & 98\% & 97\%\\ \cline{4-15}
    K=5 &  &  & 100\% & 100\% & 100\% &  100\% & 100\% & 100\% & 100\% & 100\% & 100\% & 100\% & 100\%& 98\%\\ \cline{4-15}
    K=10 &  &  &  100\% & 100\% &  100\% & 100\% & 100\% & 100\% & 100\% &100\%&100\%&100\%&99\%&97\%\\ \cline{4-15}
    K=50 &  &  & 100\% & 100\% & 100\% & 100\% & 100\% & 100\% & 100\% & 100\%& 100\%& 100\%& 99\%& 97\%\\ \hline

     K=1 & \multirow{4}{*}{\shortstack{DLA\\(18.1M)} }& \multirow{4}{*}{46.29} &  99\% & 99\% &  99\% & 98\% & 97\% & 98\% & 99\% & 99\% &  99\% & 98\% & 97\% & 99\%\\ \cline{4-15}
    K=5 &  &  & 100\% & 100\% & 100\% &  100\% & 99\% & 99\% & 100\% & 100\% & 100\% & 99\% & 99\%& 99\%\\ \cline{4-15}
    K=10 &  &  &  100\% & 100\% &  100\% & 100\% & 100\% & 100\% & 100\% &100\%&100\%&99\%&99\%&99\%\\ \cline{4-15}
    K=50 &  &  & 100\% & 100\% & 100\% & 100\% & 100\% & 100\% & 100\% & 100\%& 100\%& 99\%& 99\%& 99\%\\ \hline 
   
        \bottomrule
    \end{tabular}
    }
    \caption{ Detection results (AUC) of  spatial consistency (Spatial) based method on BDD dataset for additional targets.
   }
    \label{tbl:bdd-additional}
    
    \end{small}
\end{table}

\subsection{Spatial Context Analysis }
\begin{figure}[h!]
\centering
\begin{minipage}{.40\textwidth}
 \begin{subfigure}{\textwidth}
 \centering
 \includegraphics[width=\textwidth]{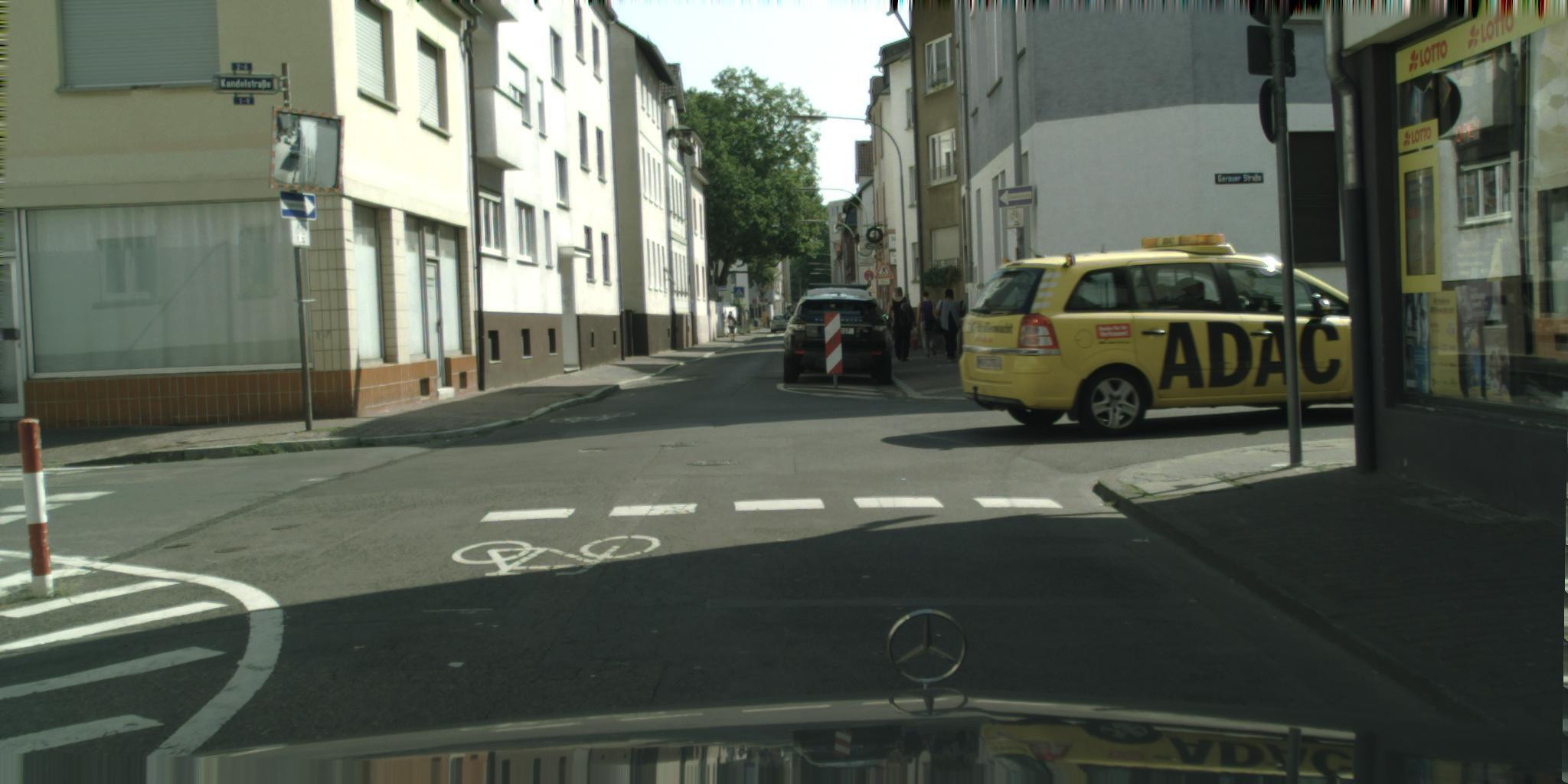}
 \caption{ Original example | DLA | Cityscapes}
 \label{fig:attention-a}
 \end{subfigure}
\end{minipage}
\begin{minipage}{.42\textwidth}
 \begin{subfigure}{\textwidth}
 \centering
 \includegraphics[width=\textwidth]{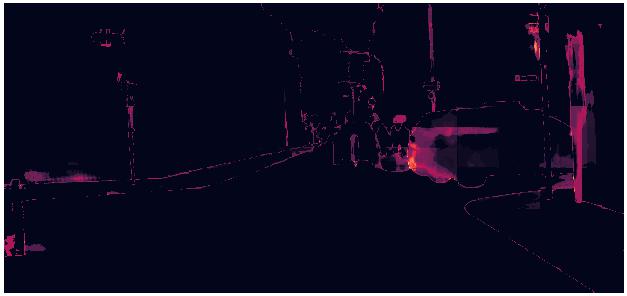}
 \caption{Benign example | DLA | Cityscapes }
 \label{fig:attention-a}
 \end{subfigure}
\end{minipage}
\begin{minipage}{.24\textwidth}
 \begin{subfigure}{\textwidth}
 \centering
 \includegraphics[width=\textwidth]{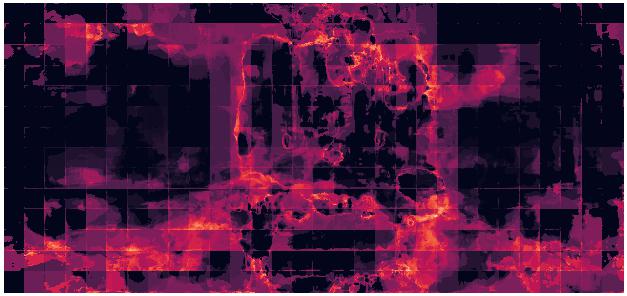}
 \caption{DAG | Kitty | DLA | Cityscapes  }
 \label{fig:attention-a}
 \end{subfigure}
\end{minipage}
\begin{minipage}{.24\textwidth}
 \begin{subfigure}{\textwidth}
 \centering
 \includegraphics[width=\textwidth]{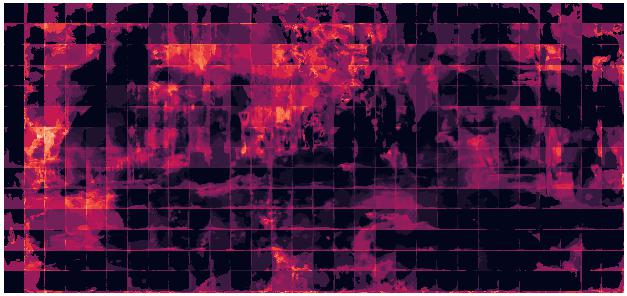}
 \caption{DAG | Pure | DLA | Cityscapes}
 \label{fig:attention-a}
 \end{subfigure}
\end{minipage}
\begin{minipage}{.24\textwidth}
 \begin{subfigure}{\textwidth}
 \centering
 \includegraphics[width=\textwidth]{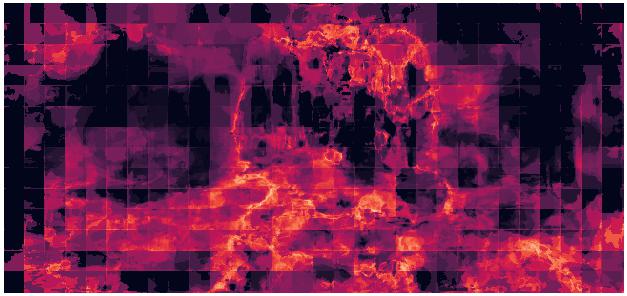}
 \caption{Houdini | Kitty | DLA | Cityscapes }
 \label{fig:attention-a}
 \end{subfigure}
\end{minipage}
\begin{minipage}{.24\textwidth}
 \begin{subfigure}{\textwidth}
 \centering
 \includegraphics[width=\textwidth]{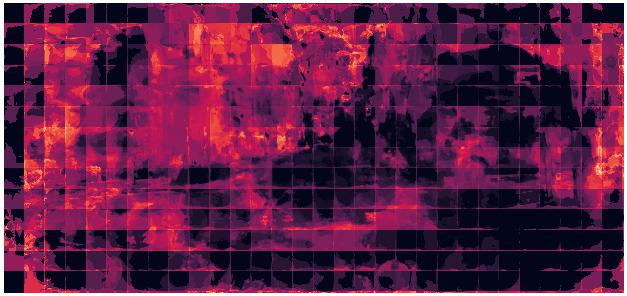} 
 \caption{Houdini |  Pure | DLA | Cityscapes}
 \label{fig:attention-a}
 \end{subfigure}
\end{minipage}
\begin{minipage}{.40\textwidth}
 \begin{subfigure}{\textwidth}
 \centering
 \includegraphics[width=\textwidth]{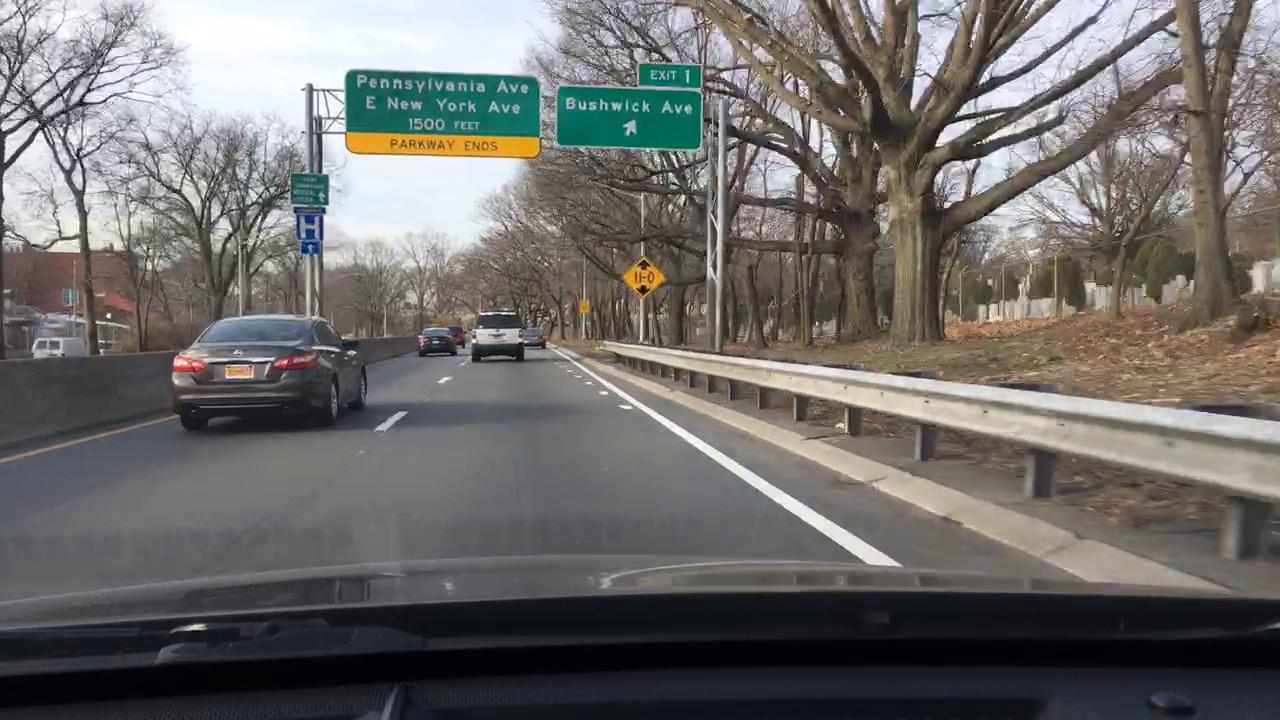}
 \caption{ Original example | DRN | BDD}
 \label{fig:attention-a}
 \end{subfigure}
\end{minipage}
\begin{minipage}{.42\textwidth}
 \begin{subfigure}{\textwidth}
 \centering
 \includegraphics[width=\textwidth]{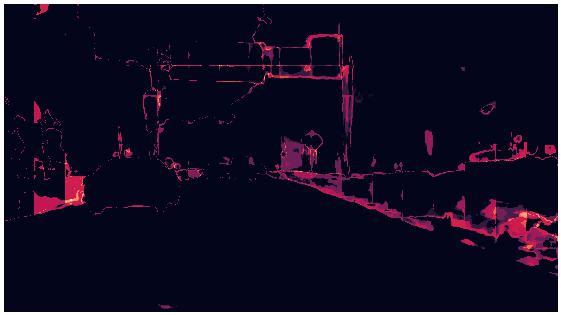}
 \caption{Benign example | DRN | BDD }
 \label{fig:attention-a}
 \end{subfigure}
\end{minipage}
\begin{minipage}{.24\textwidth}
 \begin{subfigure}{\textwidth}
 \centering
 \includegraphics[width=\textwidth]{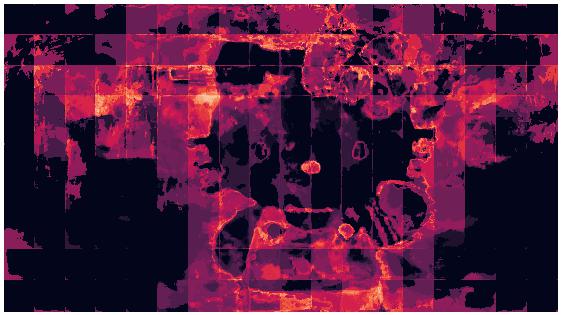}
 \caption{DAG | Kitty | DRN | BDD }
 \label{fig:attention-a}
 \end{subfigure}
\end{minipage}
\begin{minipage}{.24\textwidth}
 \begin{subfigure}{\textwidth}
 \centering
 \includegraphics[width=\textwidth]{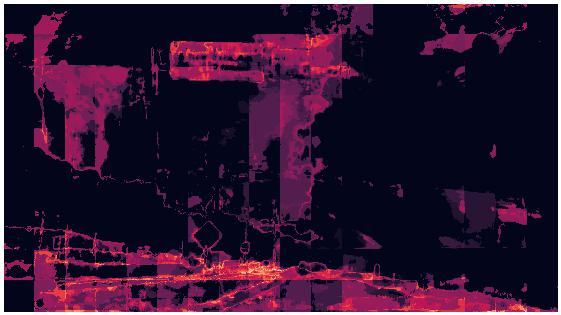}
 \caption{DAG | Scene | DRN | BDD }
 \label{fig:attention-a}
 \end{subfigure}
\end{minipage}
\begin{minipage}{.24\textwidth}
 \begin{subfigure}{\textwidth}
 \centering
 \includegraphics[width=\textwidth]{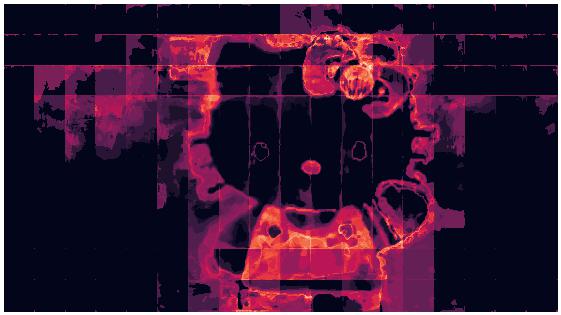}
 \caption{Houdini | Kitty | DRN | BDD }
 \label{fig:attention-a}
 \end{subfigure}
\end{minipage}
\begin{minipage}{.24\textwidth}
 \begin{subfigure}{\textwidth}
 \centering
 \includegraphics[width=\textwidth]{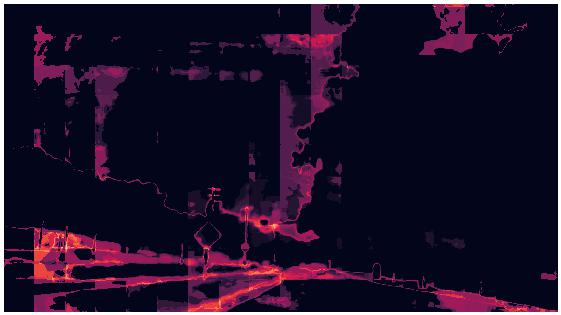} 
 \caption{Houdini | Scene | DRN | BDD}
 \label{fig:attention-a}
 \end{subfigure}
\end{minipage}
\begin{minipage}{.39\textwidth}
 \begin{subfigure}{\textwidth}
 \centering
 \includegraphics[width=\textwidth]{figure_2/entropy/dla_bdd/bdd_original_image.jpg}
 \caption{ Original example|DLA|BDD}
 \label{fig:attention-a}
 \end{subfigure}
\end{minipage}
\begin{minipage}{.40\textwidth}
 \begin{subfigure}{\textwidth}
 \centering
 \includegraphics[width=\textwidth]{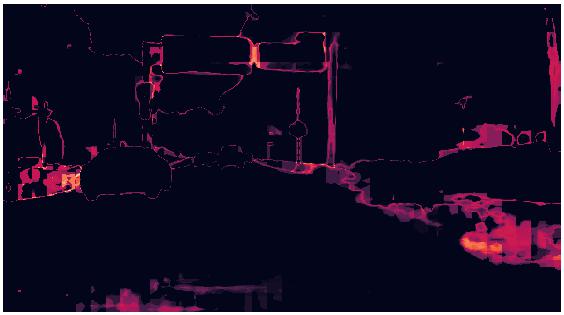}
 \caption{Benign example | DLA | BDD}
 \label{fig:attention-a}
 \end{subfigure}
\end{minipage}
\begin{minipage}{.24\textwidth}
 \begin{subfigure}{\textwidth}
 \centering
 \includegraphics[width=\textwidth]{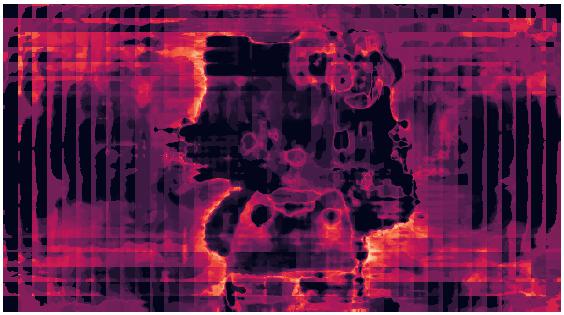}
 \caption{DAG | Kitty | DLA | BDD  }
 \label{fig:attention-a}
 \end{subfigure}
\end{minipage}
\begin{minipage}{.24\textwidth}
 \begin{subfigure}{\textwidth}
 \centering
 \includegraphics[width=\textwidth]{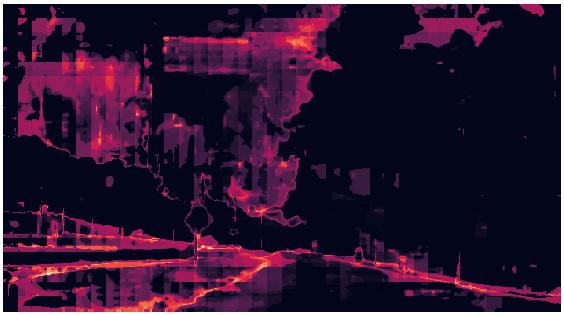}
 \caption{DAG | Scene | DLA | BDD }
 \label{fig:attention-a}
 \end{subfigure}
\end{minipage}
\begin{minipage}{.24\textwidth}
 \begin{subfigure}{\textwidth}
 \centering
 \includegraphics[width=\textwidth]{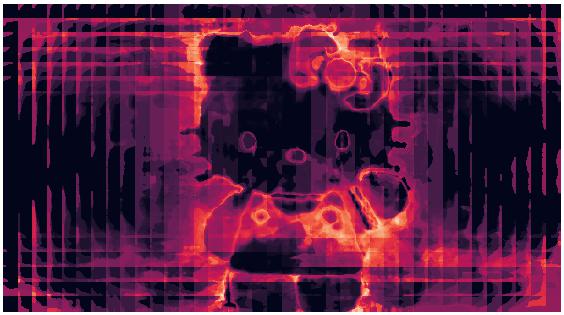}
 \caption{Houdini | Kitty | DLA | BDD}
 \end{subfigure}
\end{minipage}
\begin{minipage}{.24\textwidth}
 \begin{subfigure}{\textwidth}
 \centering
 \includegraphics[width=\textwidth]{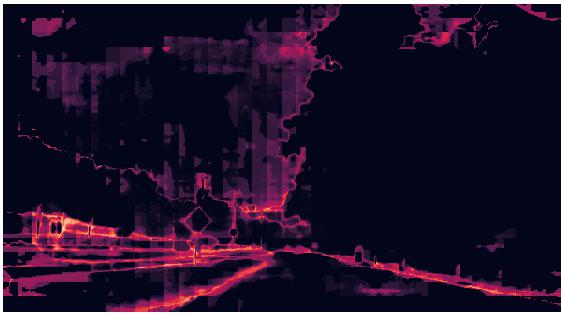} 
 \caption{Houdini | Real | DLA | BDD}
 \label{fig:attention-a}
 \end{subfigure}
\end{minipage}
\centering
\caption{
Heatmap of per-pixel self-entropy. 
(a), (b), (g), (h), (m) and (n) show benign images and its corresponding per-pixel self-entropy heatmaps. We use the format ``examples | attack model | dataset" to label them. For the rest,  we use the format ``attack method | target label | attack model | dataset'' to label each subcaption.
}
\label{fig:all_entropy}
\end{figure}
Algorithm~\ref{algo:patchselection} describes the algorithm of {\bf getOverlapPatches}. 
Figure~\ref{fig:all_entropy} shows the heatmaps of the per-pixel self-entropy on Cityscapes and BDD dataset against DRN and DLA models. It is clearly shown that the adversarial instances have higher entropy than benign ones.  
Table~\ref{tbl:bdd} shows that the detection results (AUC) based on spatial consistency method with fix patch size. It demonstrates that the spatial consistency information can help to detect adversarial examples with AUC nearly 100\% on BDD dataset. Table~\ref{tbl:city-additional}~\ref{tbl:bdd-additional} show the results on additinal targets on Cityscapes and BDD datasets.  Table~\ref{tbl:city-random}~\ref{tbl:bdd-random} show the detection results (AUC) based on spatial consistency method with random patch size. They show that random patch sizes achieve the similar detection result.

\subsection{Scale Consistency Analysis}

\begin{figure}[h!]
    \centering
    \begin{minipage}{.9\textwidth}
     \begin{subfigure}{\textwidth}
     \centering
     \includegraphics[width=\textwidth]{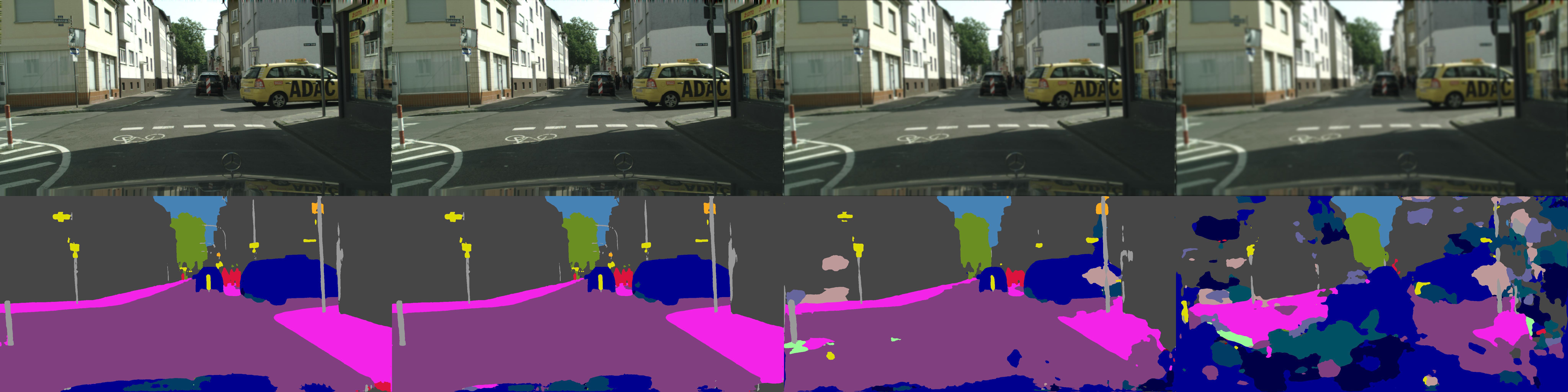}
     \caption{Benign example | DLA | Cityscapes}
     \end{subfigure}
    \end{minipage}
    \begin{minipage}{.44\textwidth}
     \begin{subfigure}{\textwidth}
     \centering
     \includegraphics[width=\textwidth]{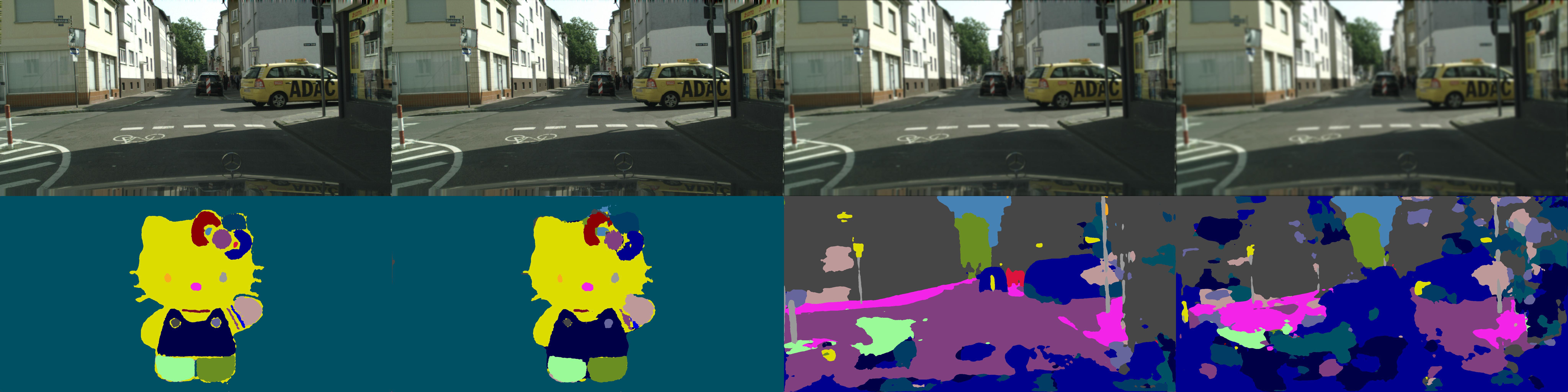}
     \caption{DAG | Kitty | DLA | Cityscapes}
     \end{subfigure}
    \end{minipage}
    \begin{minipage}{.44\textwidth}
     \begin{subfigure}{\textwidth}
     \centering
     \includegraphics[width=\textwidth]{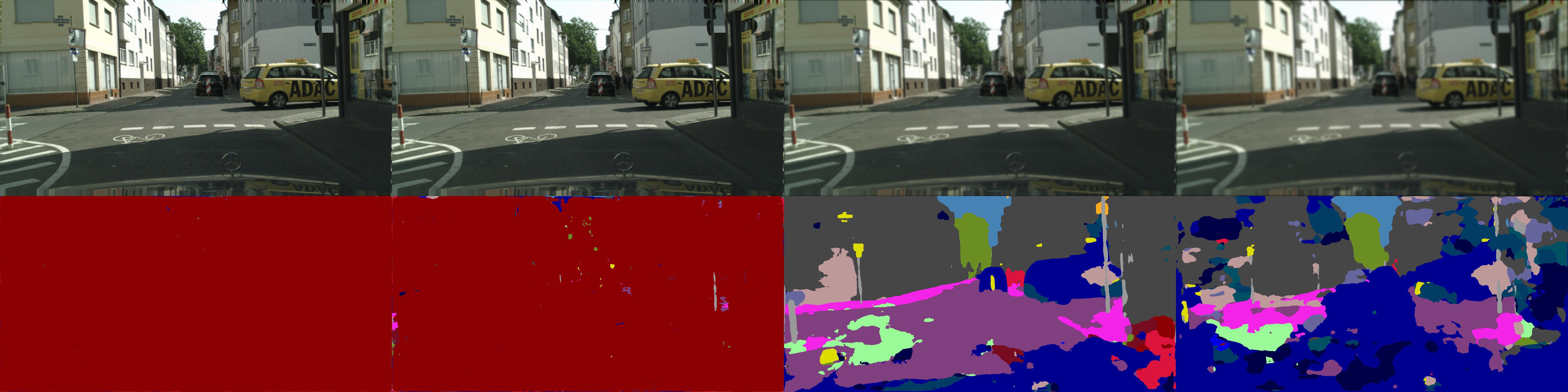}
     \caption{DAG | Pure | DLA | Cityscapes}
     \end{subfigure}
    \end{minipage}
    \begin{minipage}{.44\textwidth}
     \begin{subfigure}{\textwidth}
     \centering
     \includegraphics[width=\textwidth]{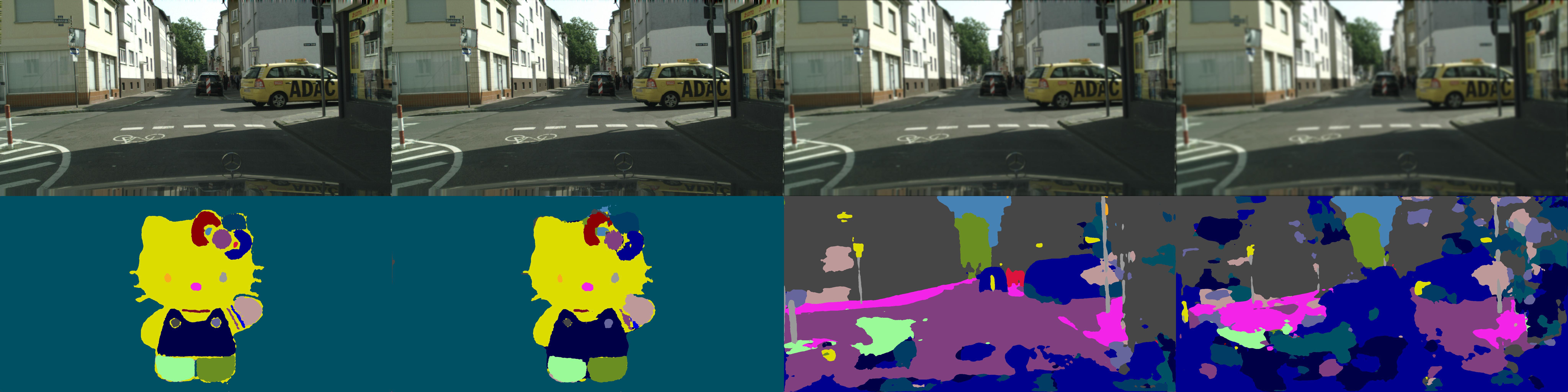}
     \caption{Houdini | Kitty | DLA | Cityscapes}
     \end{subfigure}
    \end{minipage}
    \begin{minipage}{.44\textwidth}
     \begin{subfigure}{\textwidth}
     \centering
     \includegraphics[width=\textwidth]{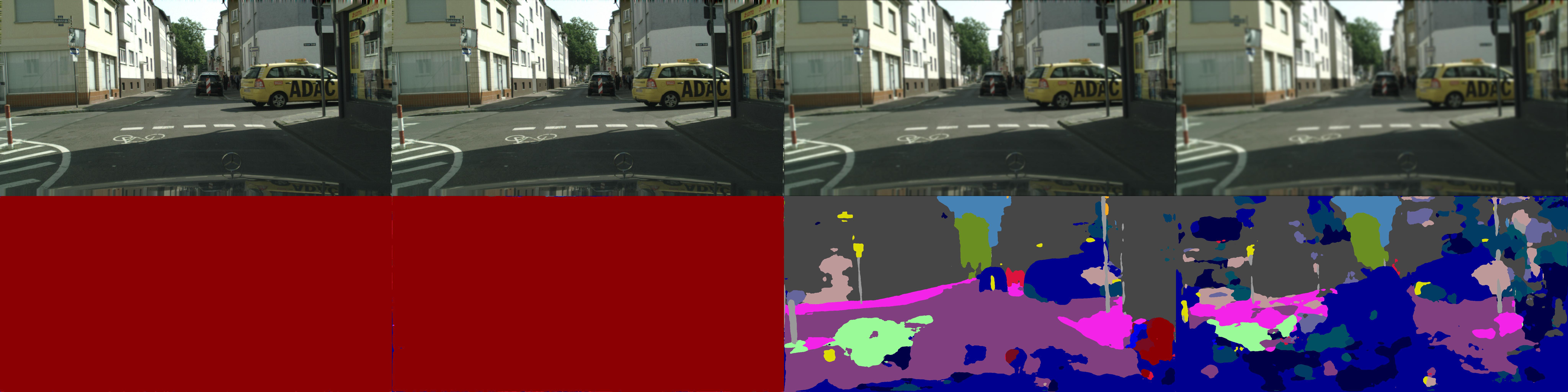}
     \caption{Houdini | Pure | DLA | Cityscapes}
     \end{subfigure}
    \end{minipage}
       \begin{minipage}{.9\textwidth}
     \begin{subfigure}{\textwidth}
     \centering
     \includegraphics[width=\textwidth]{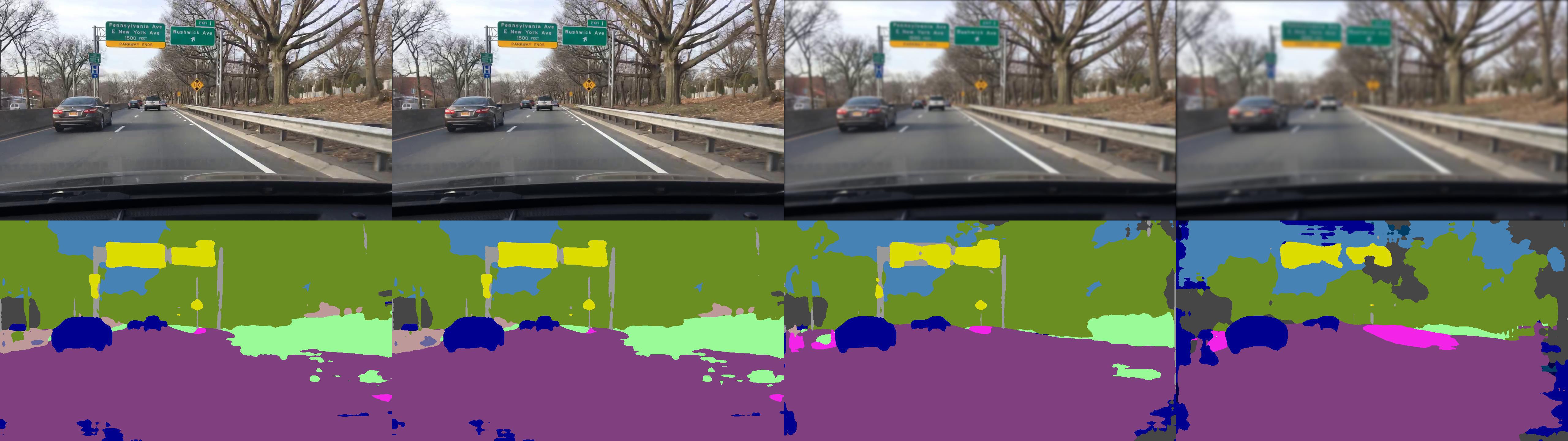}
     \caption{Benign example | DRN | BDD}
     \end{subfigure}
    \end{minipage}
    \begin{minipage}{.44\textwidth}
     \begin{subfigure}{\textwidth}
     \centering
     \includegraphics[width=\textwidth]{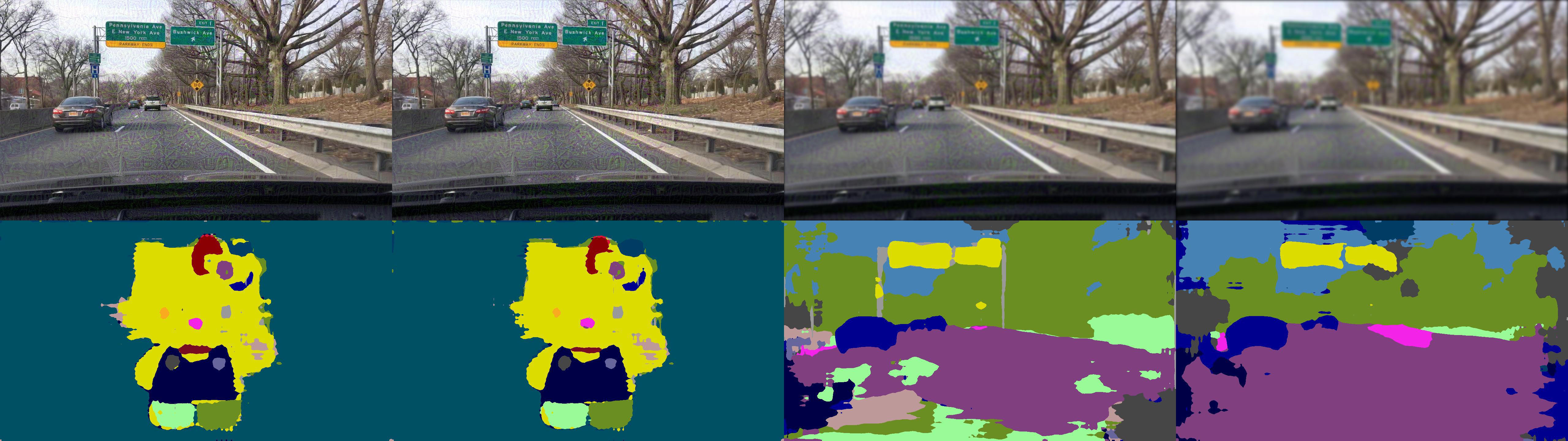}
     \caption{DAG | Kitty | DRN | BDD}
     \end{subfigure}
    \end{minipage}
    \begin{minipage}{.44\textwidth}
     \begin{subfigure}{\textwidth}
     \centering
     \includegraphics[width=\textwidth]{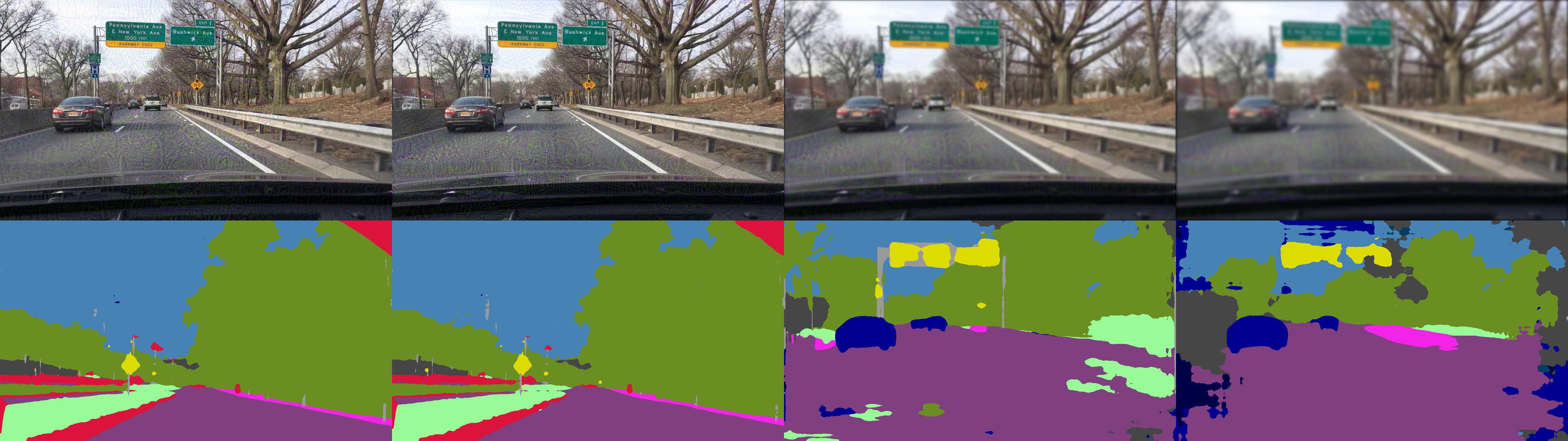}
     \caption{DAG | Scene | DRN | BDD}
     \end{subfigure}
    \end{minipage}
    \begin{minipage}{.44\textwidth}
     \begin{subfigure}{\textwidth}
     \centering
     \includegraphics[width=\textwidth]{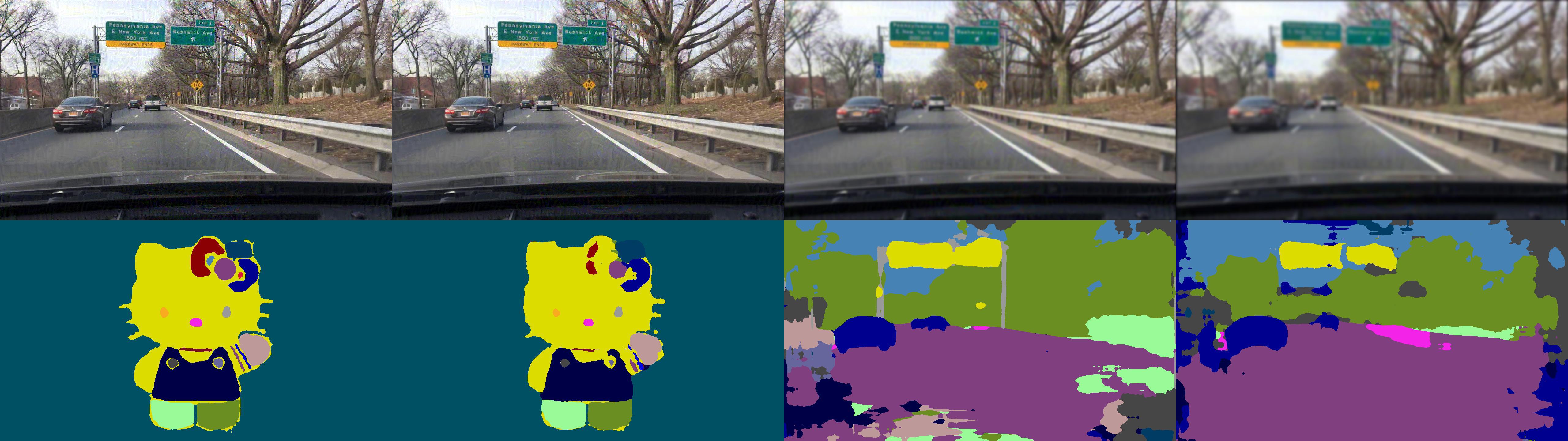}
     \caption{Houdini | Kitty | DRN | BDD}
     \end{subfigure}
    \end{minipage}
    \begin{minipage}{.44\textwidth}
     \begin{subfigure}{\textwidth}
     \centering
     \includegraphics[width=\textwidth]{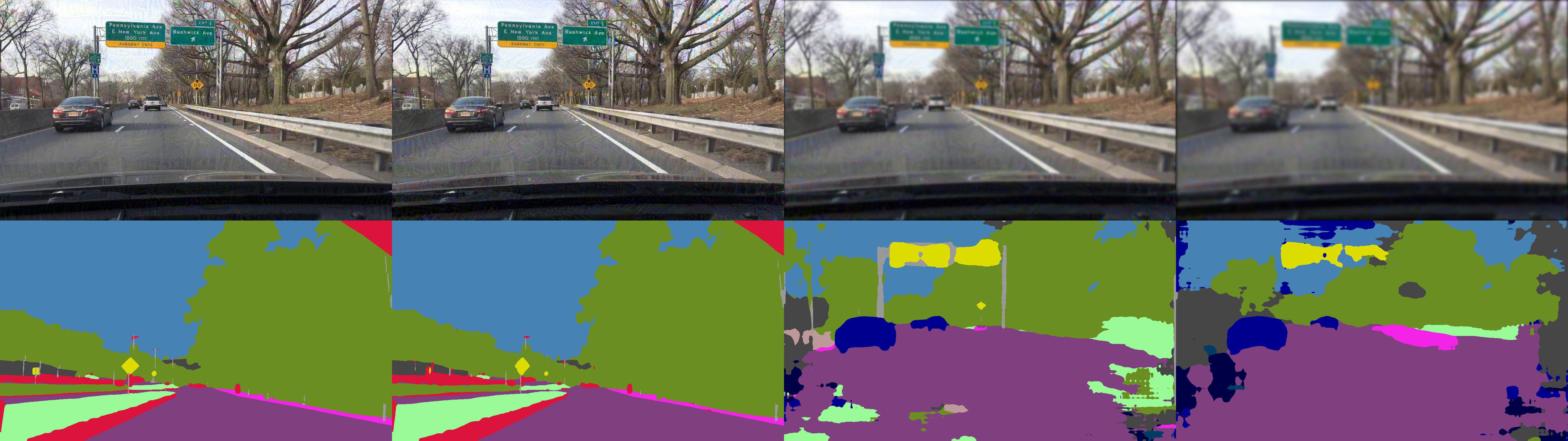}
     \caption{Houdini | Scene | DRN | BDD}
     \end{subfigure}
    \end{minipage}
\end{figure}
\begin{figure}[h!]
    \ContinuedFloat
    \centering
    \begin{minipage}{.9\textwidth}
     \begin{subfigure}{\textwidth}
     \centering
     \includegraphics[width=\textwidth]{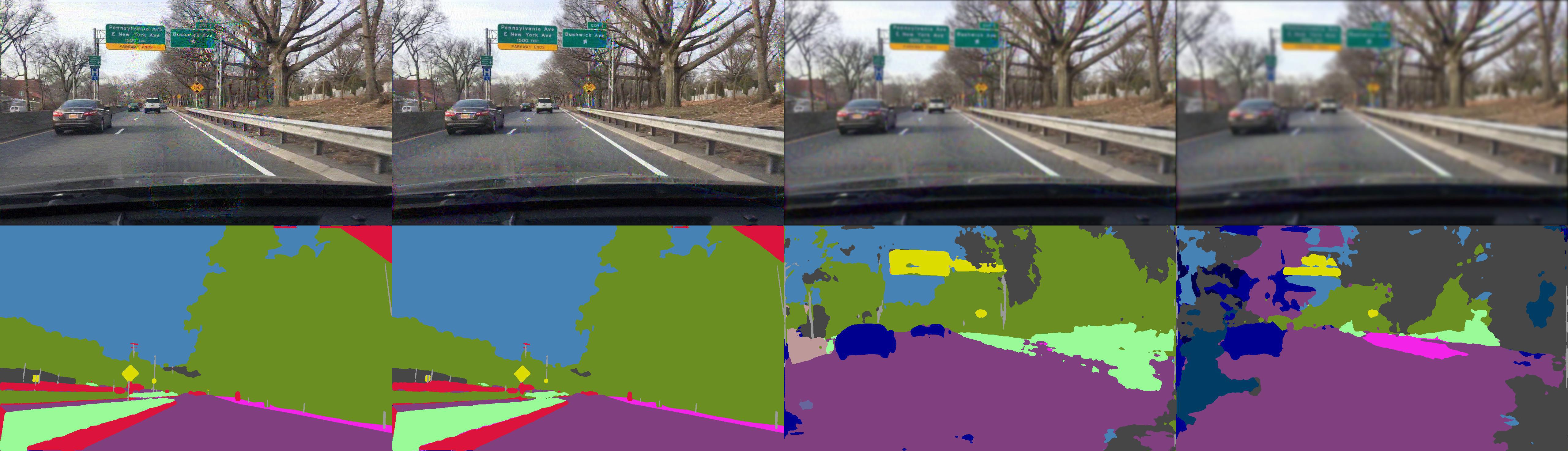}
     \caption{Benign example | DLA | BDD}
     \end{subfigure}
    \end{minipage}
    \begin{minipage}{.44\textwidth}
     \begin{subfigure}{\textwidth}
     \centering
     \includegraphics[width=\textwidth]{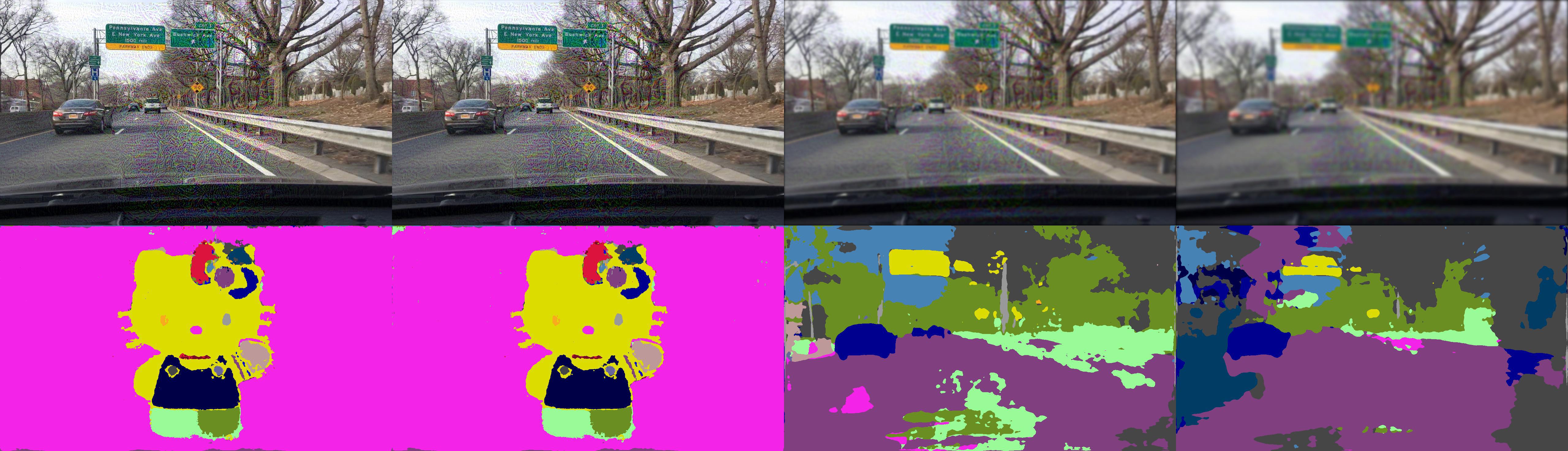}
     \caption{DAG | Kitty | DLA | BDD}
     \end{subfigure}
    \end{minipage}
    \begin{minipage}{.44\textwidth}
     \begin{subfigure}{\textwidth}
     \centering
     \includegraphics[width=\textwidth]{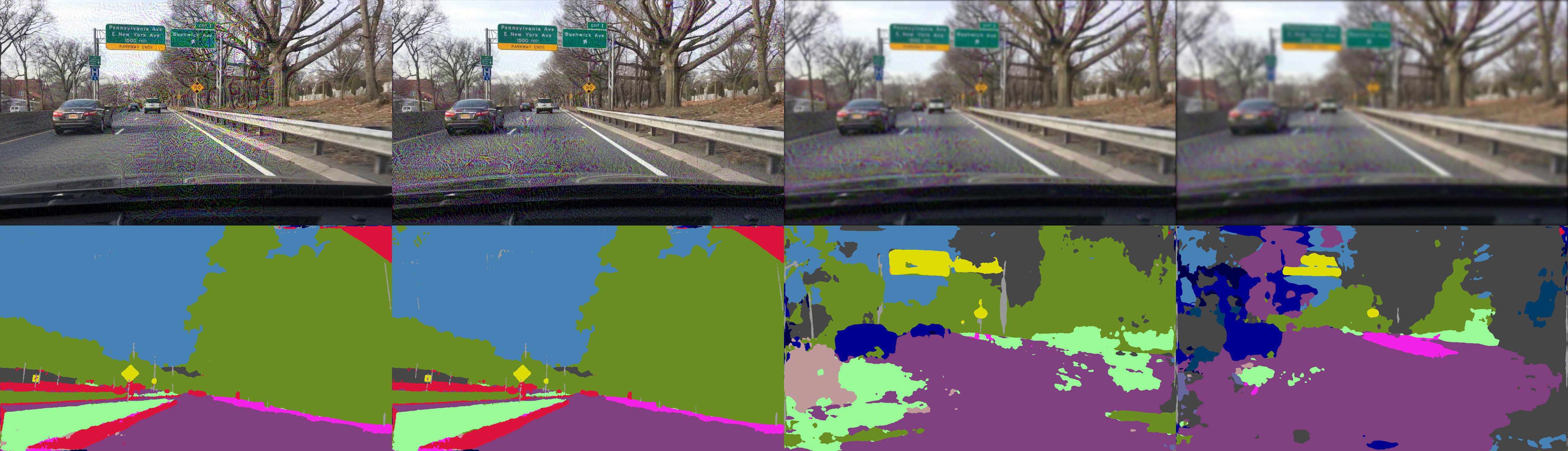}
     \caption{DAG | Scene | DLA | BDD}
     \end{subfigure}
    \end{minipage}
    \begin{minipage}{.44\textwidth}
     \begin{subfigure}{\textwidth}
     \centering
     \includegraphics[width=\textwidth]{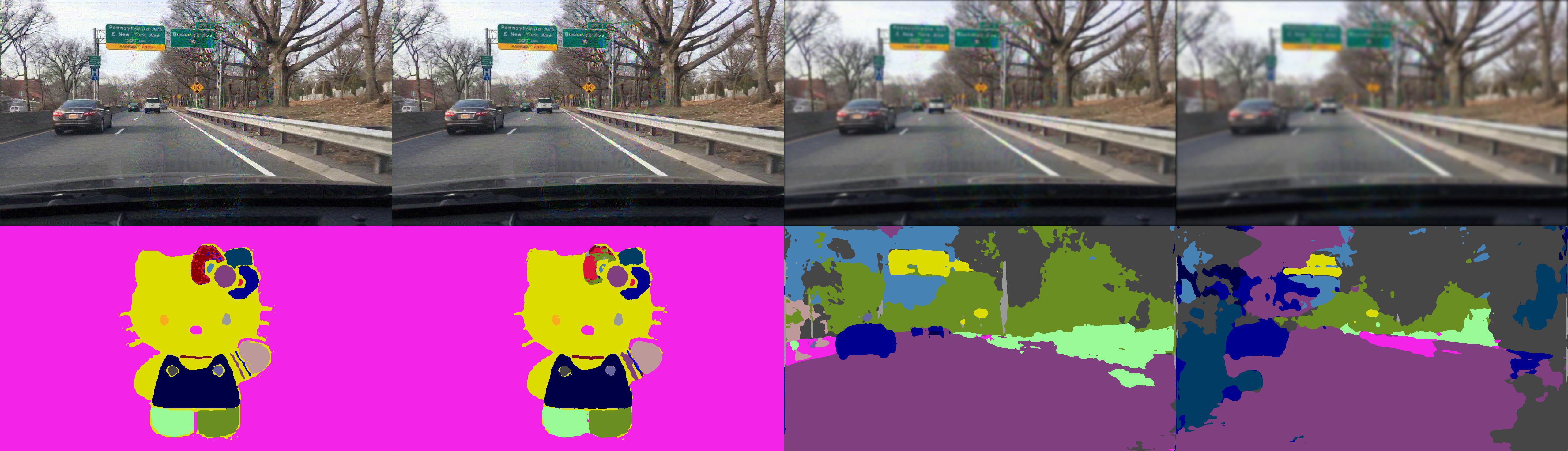}
     \caption{Houdini | Kitty | DLA | BDD}
     \end{subfigure}
    \end{minipage}
    \begin{minipage}{.44\textwidth}
     \begin{subfigure}{\textwidth}
     \centering
     \includegraphics[width=\textwidth]{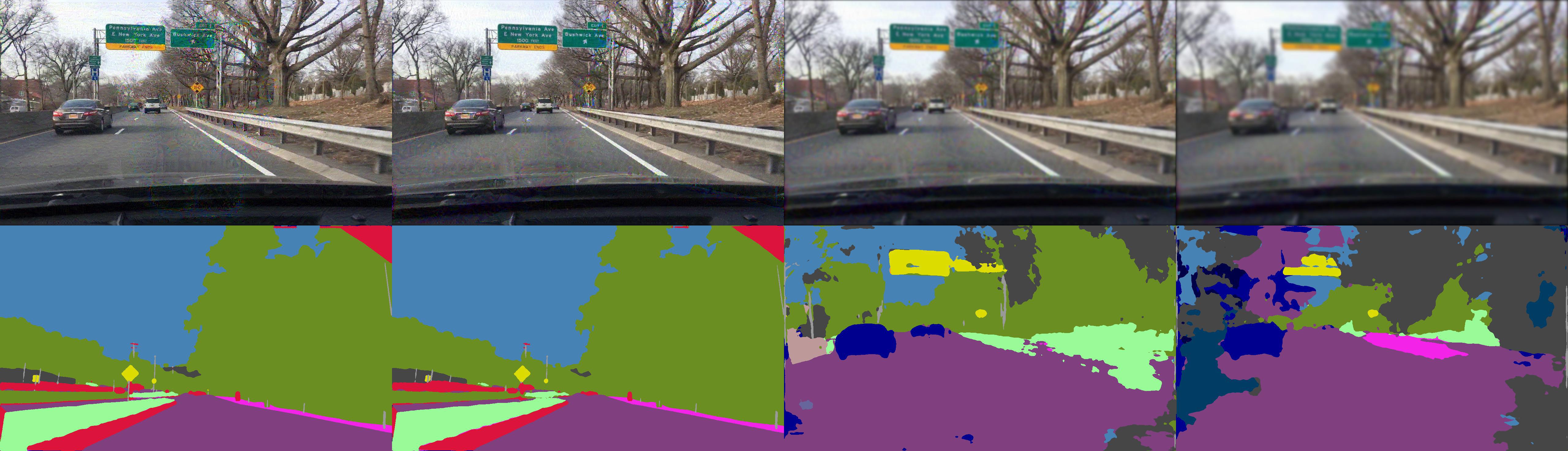}
     \caption{Houdini | Scene | DLA | BDD}
     \end{subfigure}
    \end{minipage}
    \caption{Examples of images and corresponding segmentation results before/after image scaling.
    For each subfigure, the first column shows benign/adversarial images, while the following columns represent images after scaling by applying Gaussian kernel with std as 0.5, 3, and 5, respectively. 
    (a),(f) and (k) show benign images before/after image scaling and the corresponding segmentation results and  we use the format ``example | attack model | dataset'' to identify the corresponding model and dataset; 
    (b)-(e), (g)-(j) and (l)-(o) present similar results for adversarial images and  we use the format ``attack method | target label | attack model | dataset'' to label the settings of each image.  
    }
    \label{fig:allblur}
\end{figure}
\begin{figure}[h!]
    \centering
    \begin{minipage}{.9\textwidth}
     \begin{subfigure}{\textwidth}
     \centering
     \includegraphics[width=\textwidth]{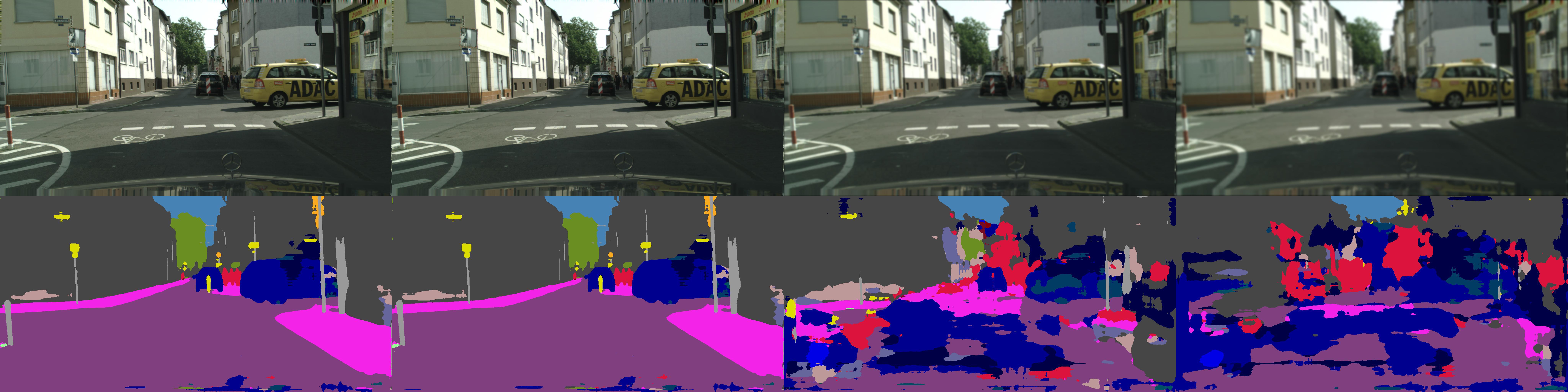}
     \caption{Benign example | DRN | Cityscapes}
     \end{subfigure}
    \end{minipage}
    \begin{minipage}{.45\textwidth}
     \begin{subfigure}{\textwidth}
     \centering
     \includegraphics[width=\textwidth]{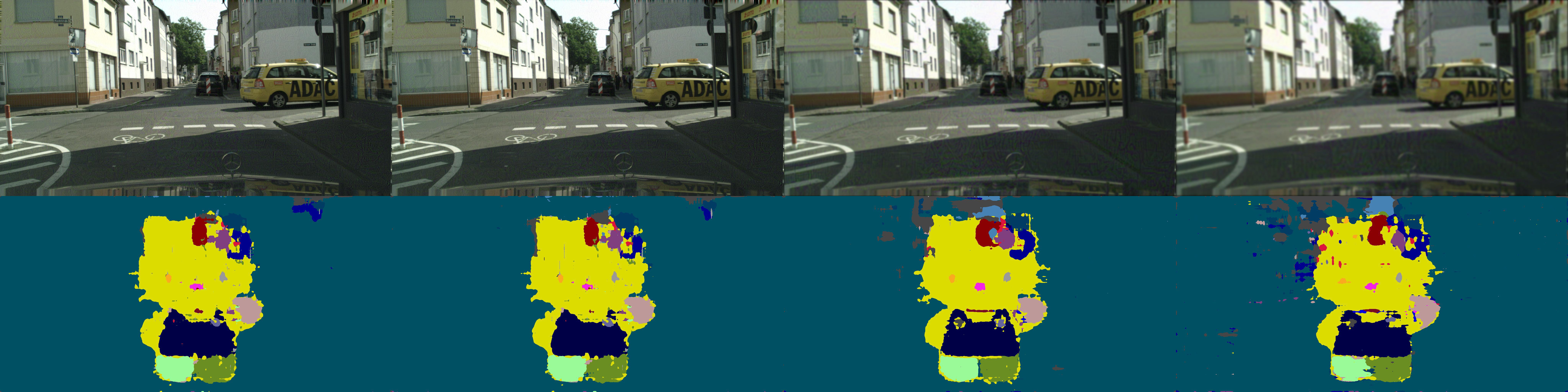}
     \caption{DAG | Kitty | DRN | Cityscapes}
     \end{subfigure}
    \end{minipage}
    \begin{minipage}{.45\textwidth}
     \begin{subfigure}{\textwidth}
     \centering
     \includegraphics[width=\textwidth]{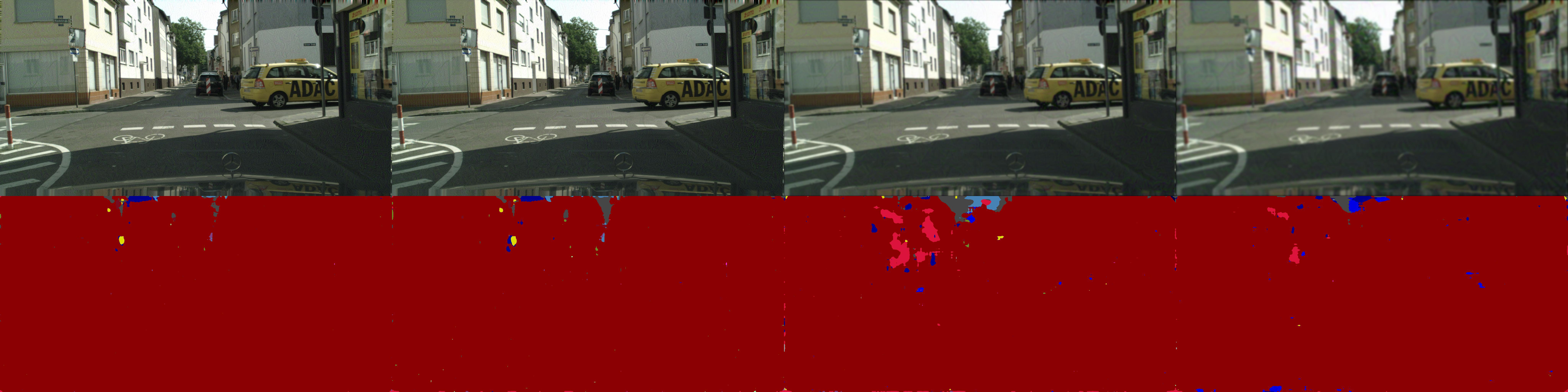}
     \caption{DAG | Pure | DRN | Cityscapes}
     \end{subfigure}
    \end{minipage}
    \begin{minipage}{.45\textwidth}
     \begin{subfigure}{\textwidth}
     \centering
     \includegraphics[width=\textwidth]{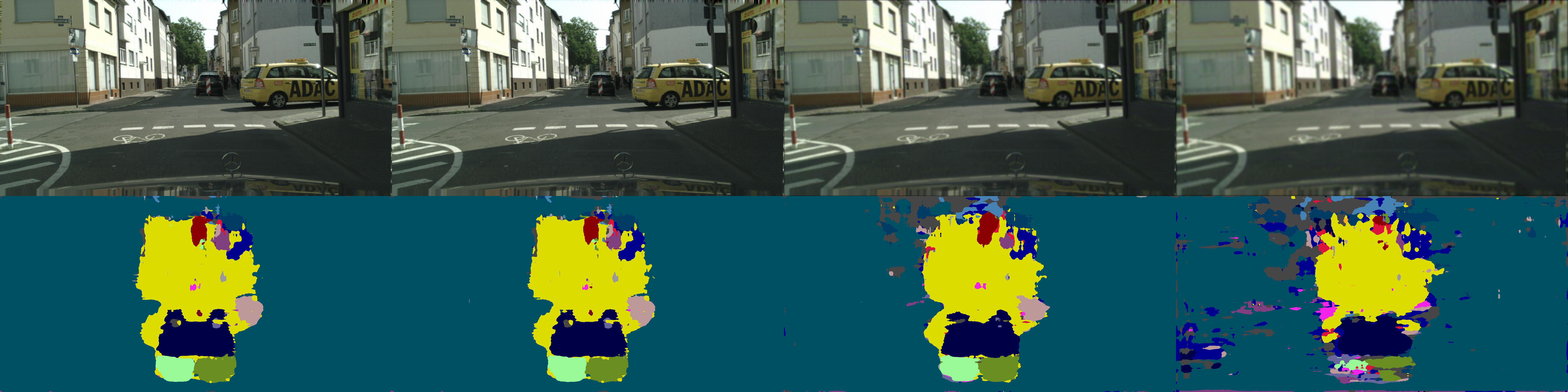}
     \caption{Houdini | Kitty | DRN | Cityscapes}
     \end{subfigure}
    \end{minipage}
    \begin{minipage}{.45\textwidth}
     \begin{subfigure}{\textwidth}
     \centering
     \includegraphics[width=\textwidth]{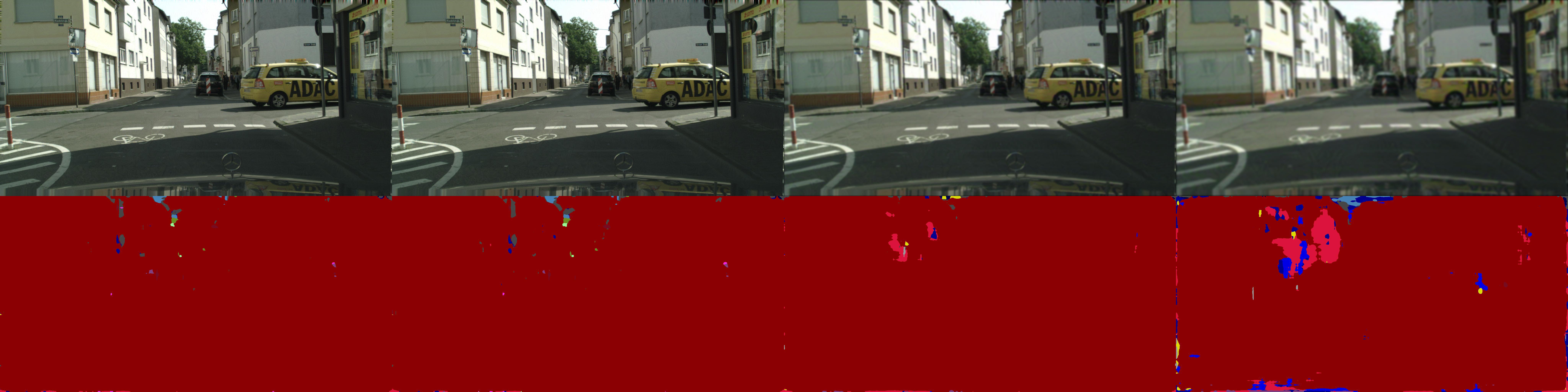}
     \caption{Houdini | Pure | DRN | Cityscapes}
     \end{subfigure}
    \end{minipage}
    \begin{minipage}{.9\textwidth}
     \begin{subfigure}{\textwidth}
     \centering
     \includegraphics[width=\textwidth]{figure_2/blur/benign_blur_dla_city.jpg}
     \caption{Benign example | DLA | Cityscapes}
     \end{subfigure}
    \end{minipage}
    \begin{minipage}{.45\textwidth}
     \begin{subfigure}{\textwidth}
     \centering
     \includegraphics[width=\textwidth]{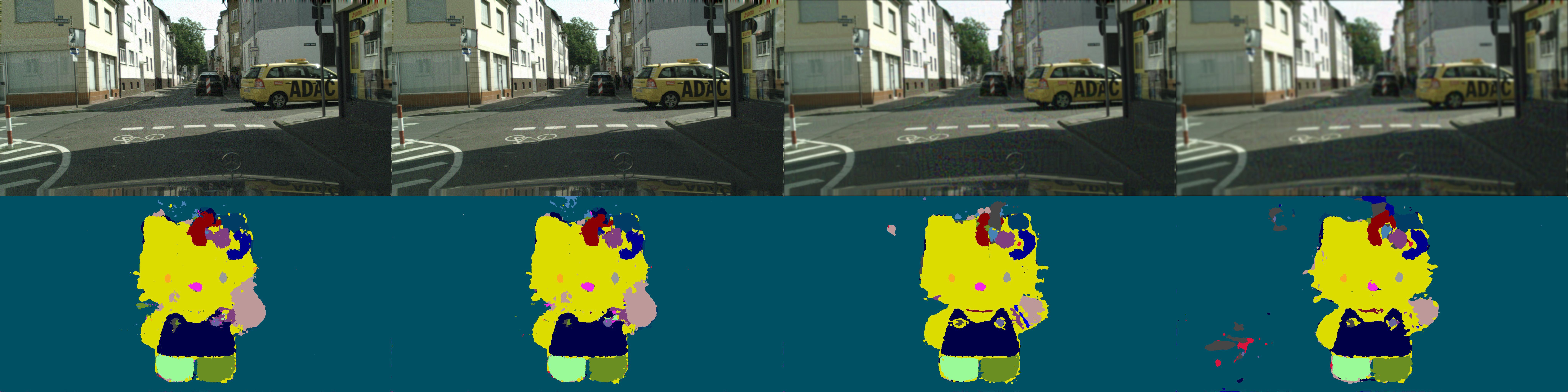}
     \caption{DAG | Kitty | DLA | Cityscapes}
     \end{subfigure}
    \end{minipage}
    \begin{minipage}{.45\textwidth}
     \begin{subfigure}{\textwidth}
     \centering
     \includegraphics[width=\textwidth]{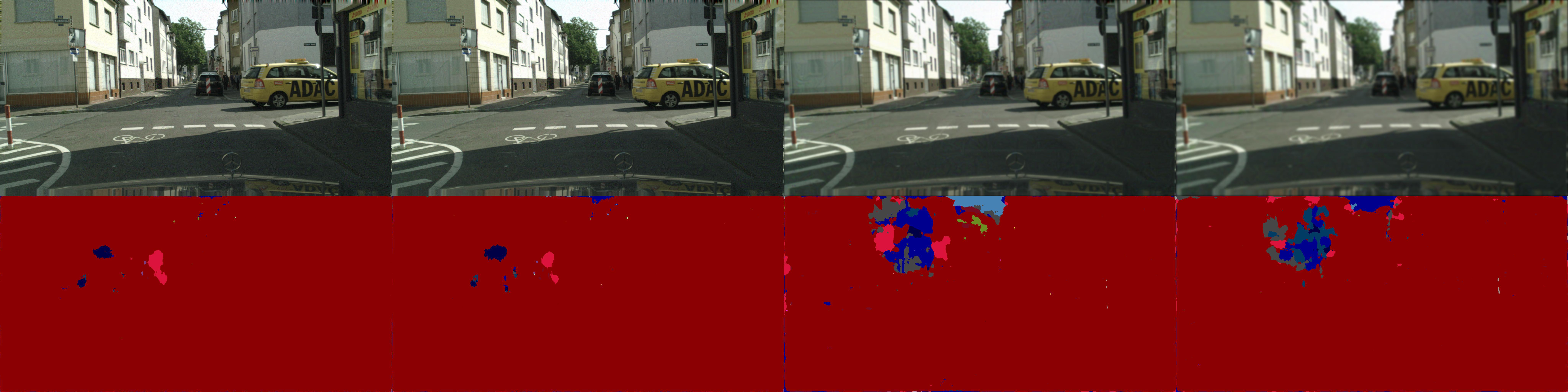}
     \caption{DAG | Pure | DLA | Cityscapes}
     \end{subfigure}
    \end{minipage}
    \begin{minipage}{.45\textwidth}
     \begin{subfigure}{\textwidth}
     \centering
     \includegraphics[width=\textwidth]{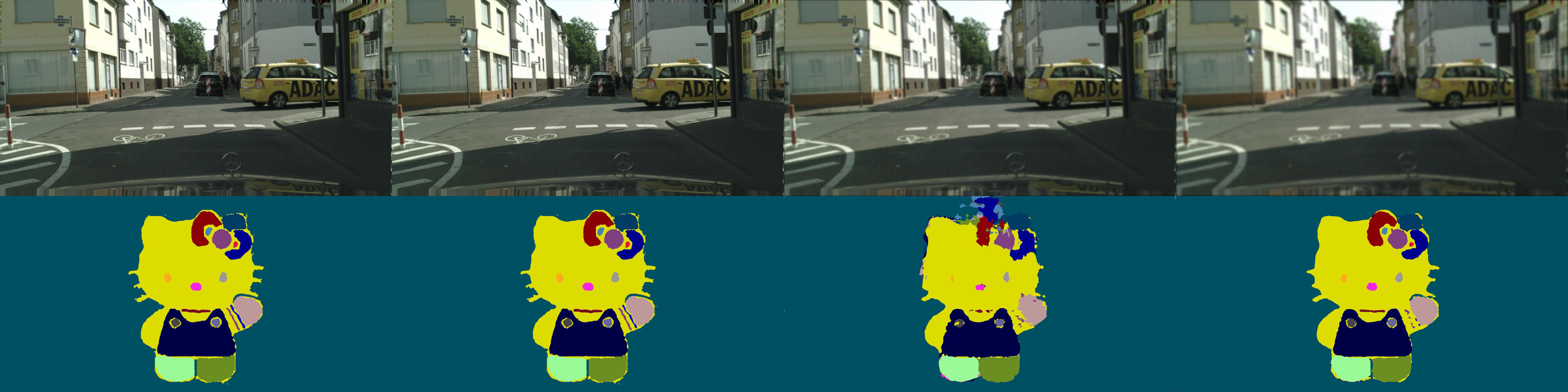}
     \caption{Houdini | Kitty | DLA | Cityscapes}
     \end{subfigure}
    \end{minipage}
    \begin{minipage}{.45\textwidth}
     \begin{subfigure}{\textwidth}
     \centering
     \includegraphics[width=\textwidth]{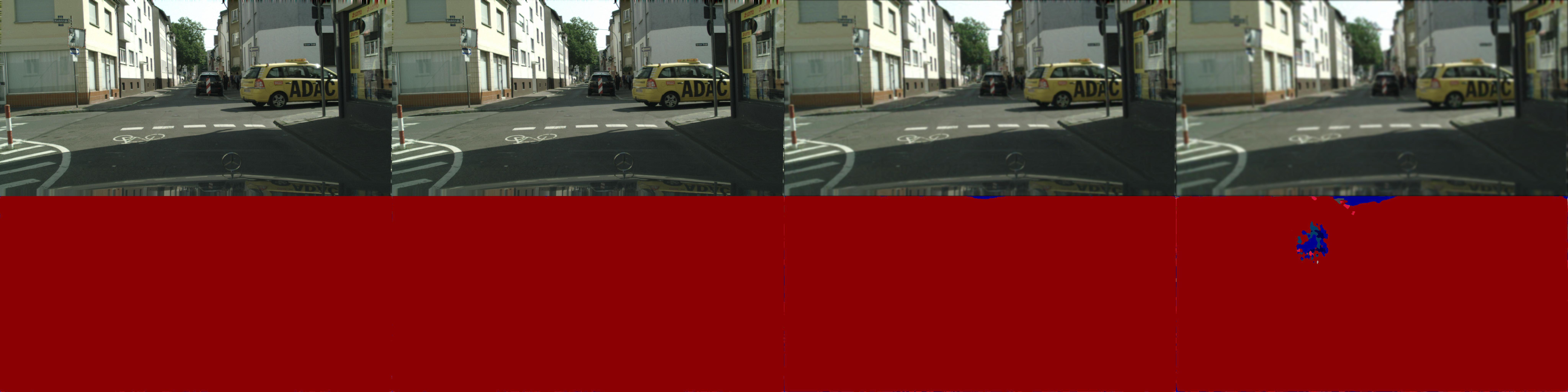}
     \caption{Houdini | Pure | DLA | Cityscapes}
     \end{subfigure}
    \end{minipage}
\end{figure}
\begin{figure}[h!]
    \ContinuedFloat
    \centering
    \begin{minipage}{.9\textwidth}
     \begin{subfigure}{\textwidth}
     \centering
     \includegraphics[width=\textwidth]{figure_2/blur/benign_blur_drn_bdd.jpg}
     \caption{Benign example | DRN | BDD}
     \end{subfigure}
    \end{minipage}
    \begin{minipage}{.45\textwidth}
     \begin{subfigure}{\textwidth}
     \centering
     \includegraphics[width=\textwidth]{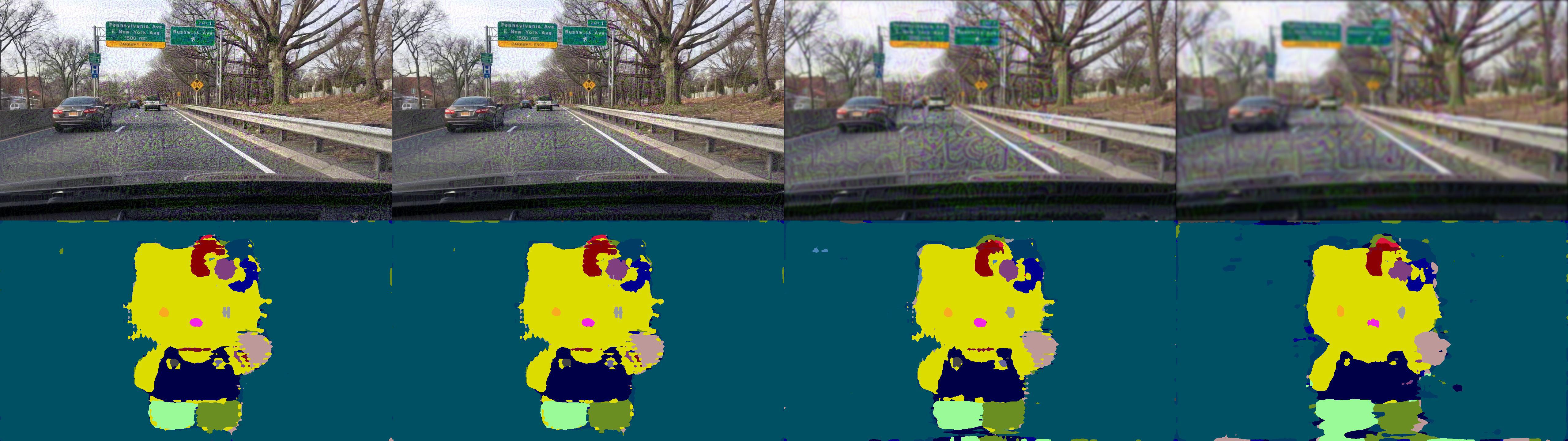}
     \caption{DAG | Kitty | DRN | BDD}
     \end{subfigure}
    \end{minipage}
    \begin{minipage}{.45\textwidth}
     \begin{subfigure}{\textwidth}
     \centering
     \includegraphics[width=\textwidth]{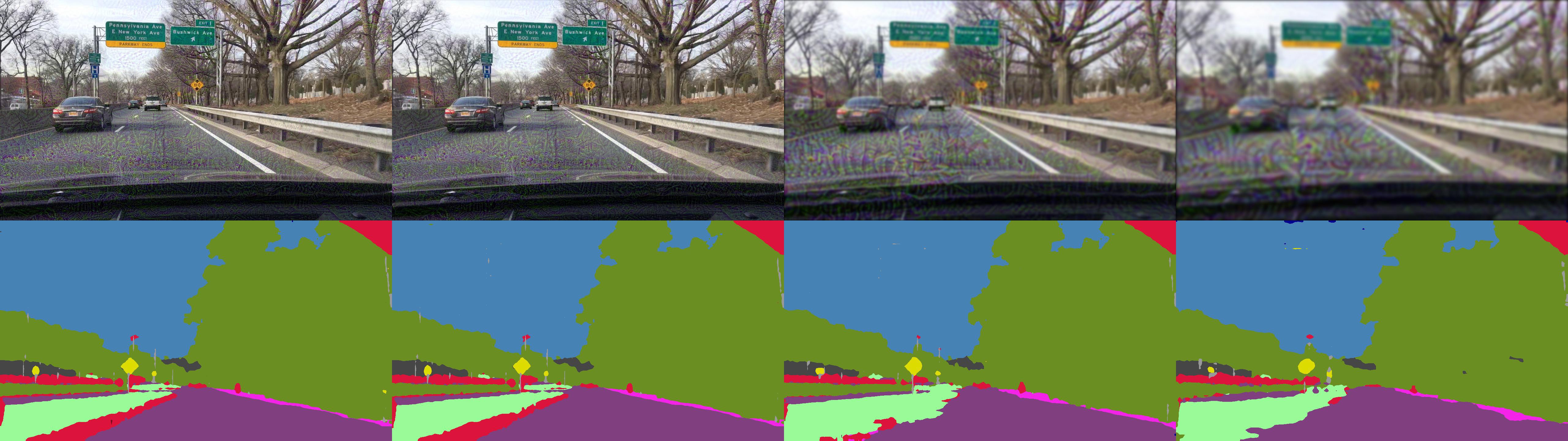}
     \caption{DAG | Scene | DRN | BDD}
     \end{subfigure}
    \end{minipage}
    \begin{minipage}{.45\textwidth}
     \begin{subfigure}{\textwidth}
     \centering
     \includegraphics[width=\textwidth]{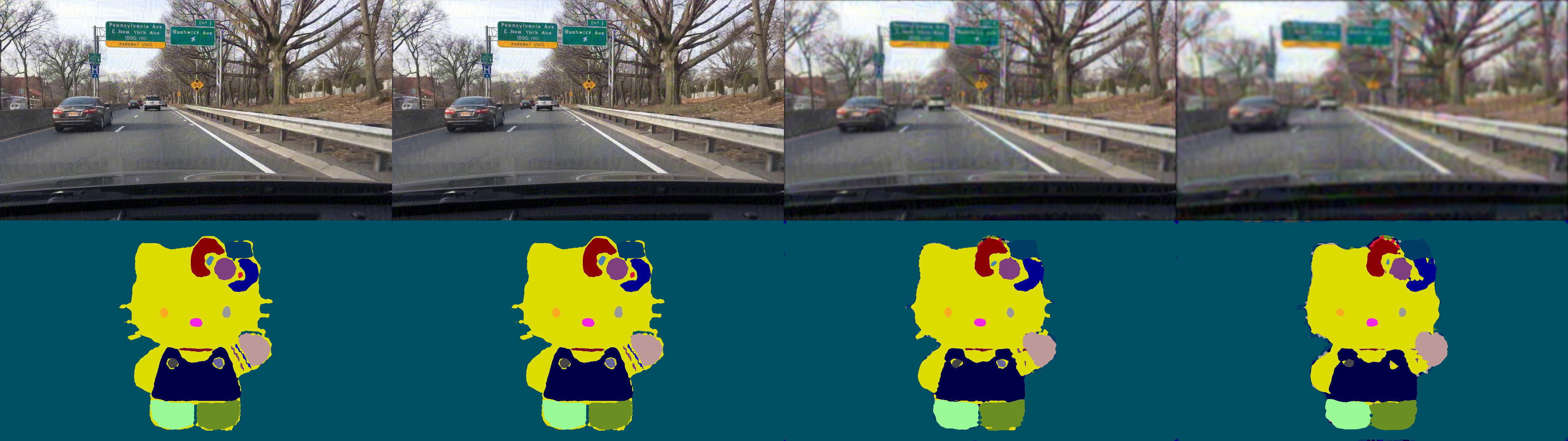}
     \caption{Houdini | Kitty | DRN | BDD}
     \end{subfigure}
    \end{minipage}
    \begin{minipage}{.45\textwidth}
     \begin{subfigure}{\textwidth}
     \centering
     \includegraphics[width=\textwidth]{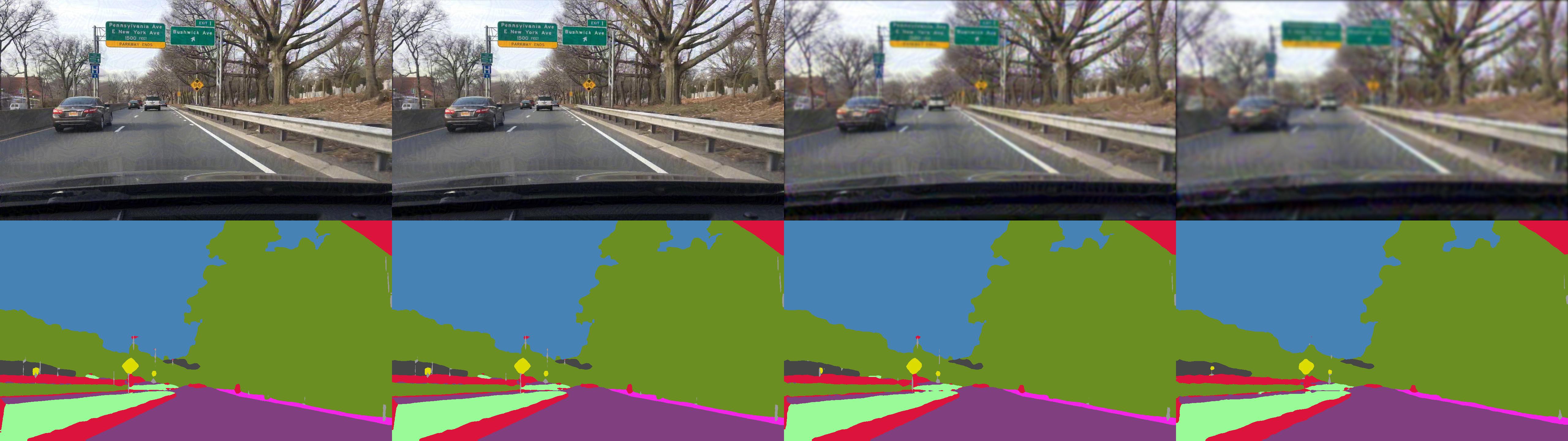}
     \caption{Houdini | Scene | DRN | BDD}
     \end{subfigure}
    \end{minipage}
    \begin{minipage}{.9\textwidth}
     \begin{subfigure}{\textwidth}
     \centering
     \includegraphics[width=\textwidth]{figure_2/blur/benign_blur_dla_bdd.jpg}
     \caption{Benign example | DLA | BDD}
     \end{subfigure}
    \end{minipage}
    \begin{minipage}{.45\textwidth}
     \begin{subfigure}{\textwidth}
     \centering
     \includegraphics[width=\textwidth]{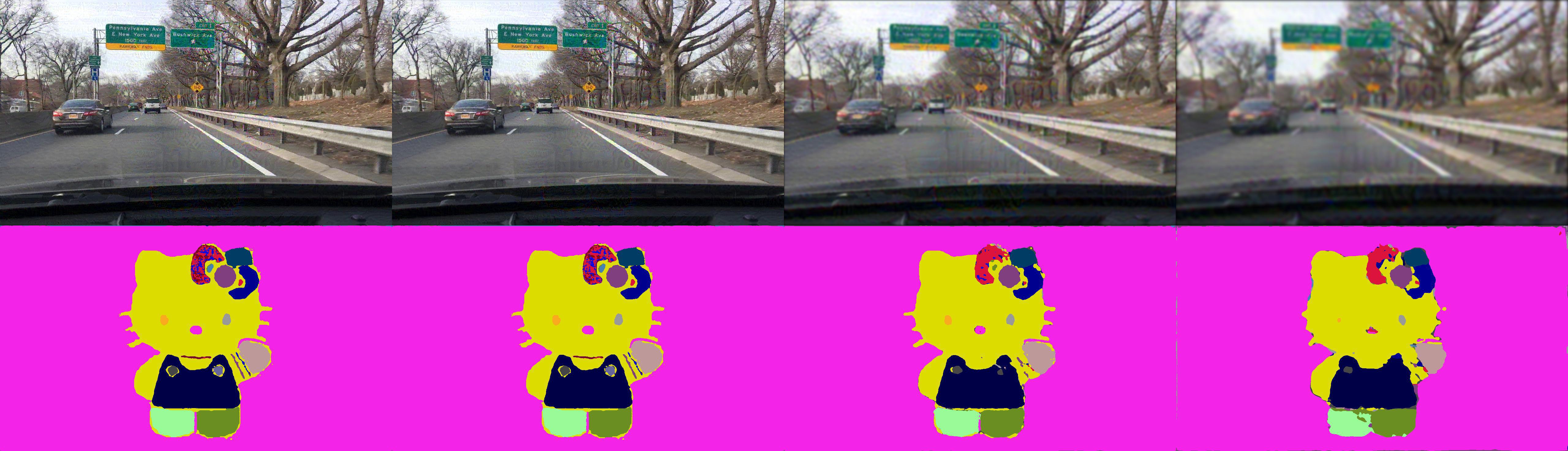}
     \caption{DAG | Kitty | DLA | BDD }
     \end{subfigure}
    \end{minipage}
    \begin{minipage}{.45\textwidth}
     \begin{subfigure}{\textwidth}
     \centering
     \includegraphics[width=\textwidth]{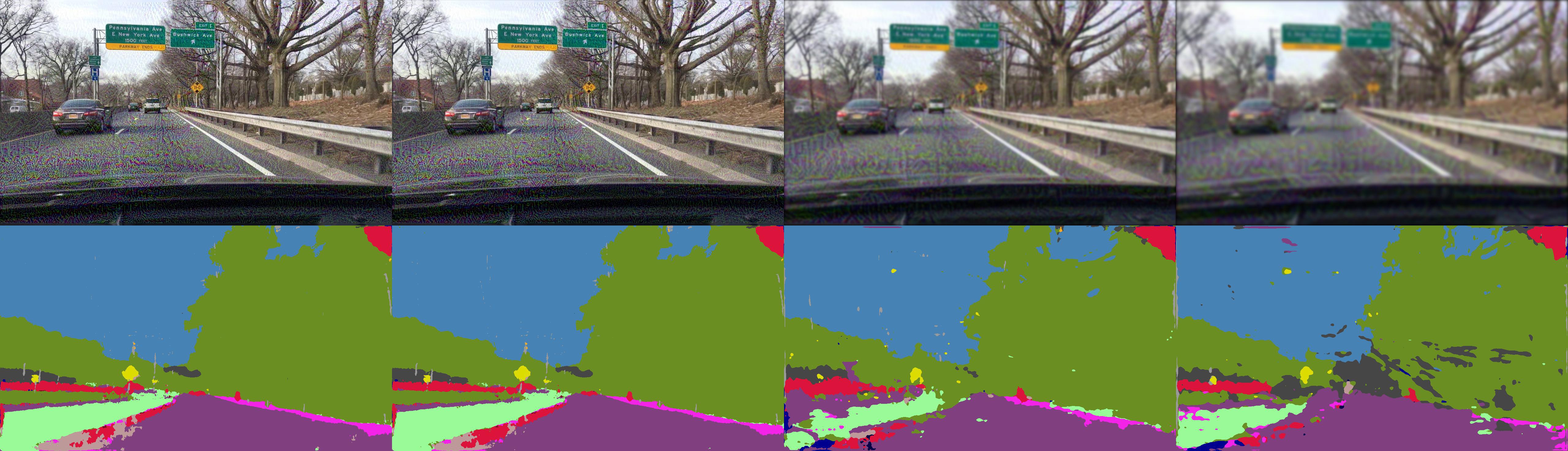}
     \caption{DAG | Scene | DLA | BDD}
     \end{subfigure}
    \end{minipage}
    \begin{minipage}{.45\textwidth}
     \begin{subfigure}{\textwidth}
     \centering
     \includegraphics[width=\textwidth]{figure_2/adaptive_blur/houdini_kitty_adap_blur_dla_bdd.jpg}
     \caption{Houdini | Kitty | DLA | BDD}
     \end{subfigure}
    \end{minipage}
    \begin{minipage}{.45\textwidth}
     \begin{subfigure}{\textwidth}
     \centering
     \includegraphics[width=\textwidth]{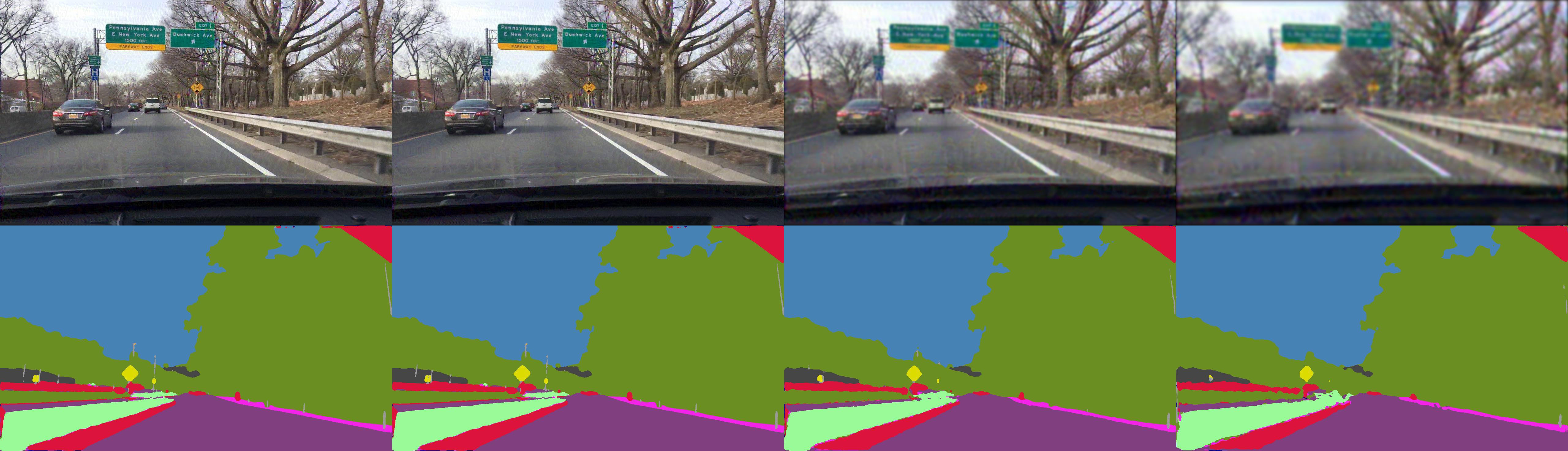}
     \caption{Houdini | Scene | DLA | BDD}
     \end{subfigure}
    \end{minipage}
    \caption{Examples of images and corresponding segmentation results for adaptive attack against image scaling. For each subfigure, the first column shows benign/adversarial images, while the following columns show images after scaling by applying Gaussian kernel with std as 0.5, 3, and 5, respectively. 
    (a), (f), (k) and (p) show benign images before/after image scaling and the corresponding segmentation results and The format ``example | attack model | dataset'' uses to identify the corresponding model and dataset; 
    (b)-(e), (g)-(j), (l)-(o) and (q)-(t) present similar results for adaptive adversarial images and we describe by the format ``attack method | target label | attack model | dataset''.
    }
    \label{fig:alladapblur}
\end{figure}

We applied image scaling to the adversarial examples generated by Houdini~\cite{cisse2017houdini} and DAG~\cite{xie2017adversarial} on Cityscapes and BDD datasets against DRN and DLA models. The result shows in Figure~\ref{fig:allblur}. We can find the same phenomenon that when we applying Gaussian blurring with high std (3 and 4), adversarial perturbation is harmed and segmentation result are no longer adversarial targets.

Table~\ref{tbl:bdd} shows that the method based on image scale information can achieve similarly AUC compared with spatial consistency based method on BDD. 
\clearpage
\subsection{Adaptive Attack Evaluation}
\begin{algorithm}[h!]
\SetAlgoLined
\SetKwInput{Input}{input}
\SetKwInput{Output}{output}
\SetKwInput{Ret}{Return}
\SetKwInput{Init}{Initialization}
\SetKwInput{Blank}{}

\tabcolsep=0pt
\nonl \begin{tabular}{@{}ll}
    \Input{}&Input image $\V{X}$ \;\\
 &Number of attack patches $K$\;\\
&Patch size $s$\; \\
&Segmentation model $f$\;\\
&$L^2$ bound $\V{b}$\;\\
& Recognition targets $\mathcal{T} = \{t_1, t_2, \cdots, t_N\}$ \;\\
&Adversarial label set $\V{Y}= \{Y_{t_1}, Y_{t_2}, \cdots, Y_{t_N}\}$\;\\
& Maximal iteration $M_{0}$ \;\\
    \Output{}& Adversarial perturbation $\V{r}$\;\\
\end{tabular}
\BlankLine
 $\V{Initialization : }  \V{X}_{0} \gets \V{X}, \V{r} \gets 0, m  \gets 0,  \mathcal{T}_{0} \gets \mathbf{Y}, w = \V{X}.width, h =\V{X}.height$\;
 \While{$m < M_{0} \And \V{l2norm}(\V{r})< \V{b}$}{
 $\mathcal{T}_{m} = \{t_n | \argmax_c \{f_{c}(\V{X}_{m}, t_{n})\} \neq Y_{t_n}\}$\;
$\V{r_{m}} \gets \sum_{t_{n} \in \mathcal{T}_{m}}  \nabla_{\V{x}_{m}}f_{Y_{t_n}}(\V{X}_{m}, t_{n})$ \;
 \tcc*[h]{Attack C random patches }\;
\For{$k\leftarrow 0$ \KwTo $K$}{

Generate two random integers $I_1,I_2$ where $0<I_1<w-s$ and $0<I_2<h-s$\;
$P_k = \V{X}[I_1:I_1+s, I_2:I_2+s]$\;
$\mathcal{T}_{k} = \{t_{n'} | \argmax_c \{f_{c}(P_k, t_{n'})\} \neq Y_{t_{n'}} \And t_{n'}\in \mathcal{T_m}\}$\;

$\V{r_{m}} \overset{+}{\leftarrow} \sum_{t_{n'} \in \mathcal{T}_{k}} \nabla_{\V{P_k}} {f_{Y_{t_{n'}}}( P_k, t_{n'})} $ 

}
$\V{r'}_{m} \gets\frac{ \gamma }{\lVert \V{r}_m \rVert}_{\infty}\V{r}_m $\;
$\V{r} \gets \V{r} + \V{r'}_m$\;
$\V{X}_{m+1} \gets \V{X}_m + \V{r'}_m$\;
$m \gets m + 1$\;
}
 \Ret{$\V{r}$}
 \caption{DAG adaptive attack against spatial consistency method }
 \label{algo:patch_attack_dag}
\end{algorithm}
\begin{algorithm}[h!]
\SetAlgoLined
\SetKwInput{Input}{input}
\SetKwInput{Output}{output}
\SetKwInput{Init}{Initialization}
\SetKwInput{Ret}{Return}
\SetKwInput{Blank}{}

\tabcolsep=0pt
\nonl \begin{tabular}{@{}ll}
    \Input{}&Input image $\V{X}$ \;\\
&Number of attack patches $K$\;\\
&Patch size $s$\; \\
&Segmentation model $f$\;\\
&$L^2$ bound $\V{b}$\;\\
&Adversarial target label $\V{Y}$\;\\
&Maximal iteration $M_{0}$ \;\\
& Houdini loss $\ell_\mathbf{H}$\;\\
    \Output{}& Adversarial perturbation $\V{r}$\;\\
\end{tabular}
\BlankLine
 $\V{Initialization : }  \V{X}_{0} \gets \V{X}, \V{r} \gets 0, m  \gets 0,  \mathcal{T_{0}} \gets \mathbf{Y}, w \gets \V{X}.width, h \gets \V{X}.height$\;
 \While{$m < M_{0} $}{
 $ \V{r_m} \gets 0$\;
 \tcc*[h]{get the gradient of the perturbation from the objective}\;
 $ \V{r_m} \gets \V{r_m} + \nabla_\V{r}\ell_\mathbf{H}(f(\mathbf{X} + \V{r}), \V{Y})$\;
 $ \V{r_m} \gets \V{r_m} + \nabla_\V{r} \max(\V{l2norm}(\V{r})-b, 0)$ \;
  \For{$k\leftarrow$ 0 \KwTo $K$}{
  Generate two random interger numbers $I_1,I_2$ where $0<I_1<w-s, 0<I_2<h-s$\;
  $ \V{r}_m \gets \V{r}_m + \nabla_\V{r}\ell_\mathbf{H}(f((\mathbf{X}+\V{r})[i:i+s, j:j+s]), \V{Y}[i:i+s, j:j+s])$\;
  }
 $ \V{r} \gets \V{r} + \V{r}_m$\;
 $m \gets m + 1$\;
}
 \Ret{$\V{r}$}
 \caption{Houdini adaptive attack against spatial consistency method }
 \label{algo:patch_attack}
\end{algorithm}
\begin{figure}[h!]
\centering
\begin{minipage}{.47\textwidth}
 \begin{subfigure}{\textwidth}
 \centering
 \includegraphics[width=\textwidth]{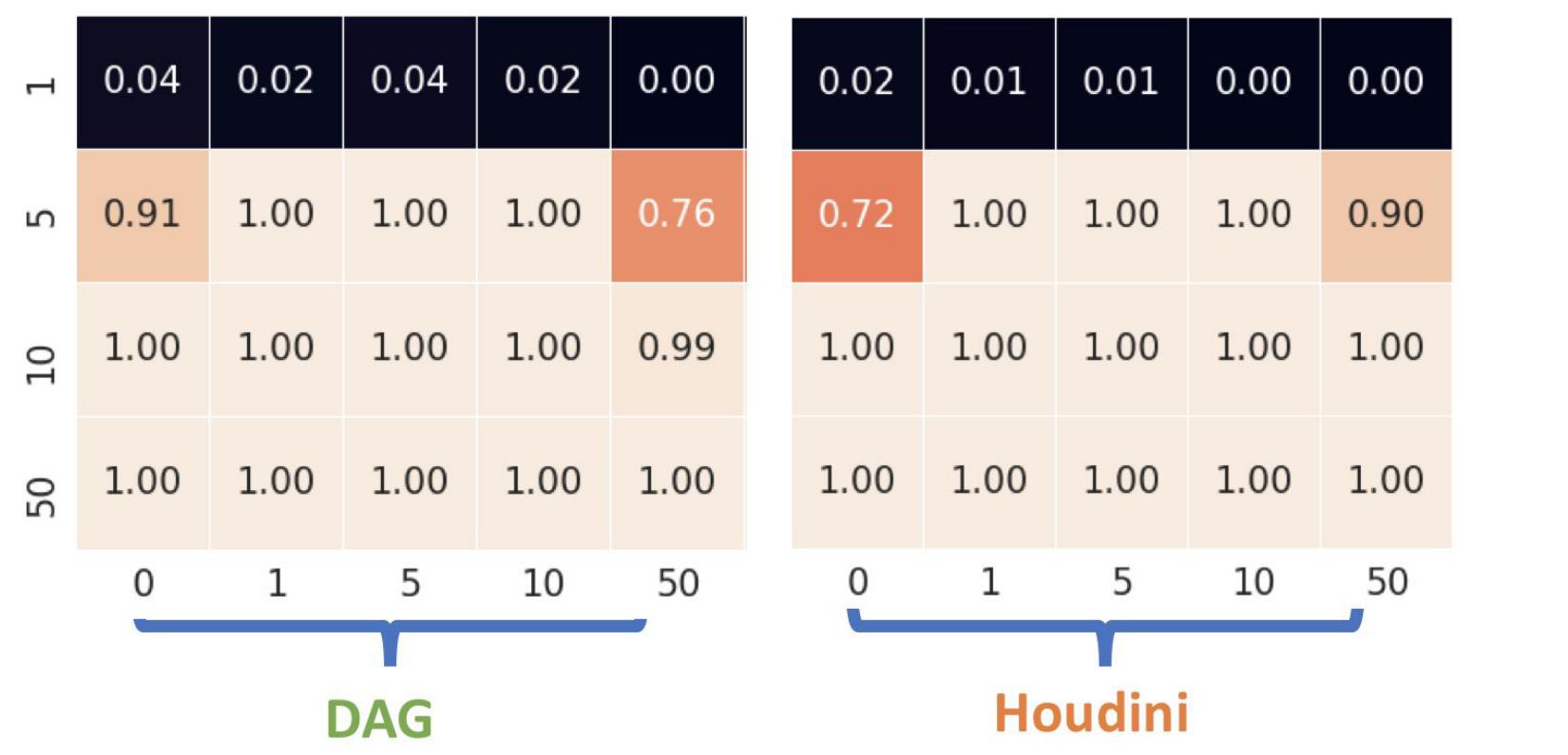}
 \caption{ DLA | Kitty | Cityscapes }
 \label{fig:attention-a}
 \end{subfigure}
\end{minipage}
\begin{minipage}{.48\textwidth}
 \begin{subfigure}{\textwidth}
 \centering
 \includegraphics[width=\textwidth]{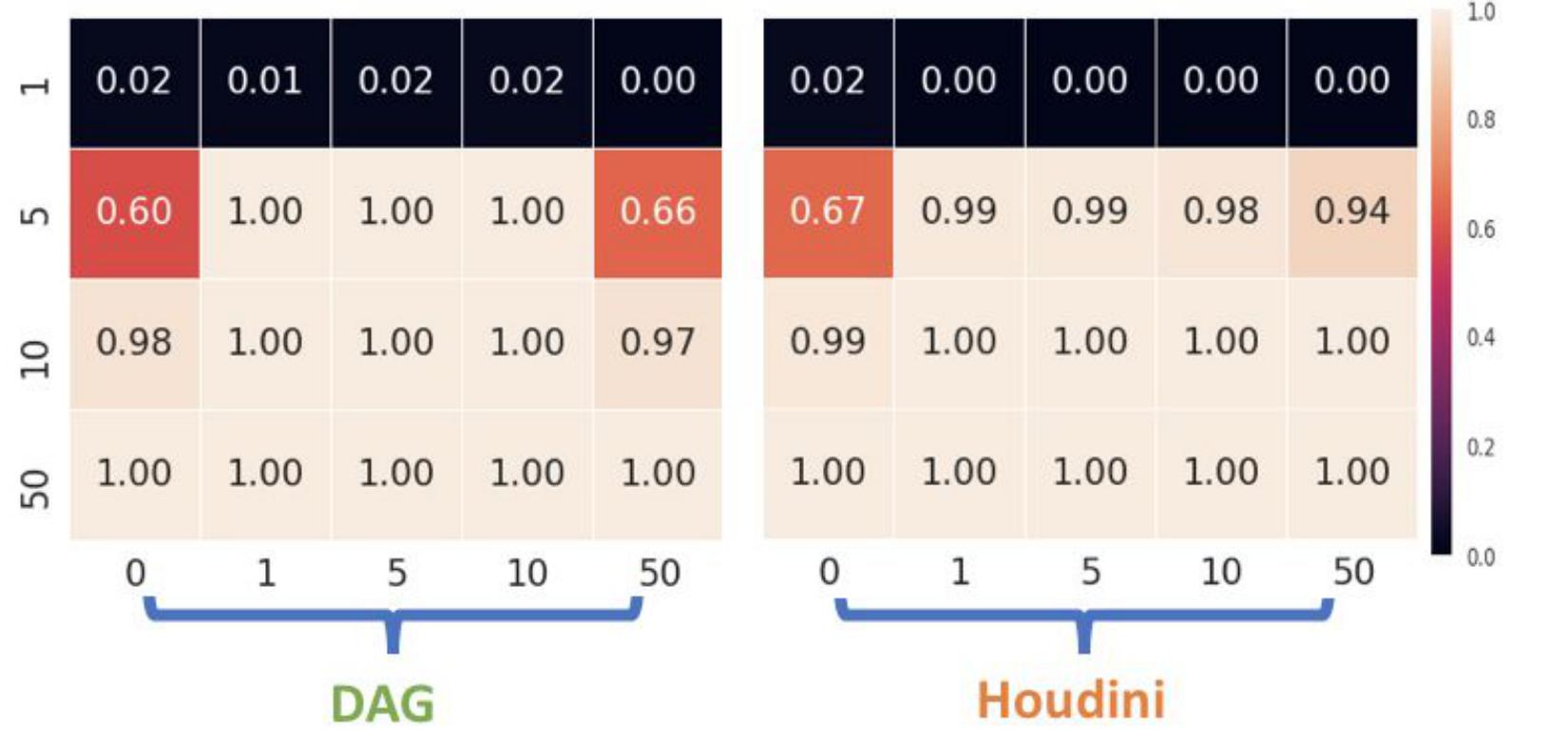}
 \caption{DLA | Pure | Citysacpes }
 \end{subfigure}
\end{minipage}
\begin{minipage}{.45\textwidth}
 \begin{subfigure}{\textwidth}
 \centering
 \includegraphics[width=\textwidth]{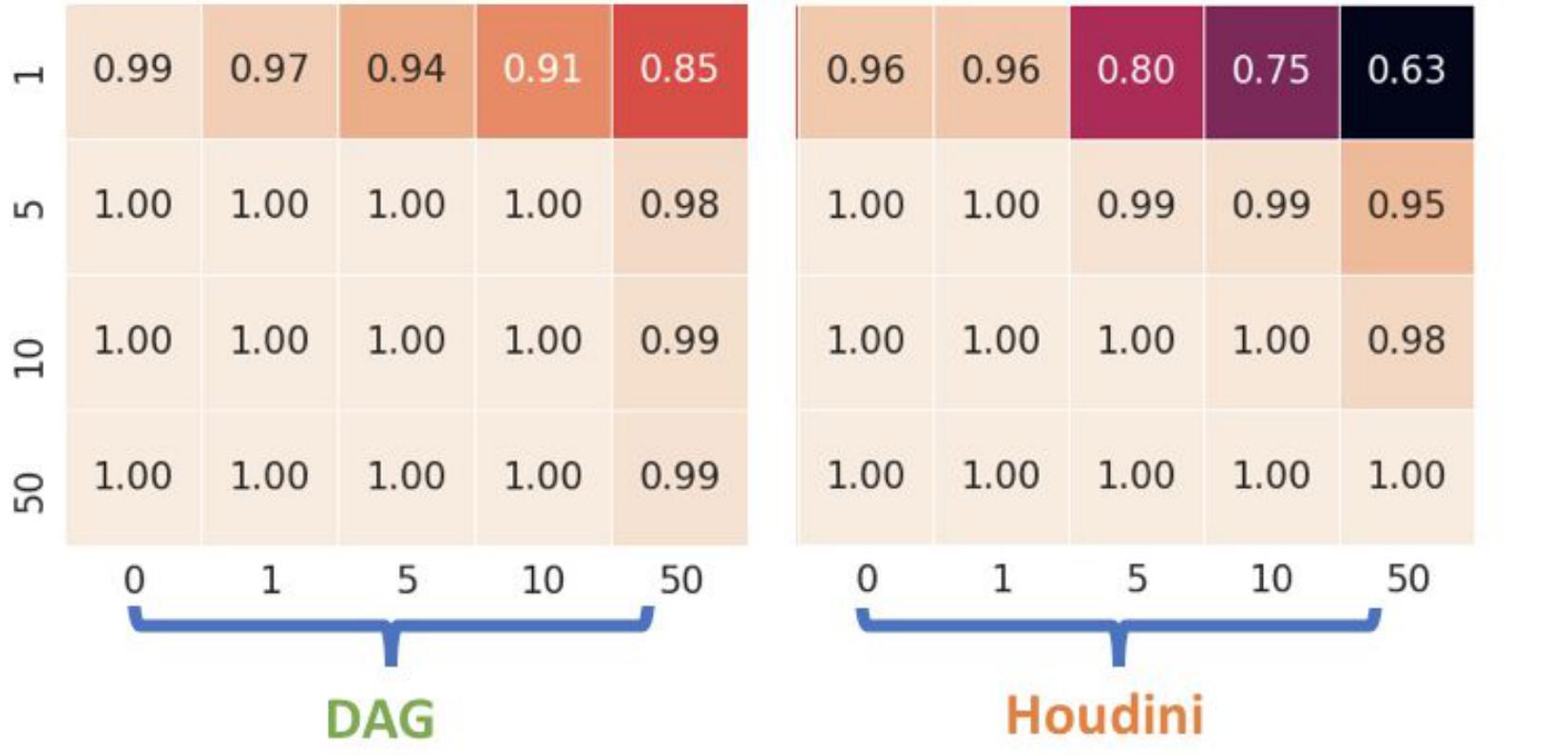}
 \caption{DRN | Kitty | BDD }
 \label{fig:attention-a}
 \end{subfigure}
\end{minipage}
\begin{minipage}{.5\textwidth}
 \begin{subfigure}{\textwidth}
 \centering
 \includegraphics[width=\textwidth]{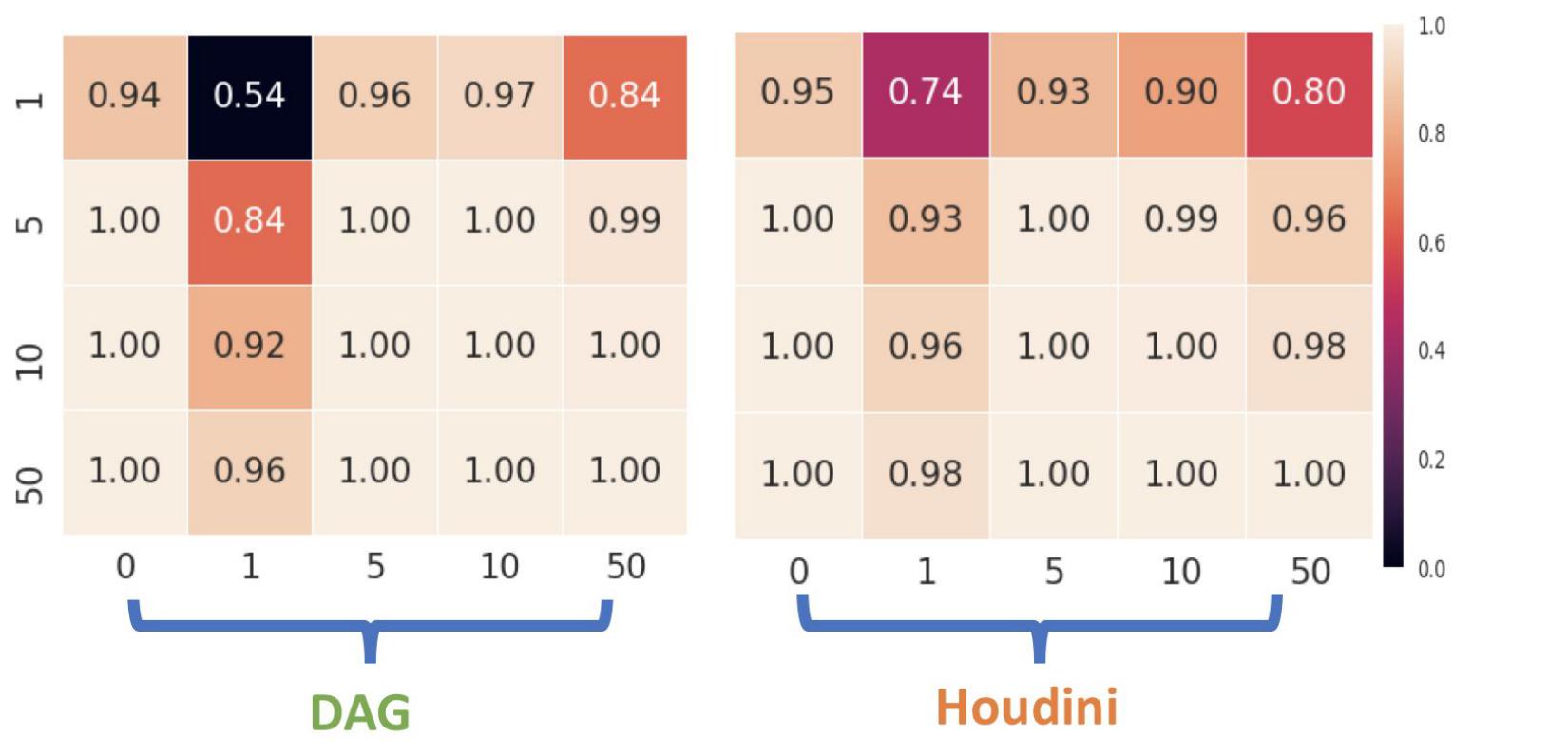}
 \caption{DRN | Scene | BDD }
 \end{subfigure}
\end{minipage}
\begin{minipage}{.45\textwidth}
 \begin{subfigure}{\textwidth}
 \centering
 \includegraphics[width=\textwidth]{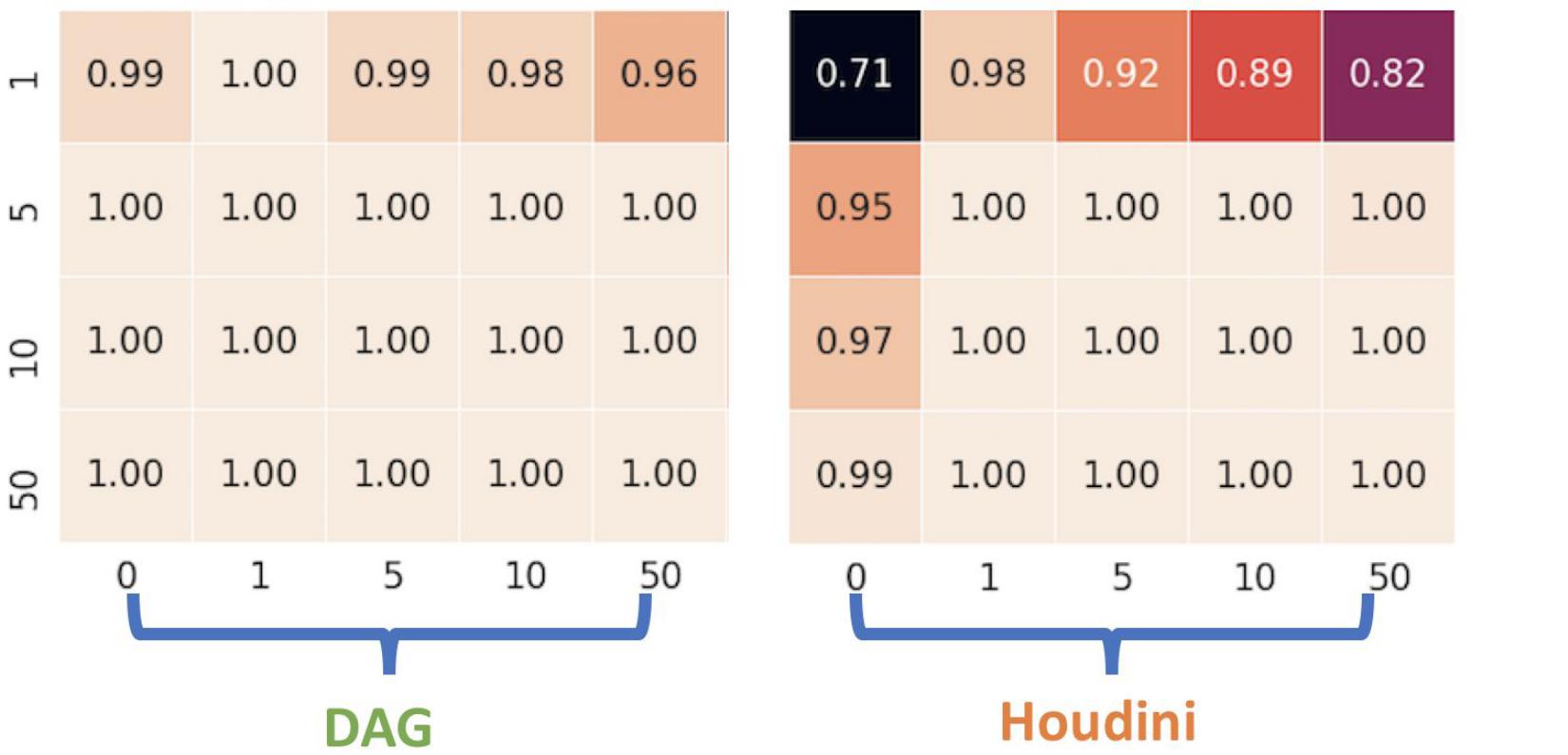}
 \caption{DLA | Kitty | BDD }
 \label{fig:attention-a}
 \end{subfigure}
\end{minipage}
\begin{minipage}{.48\textwidth}
 \begin{subfigure}{\textwidth}
 \centering
 \includegraphics[width=\textwidth]{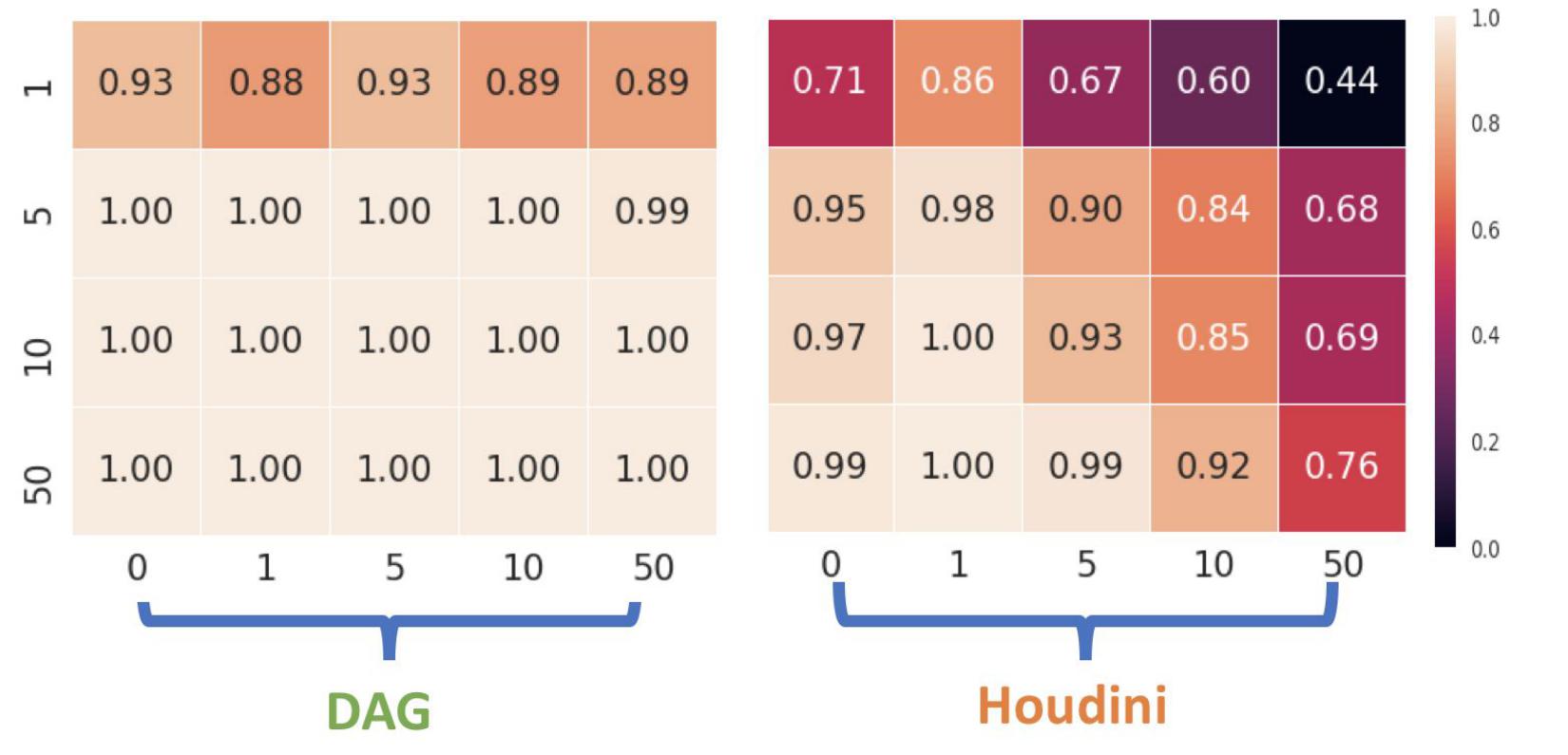}
 \caption{DLA | Scene | BDD }
 \end{subfigure}
\end{minipage}
\caption{Detection performance of \spatialconsis against adaptive attack with different $K$. We use the format ``attack model | target label | dataset'' to label the settings for each figure. X-axis indicates the number of patches selected to perform the adaptive attack (0 means regular attack). Y-axis indicates the number of overlapping regions selected during detection. We select the minimal mIOU from benign patches as our threshold on Cityscapes, and the one which guarantees accuracy as above 95\% on benign images for BDD. }
\label{fig:adapconfusion-sp}
\end{figure}
Here we illustrate the adaptive attack algorithm based on DAG and Houdini against spatial consistency method in Algorithm~\ref{algo:patch_attack_dag} and Algorithm~\ref{algo:patch_attack}. Instead of only attacking the benign image, the adaptive attack here will randomly pick some patches and attack them with the whole image together.

Let $\V{X}$ be an image and $K$ is the number of patches selected from $\V{X}$ to perform adaptive attack. We define $\mathcal{T}=\{t_1, t_2, \cdots, t_N\}$ as the set comprising the coordinates of the recognition target pixels. 
$f$ denotes the segmentation network, and we use $f(\V{X})$ to denotes the classification score of the entire image and $f(\V{X}, t_{n})$ to denote the classification score vector at pixel $n$. $\V{Y}={Y_{t_1}, Y_{t_2}, \cdots, Y_{t_N}}$ denotes the adversarial label set where $Y_{t_n}$ represents the adversarial label of pixel $n$. 
Given an input tensor $\V{r}\in \mathbb{R}^{w\times h\times c}$, the function returns its $L^2$ norm defined to be $ ||\V{r}||_2 = \sqrt{\frac{\sum_{i=1}^{w}\sum_{j=1}^{h}\sum_{k=1}^{c} \V{r}_{ijk}^2}{w\cdot h \cdot c}}$. 
In Algorithm~\ref{algo:patch_attack}, we follow the same definition of Houdini loss $\ell_H$ as proposed in the work~\cite{cisse2017houdini}.
We set the maximal number of iteration to be approximately 300\footnote{
300 is approximately three times the average number of iterations in non-adaptive attack.
} in all settings. We set $L^2$ bound to be $0.06$ for simplicity.

The detection results in term of AUC of the spatial consistency based method against such adaptive attacks on BDD and Cityscapes are shown in Table~\ref{tbl:bdd}~\ref{tbl:city-random}~\ref{tbl:bdd-random}~\ref{tbl:city-additional}~\ref{tbl:bdd-additional}. 
Even against such strong adaptive attacks, the spatial consistency based method can still achieve nearly 100\% AUC. Figure~\ref{fig:adapconfusion-sp} shows the confusion matrix of detection result for adversaries and detection method choosing various $K$. It is clear that when we choose $K=50$, it is already sufficient to detect sophisticated attacks on Cityscapes dataset against DLA model and can also achieved 100\% detection rate with  false positive rate 95\% on BDD. 

The detection results of the image scaling based method against adaptive attacks on BDD are shown in Table~\ref{tbl:bdd}. For image scaling based detection method, it can be easily attacked (AUC drops drastically). Figure~\ref{fig:alladapblur} shows the qualitative results. It is obvious that even under Gaussian kernel with large standard deviation, the adversarial example can still be fooled into predicting different malicious targets (``Kitty'', ``Pure'', ``Scene'') on Cityscapes and BDD dataset against DRN and DLA models.

\clearpage
\section{Additional Results for Transferability Analysis}
\begin{figure}[h]
    \centering
    \begin{minipage}{0.45\textwidth}
     \begin{subfigure}{\textwidth}
     \centering
     \includegraphics[width=\textwidth]{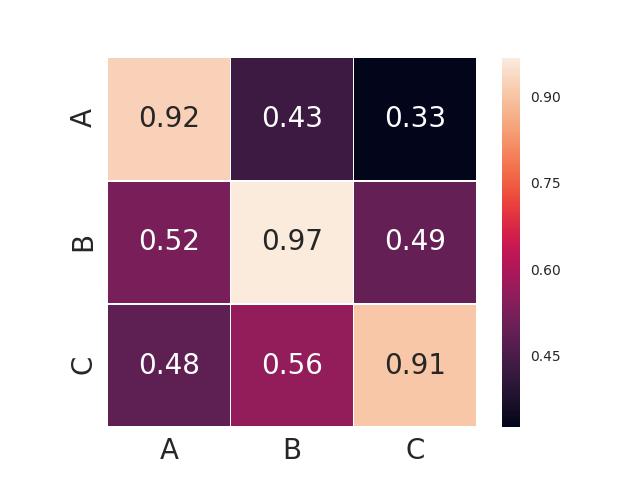}
     \caption{MNIST $|$ DRN-D-22}
     \end{subfigure}
    \end{minipage}
    \begin{minipage}{0.45\textwidth}
     \begin{subfigure}{\textwidth}
     \centering
     \includegraphics[width=\textwidth]{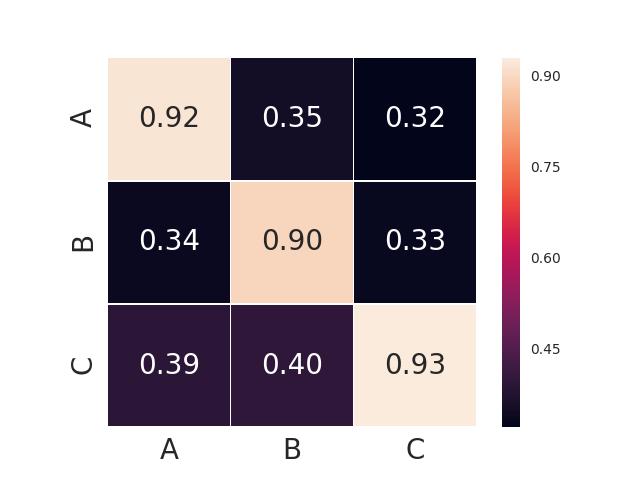}
     \caption{CIFAR10 $|$ DRN-D-22}
     \end{subfigure}
    \end{minipage}
    \begin{minipage}{0.45\textwidth}
     \begin{subfigure}{\textwidth}
     \centering
     \includegraphics[width=\textwidth]{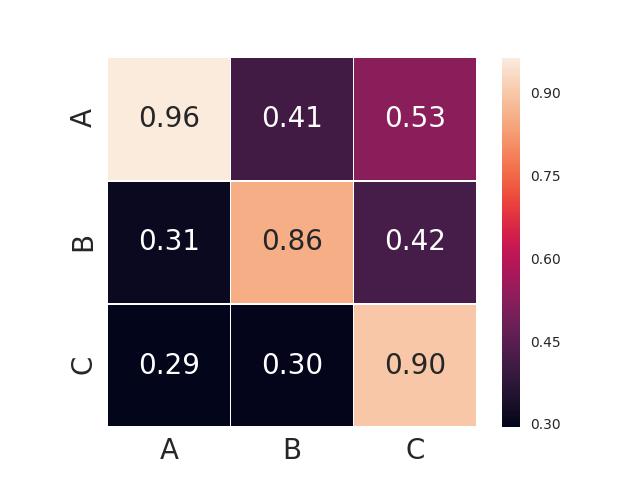}
     \caption{MNIST $|$ DRN-C-26}
     \end{subfigure}
    \end{minipage}
    \begin{minipage}{0.45\textwidth}
     \begin{subfigure}{\textwidth}
     \centering
     \includegraphics[width=\textwidth]{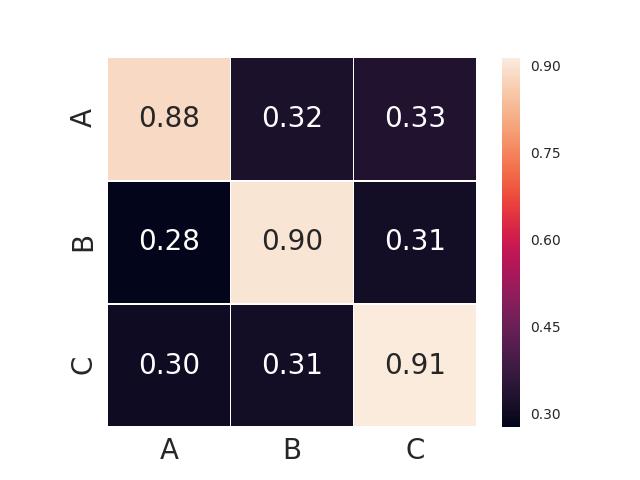}
     \caption{CIFAR10 $|$ DRN-C-26}
     \end{subfigure}
    \end{minipage}
    \begin{minipage}{0.45\textwidth}
     \begin{subfigure}{\textwidth}
     \centering
     \includegraphics[width=\textwidth]{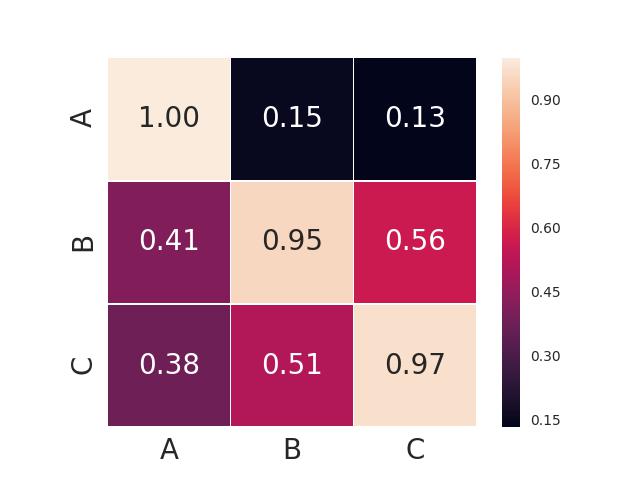}
     \caption{MNIST $|$ DLA34}
     \end{subfigure}
    \end{minipage}
    \begin{minipage}{0.45\textwidth}
     \begin{subfigure}{\textwidth}
     \centering
     \includegraphics[width=\textwidth]{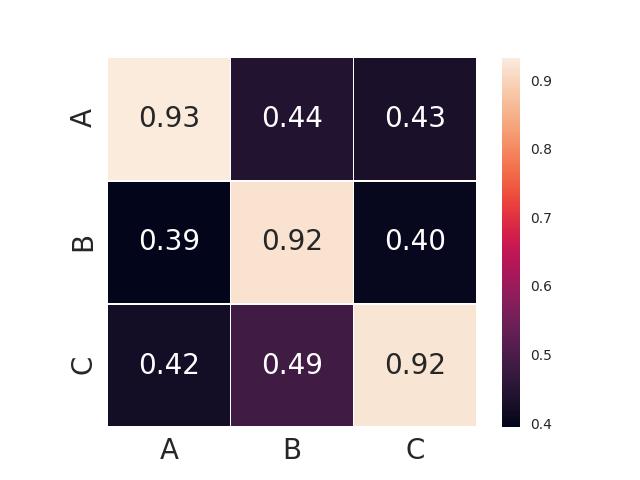}
     \caption{CIFAR10 $|$ DLA34}
     \end{subfigure}
    \end{minipage}
     \caption{Transferability analysis for classification models: cell $(i,j)$ shows the attack success rate of the adversarial examples generated against Model $j$ and evaluate on Model $i$ under targeted attack setting. Model A,B,C are model with the same architecture (DRN-D-22, DRN-C-26 or DLA34UP) and different initialization. All the adversarial examples are generated using fast iterative gradient sign method. The caption of each sub-figure bear the ``dataset $|$ model''.} 
    \label{fig:transfer-class}
\end{figure}

We present additional results for transferability analysis in Figure~\ref{fig:transfer-cityscapes} to Figure~\ref{fig:transfer-class}. 

Figure~\ref{fig:transfer-cityscapes} to Figure~\ref{fig:transfer-bdd} show the transferability analysis results for segmentation models. 
We report pixel-wise attack success rate for the pure target and normalized mIoU after eliminating K classes with the lowest IoU values for other targets. We set K to be 13 for CityScapes dataset and 5 for BDD dataset. 
Additional qualitative results are presented in Figure~\ref{fig:cityscapes-transfer-vis} to Figure~\ref{fig:bdd-transfer-vis}.

Fig.~\ref{fig:transfer-class} shows the transferability experiments results for classification models under targeted attack. The adversarial images are generated using iterative FGSM method from MNIST and CIFAR10 datasets. The caption of each sub-figure indicates the dataset and the attacked classification model.
\begin{figure}[htb]
    \centering
    \begin{minipage}{0.45\textwidth}
     \begin{subfigure}{\textwidth}
     \centering
     \includegraphics[width=\textwidth]{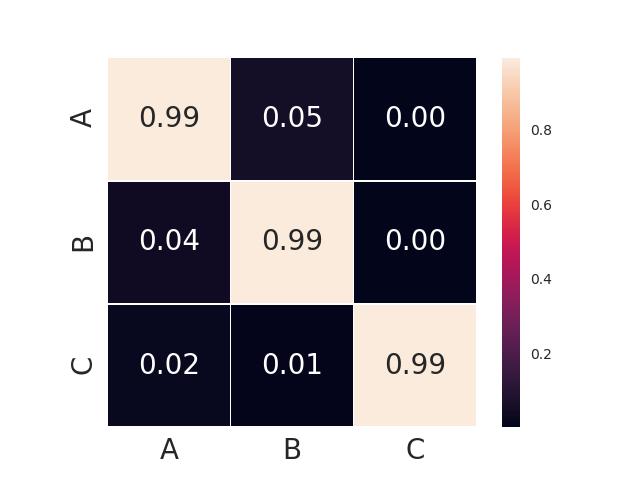}
     \caption{DAG $|$ Pure $|$ DRN-D-22}
     \label{fig:attention-a}
     \end{subfigure}
    \end{minipage}
    \begin{minipage}{0.45\textwidth}
     \begin{subfigure}{\textwidth}
     \centering
     \includegraphics[width=\textwidth]{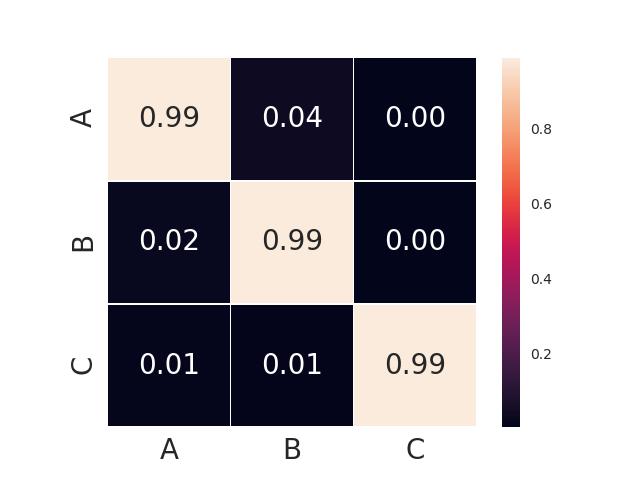}
     \caption{Houdini $|$ Pure $|$ DRN-D-22 }
     \label{fig:attention-a}
     \end{subfigure}
    \end{minipage}
    \begin{minipage}{0.45\textwidth}
     \begin{subfigure}{\textwidth}
     \centering
     \includegraphics[width=\textwidth]{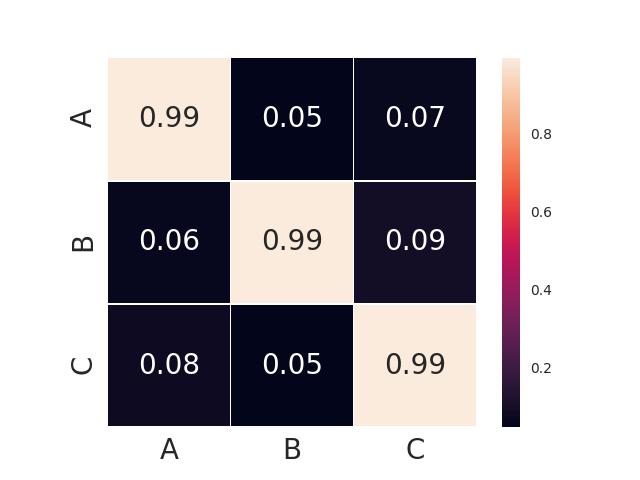}
     \caption{DAG $|$ Pure $|$ DRN-C-26}
     \label{fig:attention-a}
     \end{subfigure}
    \end{minipage}
    \begin{minipage}{0.45\textwidth}
     \begin{subfigure}{\textwidth}
     \centering
     \includegraphics[width=\textwidth]{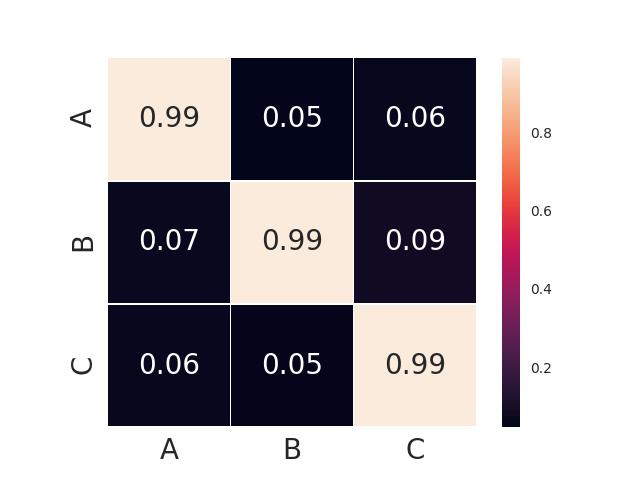}
     \caption{Houdini $|$ Pure $|$ DRN-C-26}
     \label{fig:attention-a}
     \end{subfigure}
    \end{minipage}
\end{figure}


\begin{figure}[h]
    \ContinuedFloat
    \centering
    \begin{minipage}{0.45\textwidth}
     \begin{subfigure}{\textwidth}
     \centering
     \includegraphics[width=\textwidth]{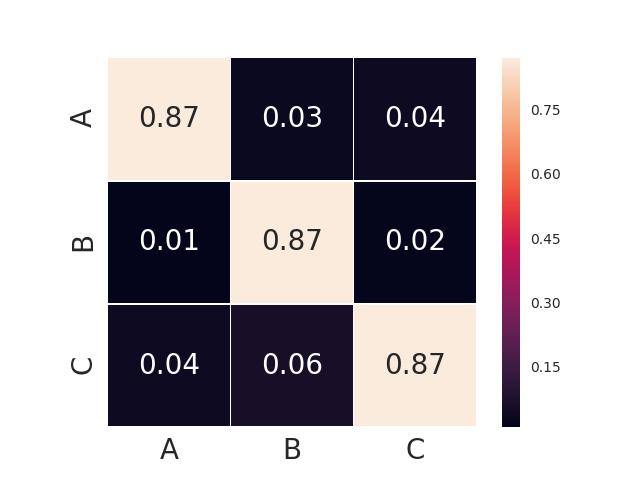}
     \caption{DAG $|$ Hello Kitty $|$ DRN-C-26}
     \label{fig:attention-a}
     \end{subfigure}
    \end{minipage}
    \begin{minipage}{0.45\textwidth}
     \begin{subfigure}{\textwidth}
     \centering
     \includegraphics[width=\textwidth]{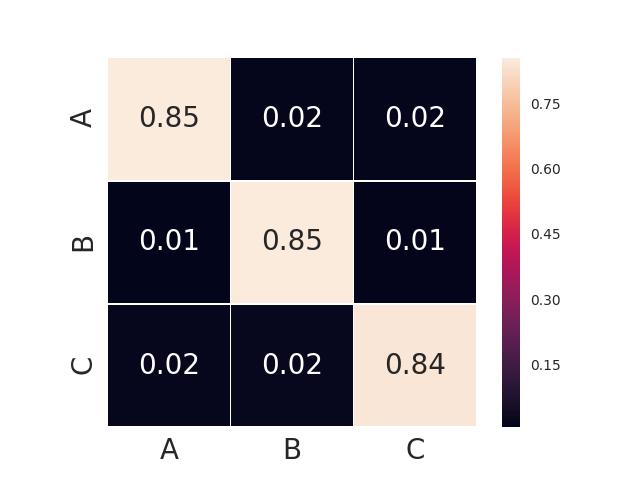}
     \caption{Houdini $|$ Hello Kitty $|$ DRN-C-26}
     \label{fig:attention-a}
     \end{subfigure}
    \end{minipage}
    \begin{minipage}{0.45\textwidth}
     \begin{subfigure}{\textwidth}
     \centering
     \includegraphics[width=\textwidth]{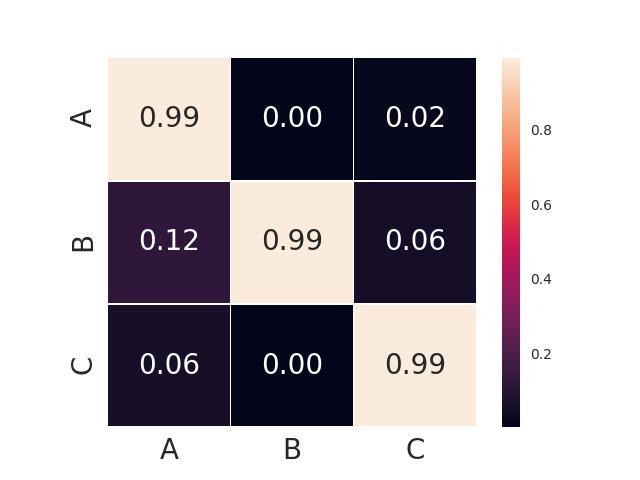}
     \caption{DAG $|$ Pure $|$ DLA34UP }
     \label{fig:attention-a}
     \end{subfigure}
    \end{minipage}
    \begin{minipage}{0.45\textwidth}
     \begin{subfigure}{\textwidth}
     \centering
     \includegraphics[width=\textwidth]{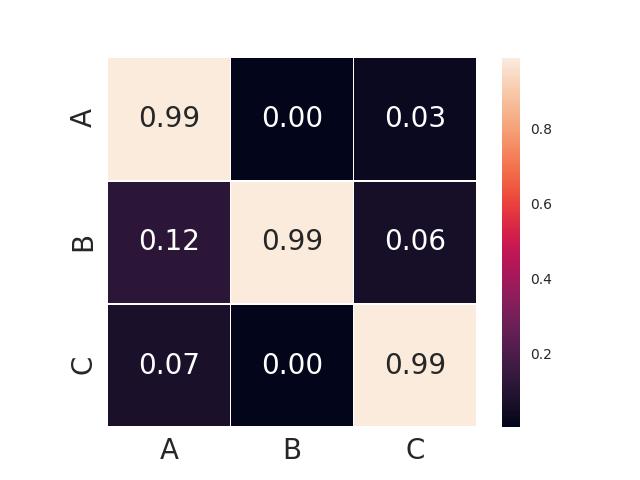}
     \caption{Houdini $|$ Pure $|$ DLA34UP }
     \label{fig:attention-a}
     \end{subfigure}
    \end{minipage}

    \begin{minipage}{0.45\textwidth}
     \begin{subfigure}{\textwidth}
     \centering
     \includegraphics[width=\textwidth]{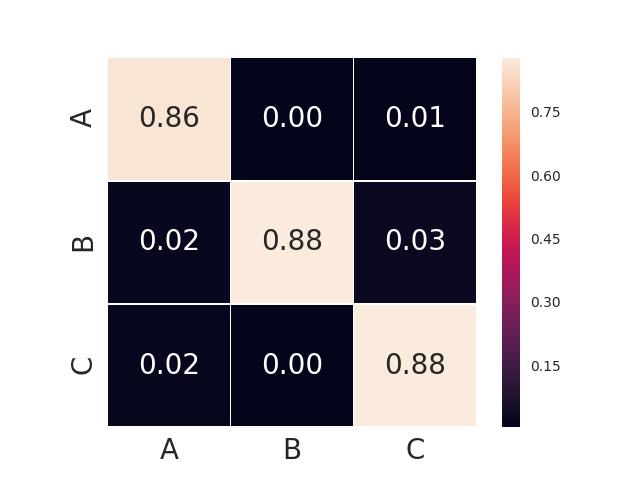}
     \caption{DAG $|$ Hello Kitty $|$ DLA34UP }
     \label{fig:attention-a}
     \end{subfigure}
    \end{minipage}
    \begin{minipage}{0.45\textwidth}
     \begin{subfigure}{\textwidth}
     \centering
     \includegraphics[width=\textwidth]{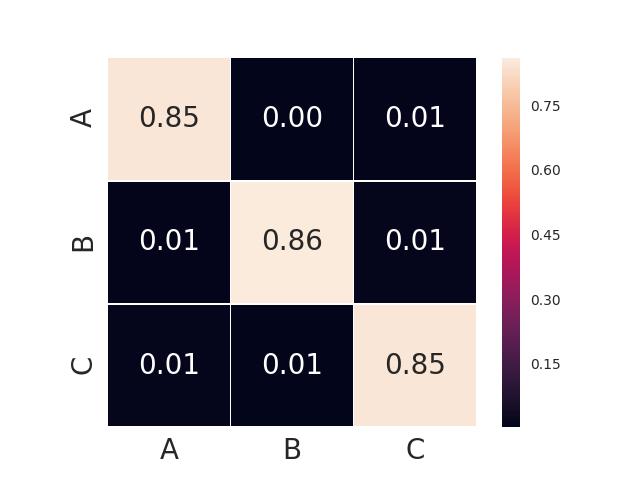}
     \caption{Houdini $|$ Hello Kitty $|$ DLA34UP }
     \label{fig:attention-a}
     \end{subfigure}
    \end{minipage}
     \caption{Transferability analysis on CityScapes dataset: cell $(i,j)$ shows the normalized mIoU value or pixel-wise attack success rate of adversarial examples generated against model $j$ and evaluate on model $i$. Model A,B,C have the same architecture (DRN-C-26 or DLA34UP) with different initialization. We use format ``attack method $|$ attack target $|$ model '' to denote the caption of each sub-figure.}
    \label{fig:transfer-cityscapes}
\end{figure}

\begin{figure}[h]
    \centering
    \begin{minipage}{0.45\textwidth}
     \begin{subfigure}{\textwidth}
     \centering
     \includegraphics[width=\textwidth]{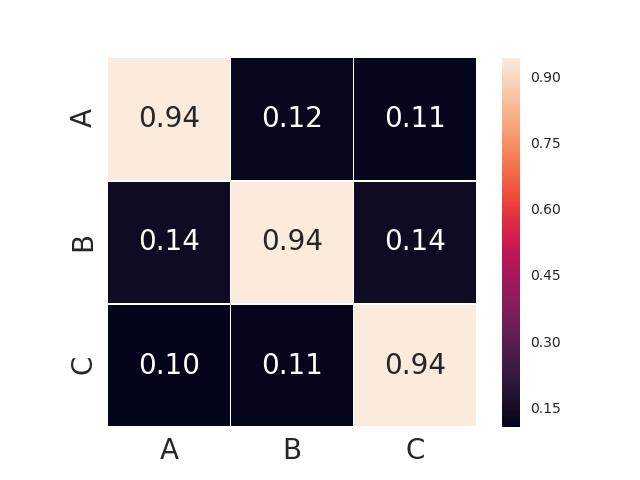}
     \caption{DAG $|$ Scene $|$ DRN-D-22}
     \label{fig:attention-a}
     \end{subfigure}
    \end{minipage}
    \begin{minipage}{0.45\textwidth}
     \begin{subfigure}{\textwidth}
     \centering
     \includegraphics[width=\textwidth]{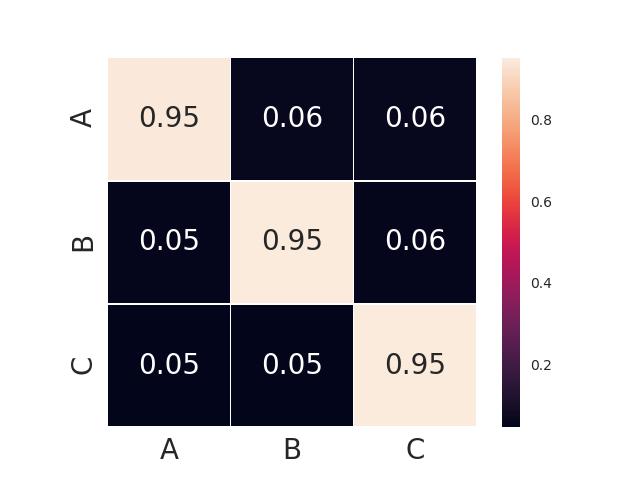}
     \caption{Houdini $|$ Scene $|$ DRN-D-22}
     \label{fig:attention-a}
     \end{subfigure}
    \end{minipage}
    \begin{minipage}{0.45\textwidth}
     \begin{subfigure}{\textwidth}
     \centering
     \includegraphics[width=\textwidth]{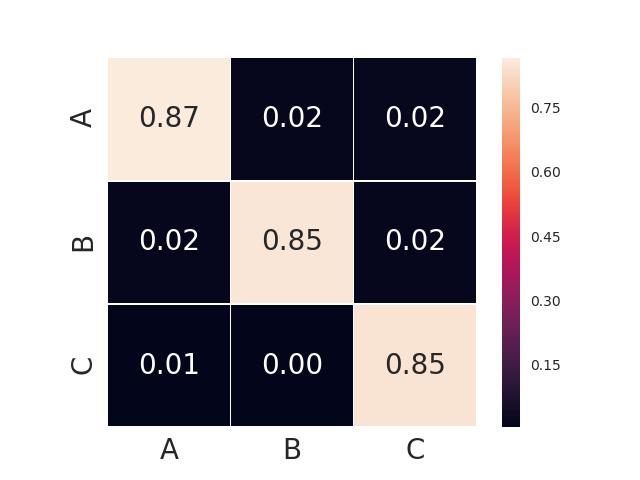}
     \caption{DAG $|$ Hello Kitty $|$ DRN-D-22}
     \label{fig:attention-a}
     \end{subfigure}
    \end{minipage}
    \begin{minipage}{0.45\textwidth}
     \begin{subfigure}{\textwidth}
     \centering
     \includegraphics[width=\textwidth]{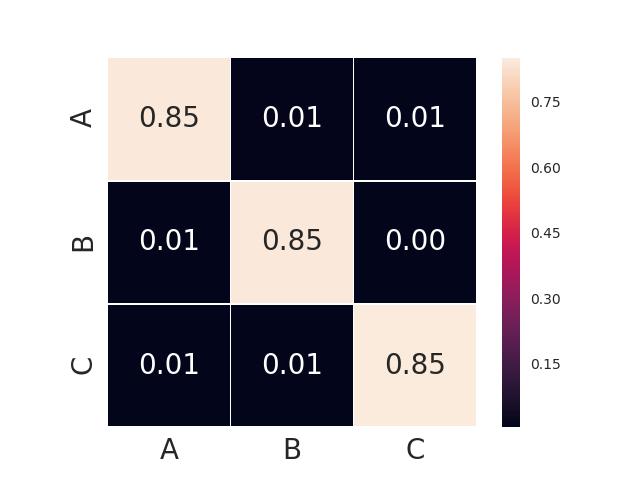}
     \caption{Houdini $|$ Hello Kitty $|$ DRN-D-22}
     \label{fig:attention-a}
     \end{subfigure}
    \end{minipage}
    \begin{minipage}{0.45\textwidth}
     \begin{subfigure}{\textwidth}
     \centering
     \includegraphics[width=\textwidth]{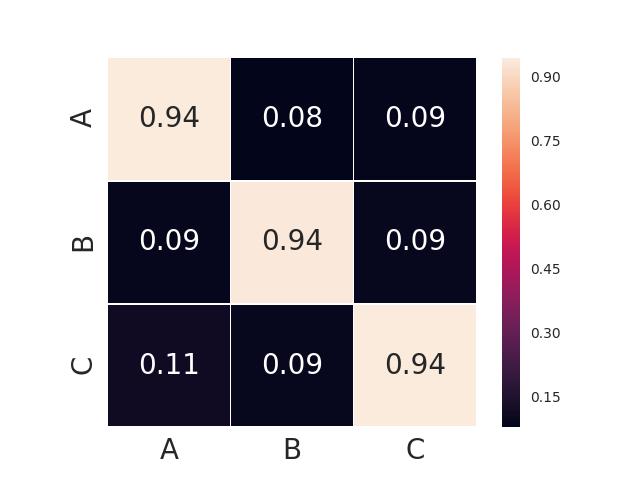}
     \caption{DAG $|$ Scene $|$ DRN-C-26}
     \label{fig:attention-a}
     \end{subfigure}
    \end{minipage}
    \begin{minipage}{0.45\textwidth}
     \begin{subfigure}{\textwidth}
     \centering
     \includegraphics[width=\textwidth]{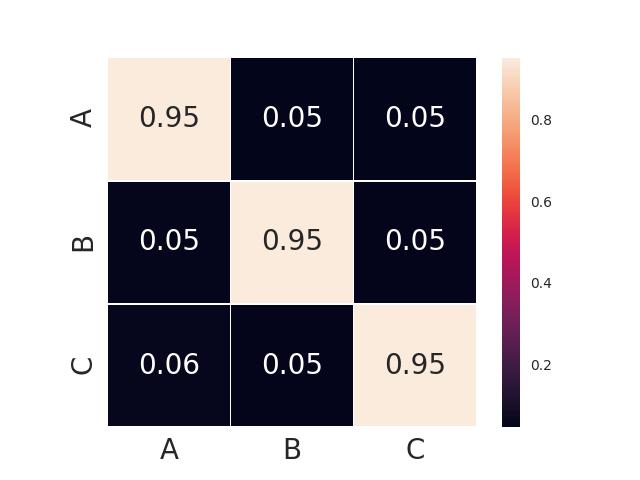}
     \caption{Houdini $|$ Scene $|$ DRN-C-26}
     \label{fig:attention-a}
     \end{subfigure}
    \end{minipage}
    \begin{minipage}{0.45\textwidth}
     \begin{subfigure}{\textwidth}
     \centering
     \includegraphics[width=\textwidth]{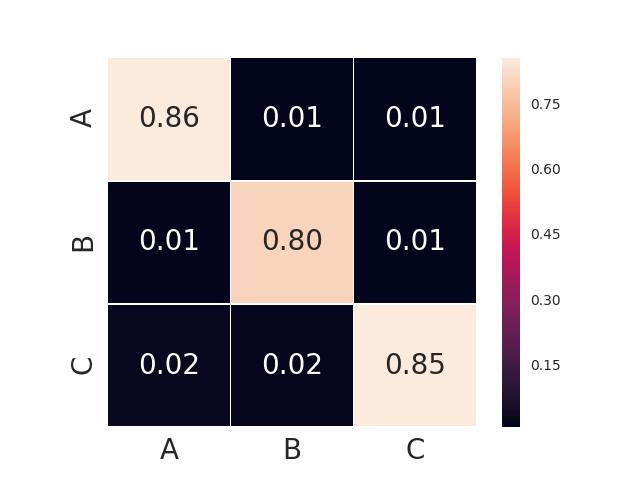}
     \caption{DAG $|$ Hello Kitty $|$ DRN-C-26}
     \label{fig:attention-a}
     \end{subfigure}
    \end{minipage}
    \begin{minipage}{0.45\textwidth}
     \begin{subfigure}{\textwidth}
     \centering
     \includegraphics[width=\textwidth]{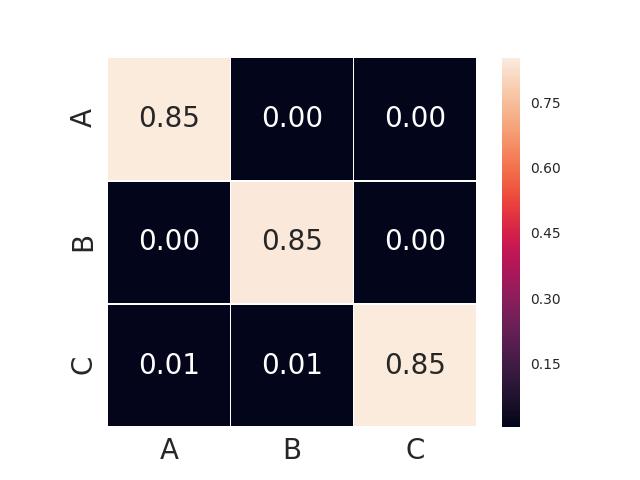}
     \caption{Houdini $|$ Hello Kitty $|$ DRN-C-26}
     \label{fig:attention-a}
     \end{subfigure}
    \end{minipage}
\end{figure}

\begin{figure}[h]
    \ContinuedFloat
    \centering
    \begin{minipage}{0.45\textwidth}
     \begin{subfigure}{\textwidth}
     \centering
     \includegraphics[width=\textwidth]{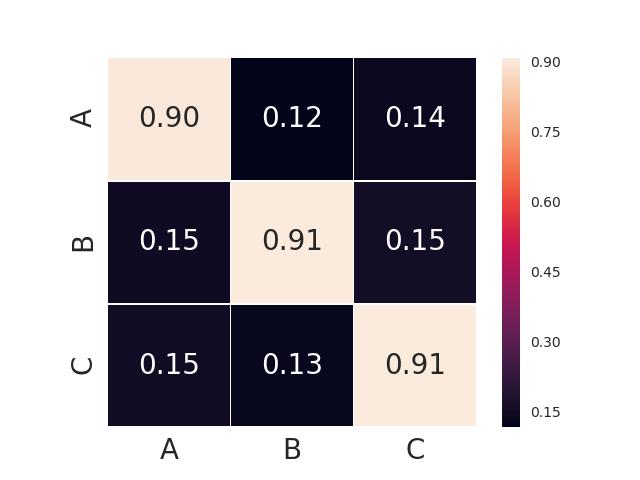}
     \caption{DAG $|$ Scene $|$ DLA34UP}
     \label{fig:attention-a}
     \end{subfigure}
    \end{minipage}
    \begin{minipage}{0.45\textwidth}
     \begin{subfigure}{\textwidth}
     \centering
     \includegraphics[width=\textwidth]{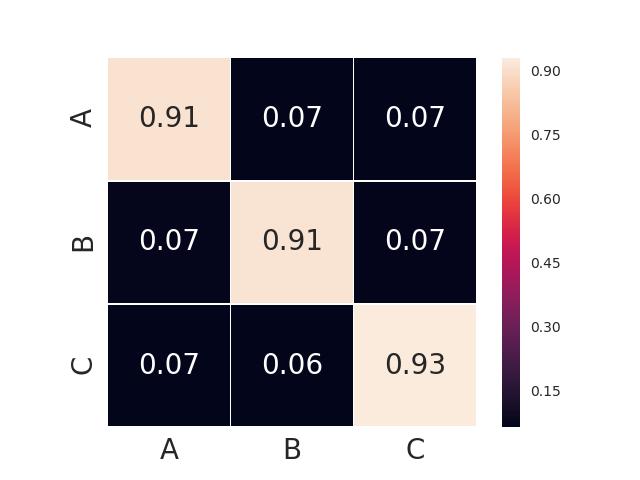}
     \caption{Houdini $|$ Scene $|$ DLA34UP}
     \label{fig:attention-a}
     \end{subfigure}
    \end{minipage}
    \begin{minipage}{0.45\textwidth}
     \begin{subfigure}{\textwidth}
     \centering
     \includegraphics[width=\textwidth]{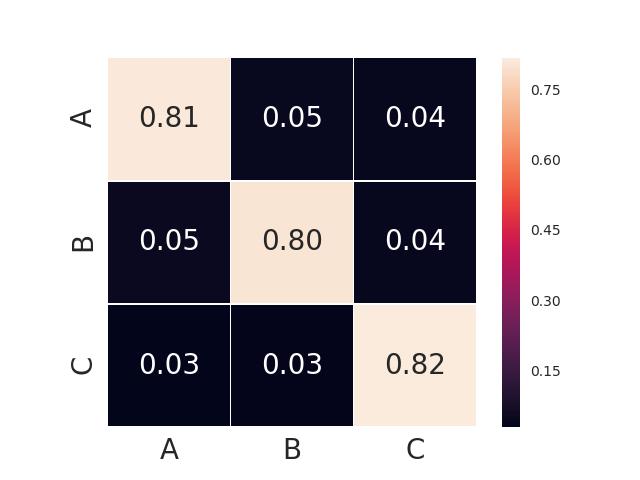}
     \caption{DAG $|$ Hello Kitty $|$ DLA34UP}
     \label{fig:attention-a}
     \end{subfigure}
    \end{minipage}
    \begin{minipage}{0.45\textwidth}
     \begin{subfigure}{\textwidth}
     \centering
     \includegraphics[width=\textwidth]{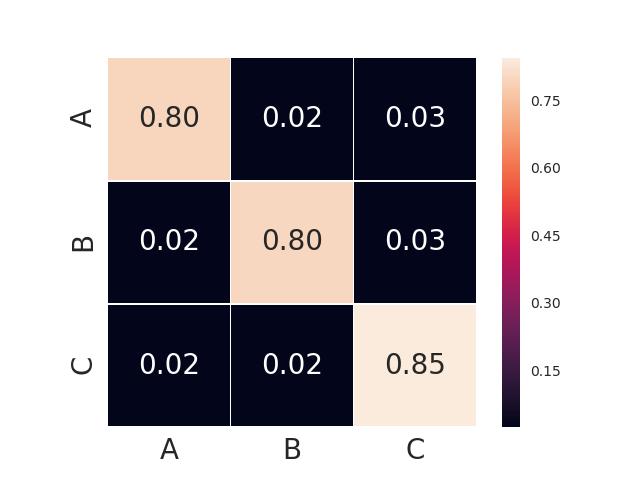}
     \caption{Houdini $|$ Hello Kitty $|$ DLA34UP}
     \label{fig:attention-a}
     \end{subfigure}
    \end{minipage}
    \caption{Transferability analysis on BDD dataset.}
    \label{fig:transfer-bdd}
\end{figure}

\begin{figure}[h]
    \centering
    \begin{minipage}{0.45\textwidth}
     \begin{subfigure}{\textwidth}
     \centering
     \includegraphics[width=\textwidth]{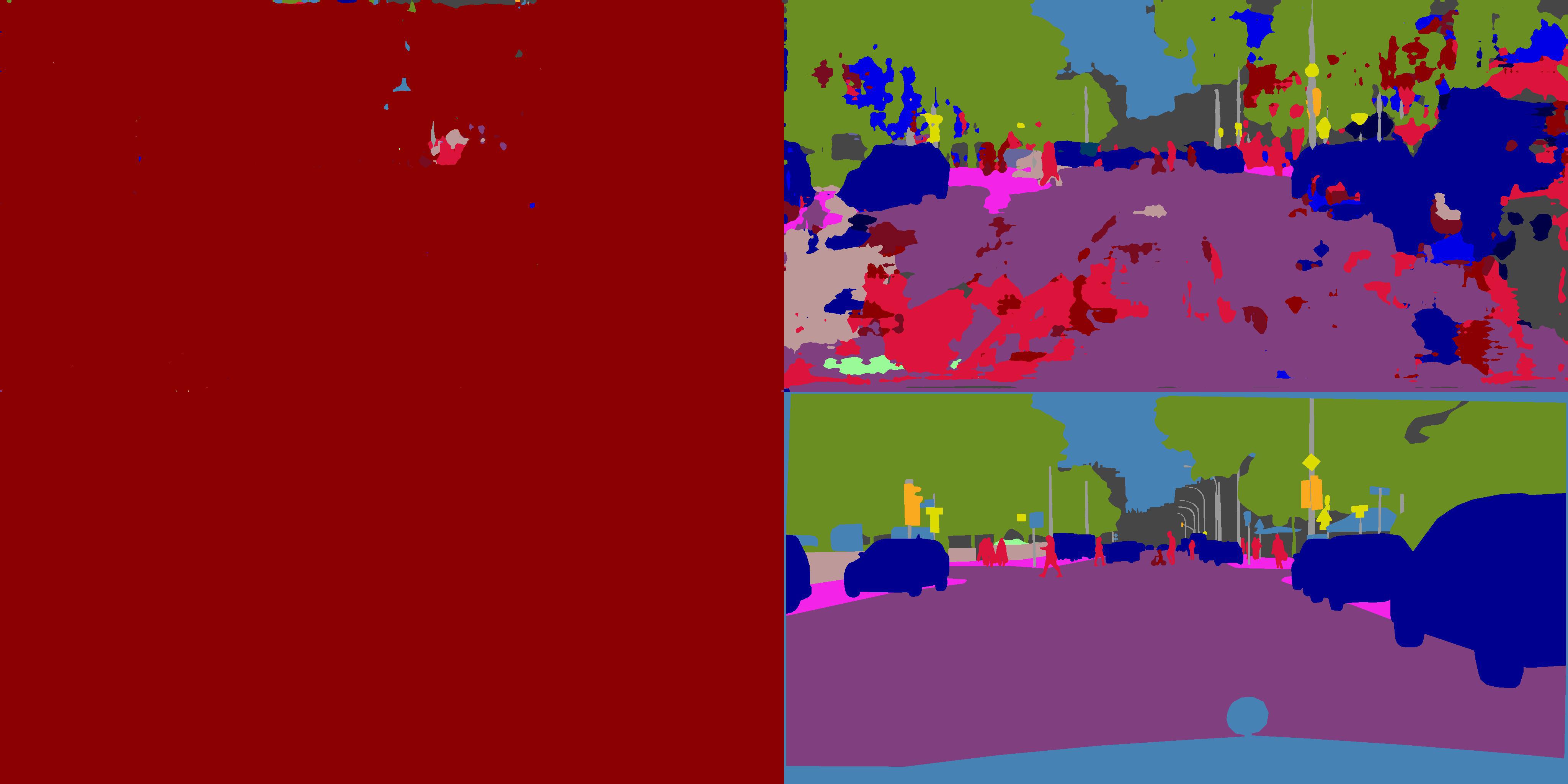}
     \caption{DAG $|$ Pure $|$ DRN-D-22}
     \label{fig:attention-a}
     \end{subfigure}
    \end{minipage}
    \begin{minipage}{0.45\textwidth}
     \begin{subfigure}{\textwidth}
     \centering
     \includegraphics[width=\textwidth]{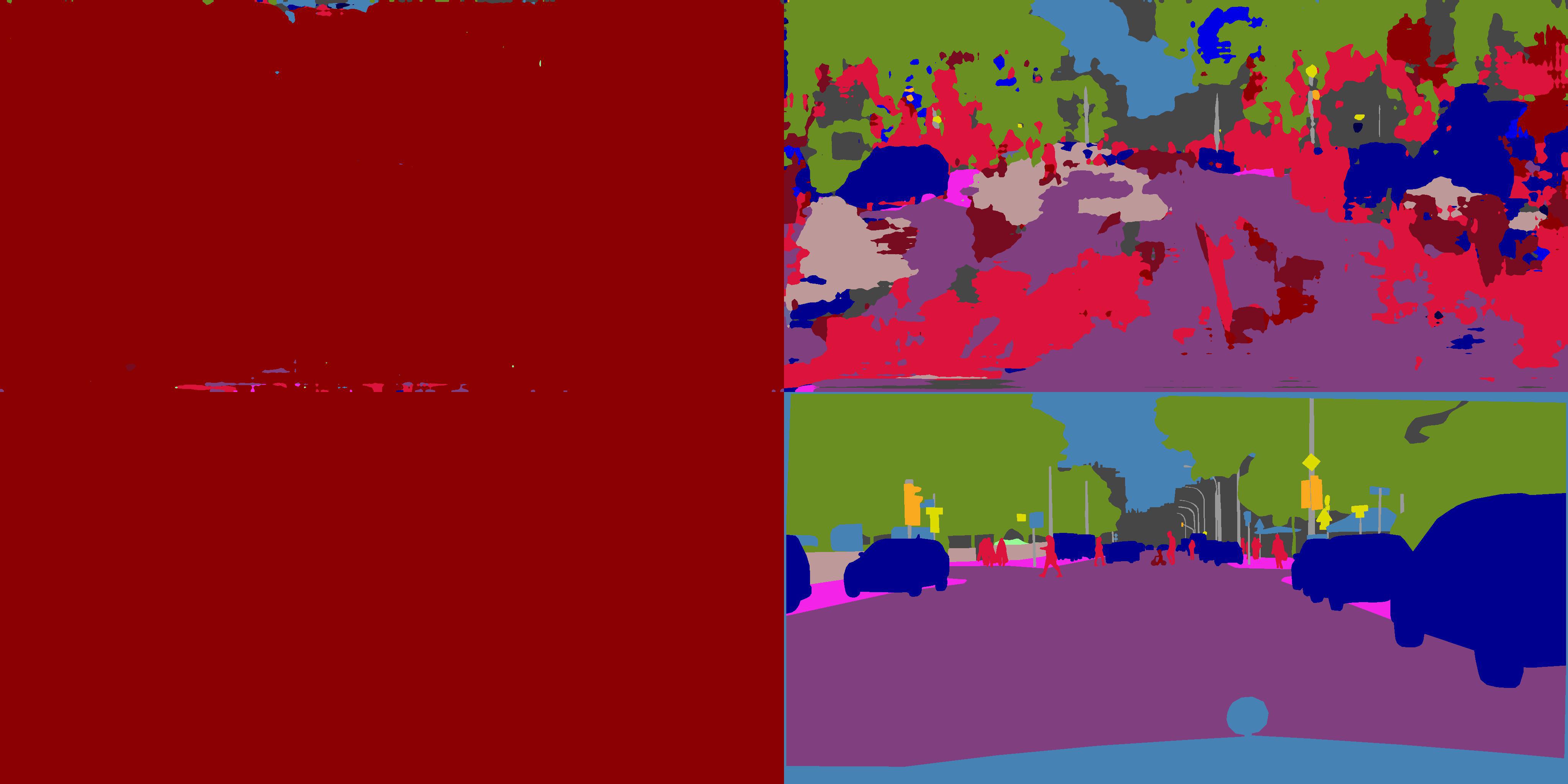}
     \caption{Houdini $|$ Pure $|$ DRN-D-22}
     \label{fig:attention-a}
     \end{subfigure}
    \end{minipage}
    \begin{minipage}{0.45\textwidth}
     \begin{subfigure}{\textwidth}
     \centering
     \includegraphics[width=\textwidth]{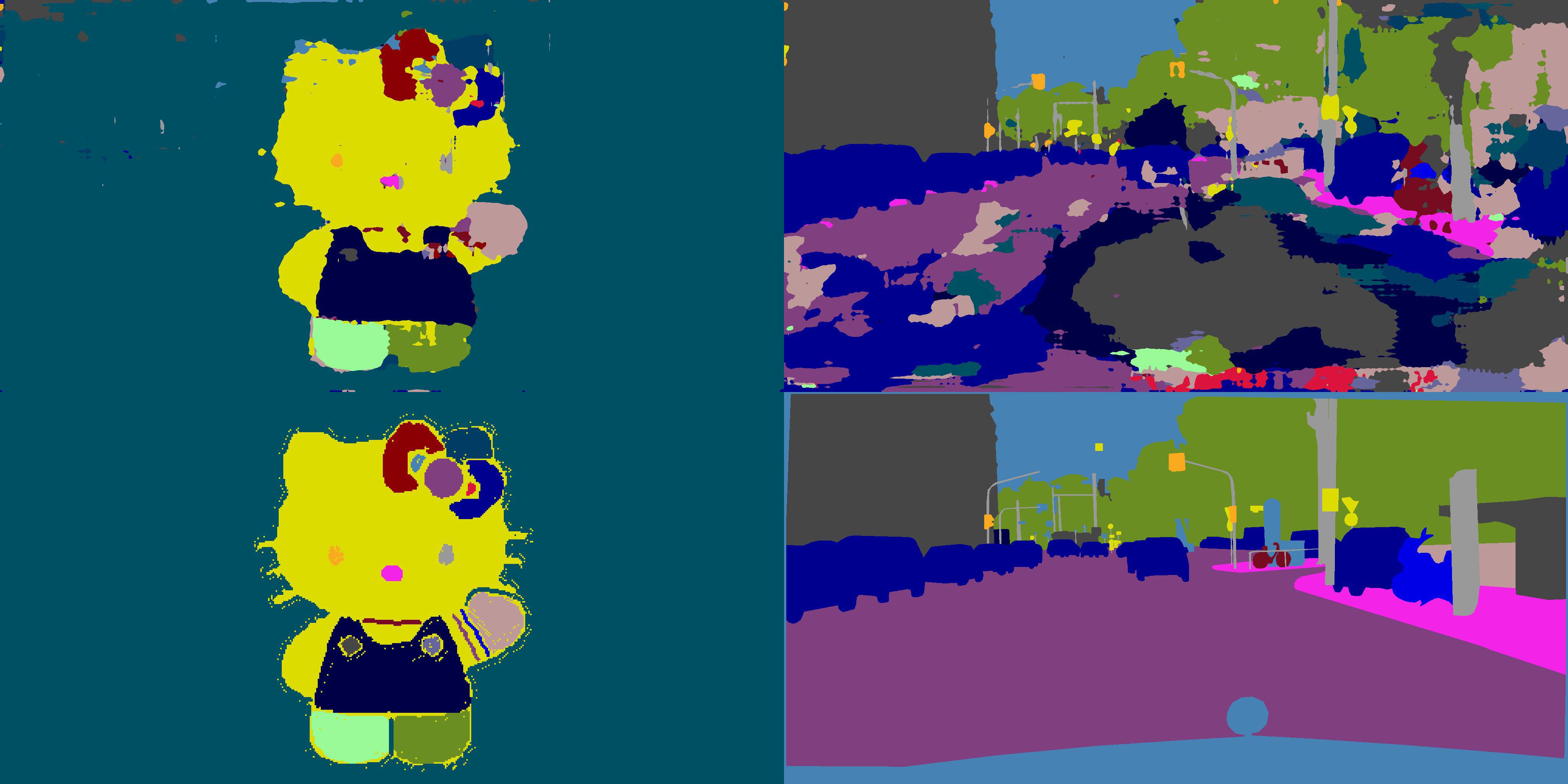}
     \caption{DAG $|$ Hello Kitty $|$ DRN-D-22}
     \label{fig:attention-a}
     \end{subfigure}
    \end{minipage}
    \begin{minipage}{0.45\textwidth}
     \begin{subfigure}{\textwidth}
     \centering
     \includegraphics[width=\textwidth]{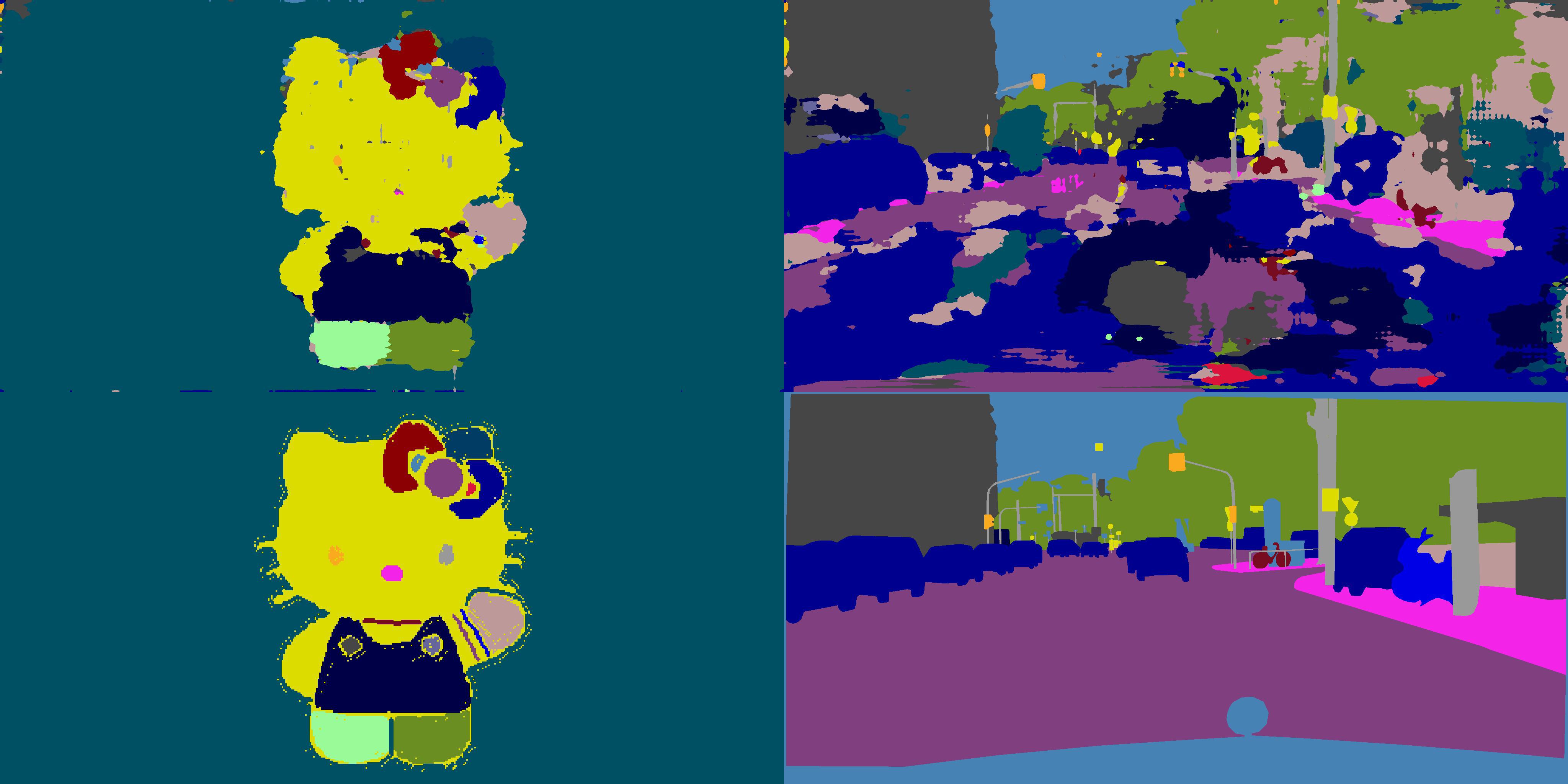}
     \caption{Houdini $|$ Hello Kitty $|$ DRN-D-22}
     \label{fig:attention-a}
     \end{subfigure}
    \end{minipage}
\end{figure}

\begin{figure}[h]
    \ContinuedFloat
    \centering
    \begin{minipage}{0.45\textwidth}
     \begin{subfigure}{\textwidth}
     \centering
     \includegraphics[width=\textwidth]{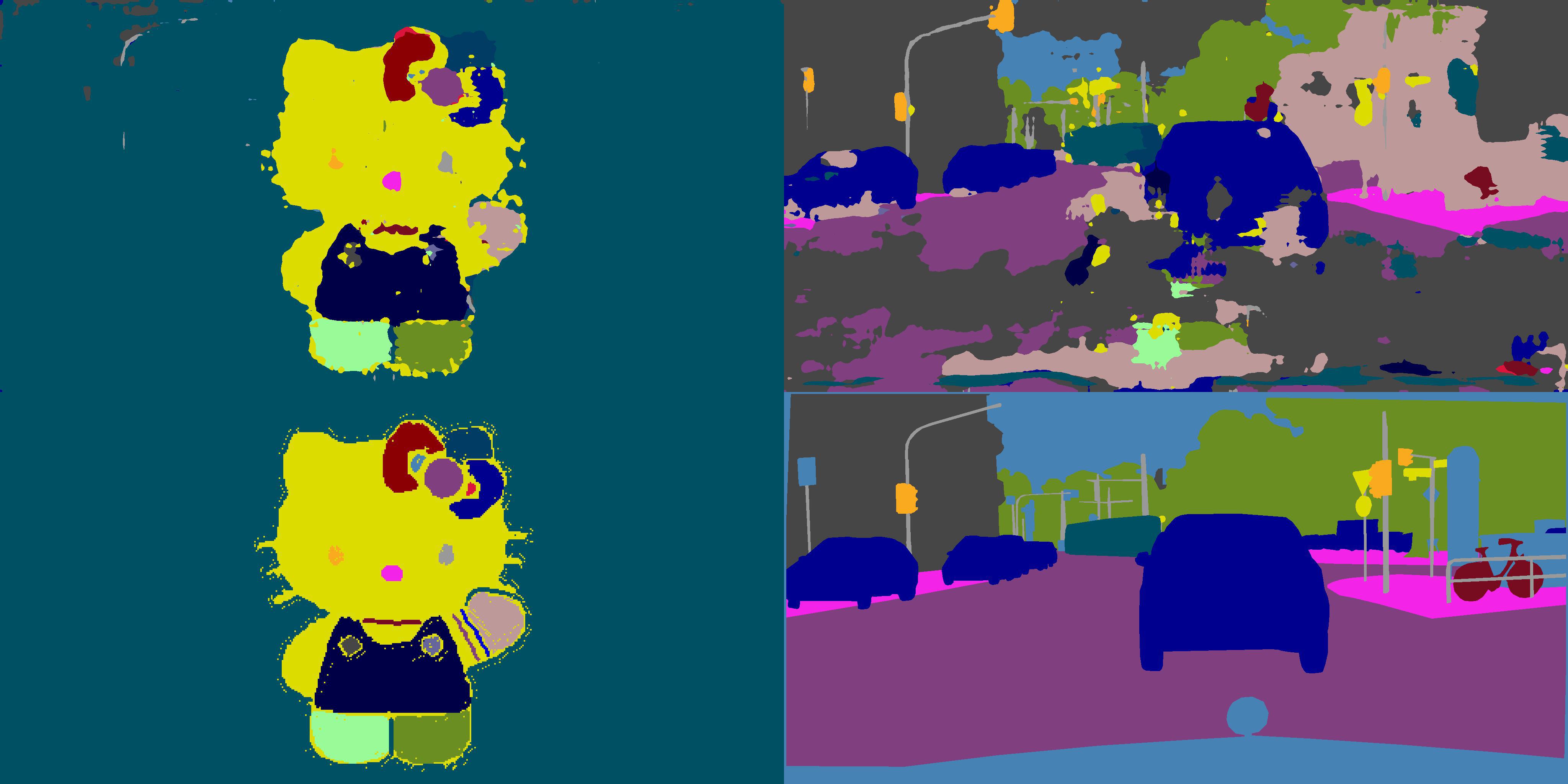}
     \caption{DAG $|$ Pure Rider $|$ DRN-C-26}
     \label{fig:attention-a}
     \end{subfigure}
    \end{minipage}
    \begin{minipage}{0.45\textwidth}
     \begin{subfigure}{\textwidth}
     \centering
     \includegraphics[width=\textwidth]{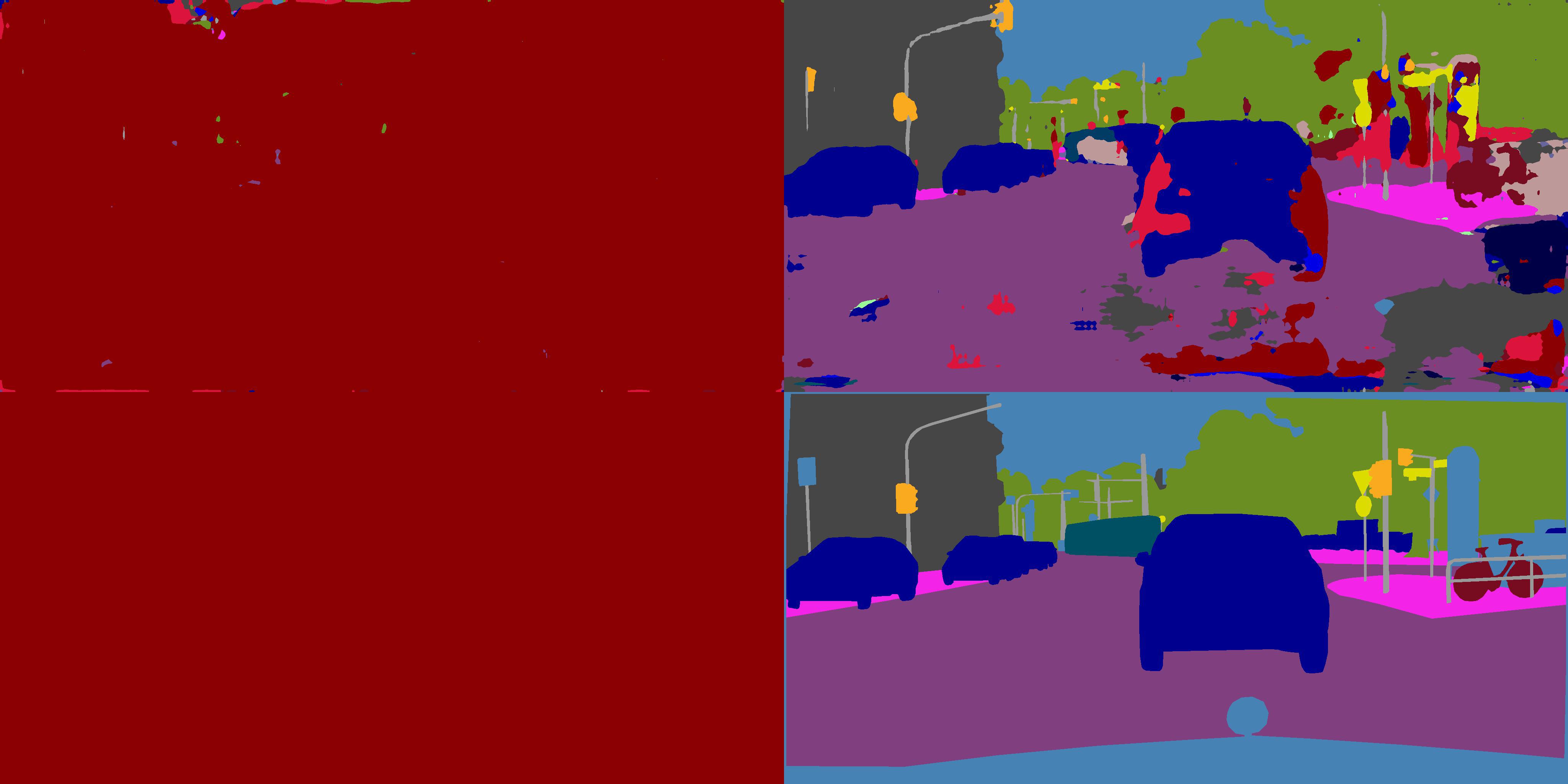}
     \caption{Houdini $|$ Pure Rider $|$ DRN-C-26}
     \label{fig:attention-a}
     \end{subfigure}
    \end{minipage}
    \begin{minipage}{0.45\textwidth}
     \begin{subfigure}{\textwidth}
     \centering
     \includegraphics[width=\textwidth]{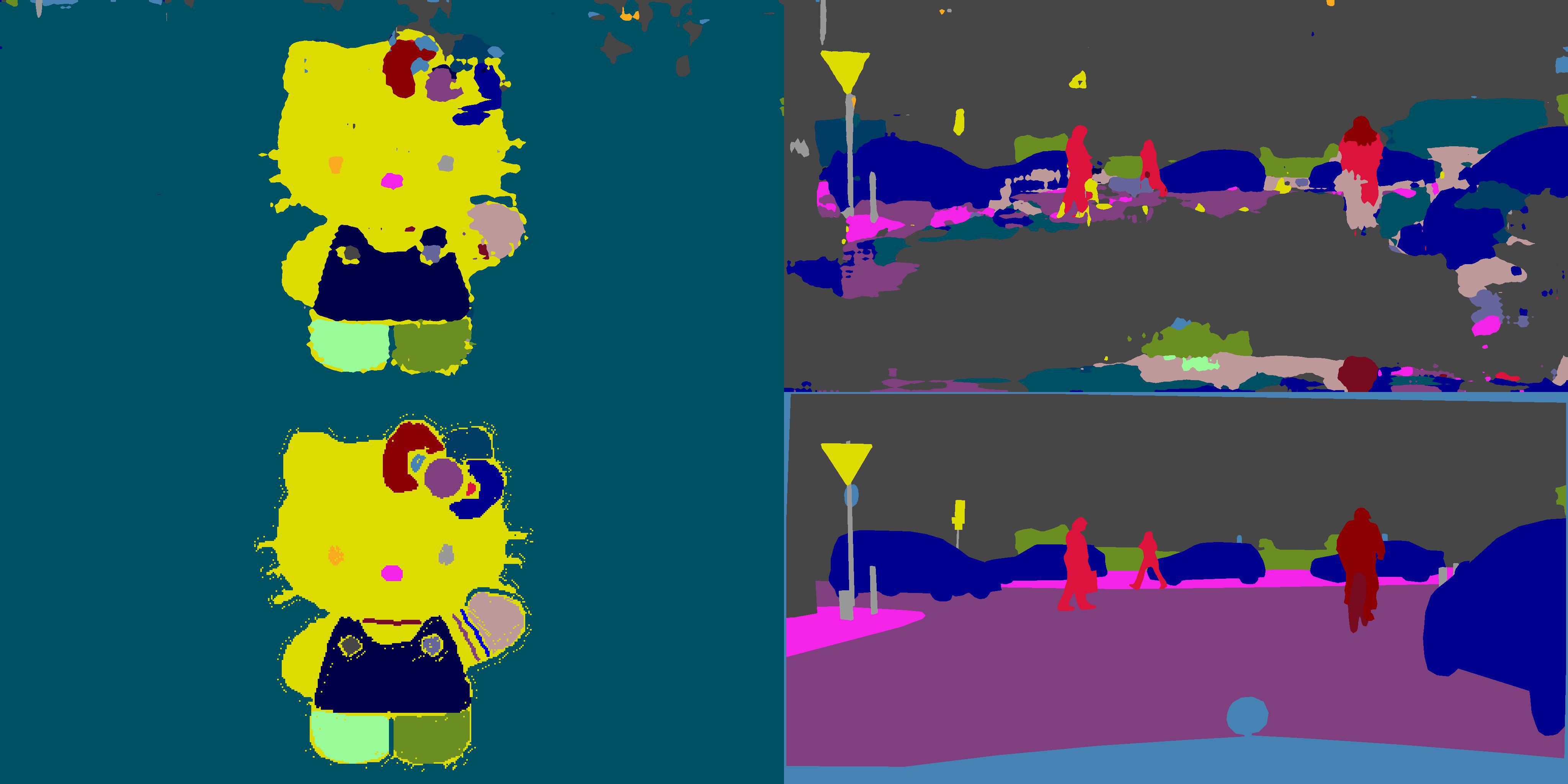}
     \caption{DAG $|$ Hello Kitty $|$ DRN-C-26}
     \label{fig:attention-a}
     \end{subfigure}
    \end{minipage}
    \begin{minipage}{0.45\textwidth}
     \begin{subfigure}{\textwidth}
     \centering
     \includegraphics[width=\textwidth]{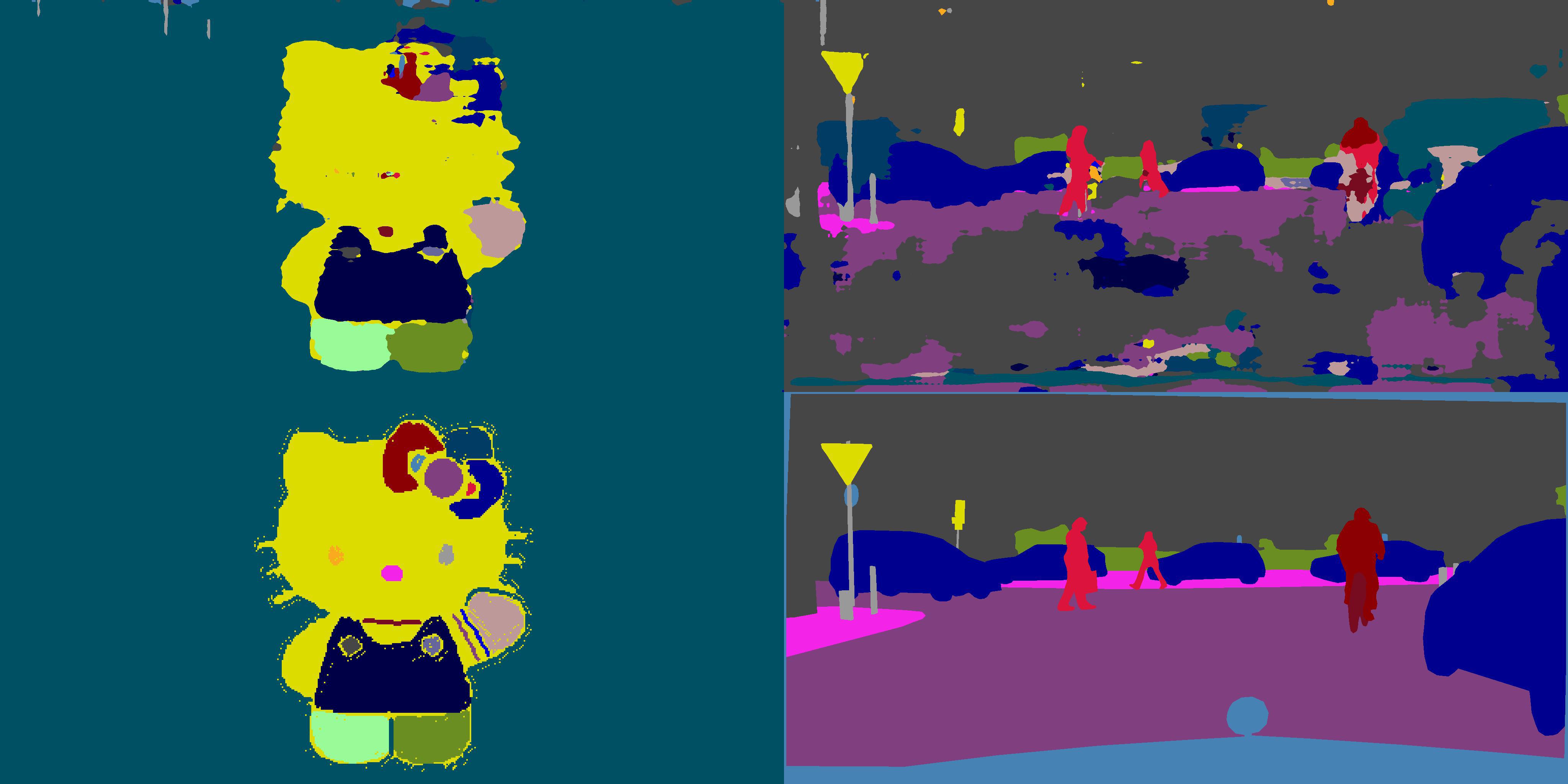}
     \caption{Houdini $|$ Hello Kitty $|$ DRN-C-26}
     \label{fig:attention-a}
     \end{subfigure}
    \end{minipage}

    \begin{minipage}{0.45\textwidth}
     \begin{subfigure}{\textwidth}
     \centering
     \includegraphics[width=\textwidth]{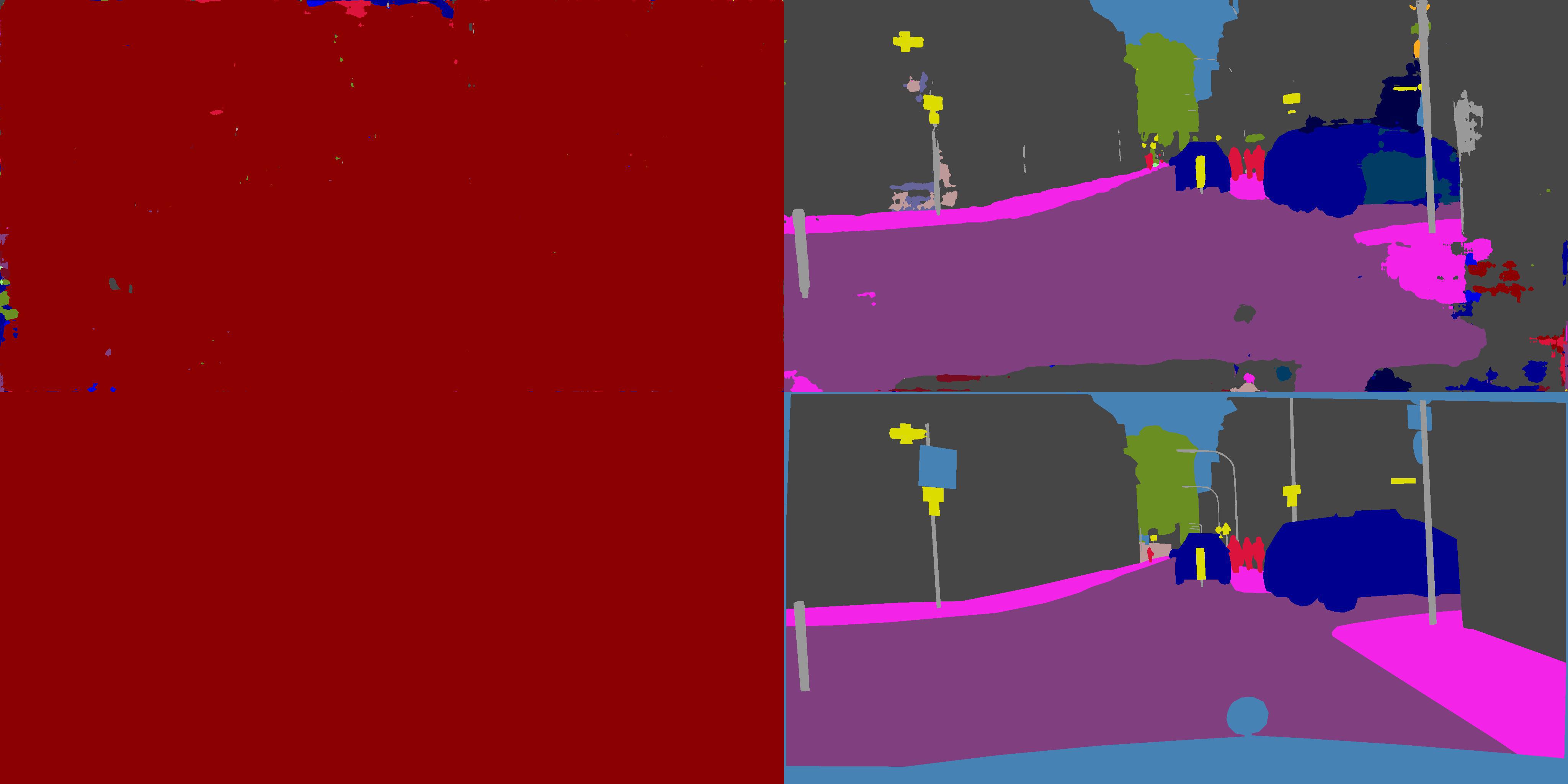}
     \caption{DAG $|$ Pure Rider $|$DLA34UP}
     \label{fig:attention-a}
     \end{subfigure}
    \end{minipage}
    \begin{minipage}{0.45\textwidth}
     \begin{subfigure}{\textwidth}
     \centering
     \includegraphics[width=\textwidth]{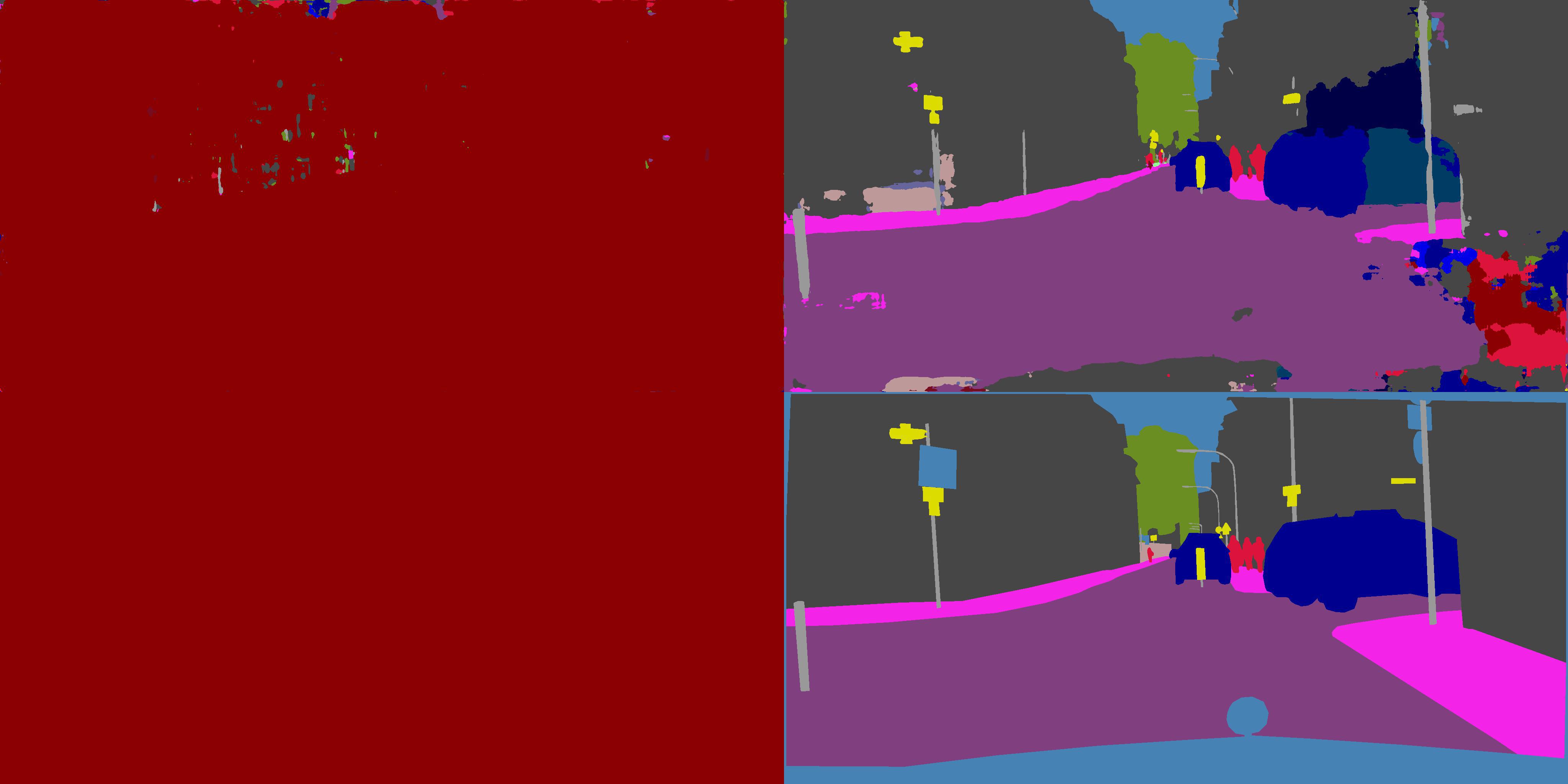}
     \caption{Houdini $|$ Pure Rider $|$DLA34UP}
     \label{fig:attention-a}
     \end{subfigure}
    \end{minipage}
    \begin{minipage}{0.45\textwidth}
     \begin{subfigure}{\textwidth}
     \centering
     \includegraphics[width=\textwidth]{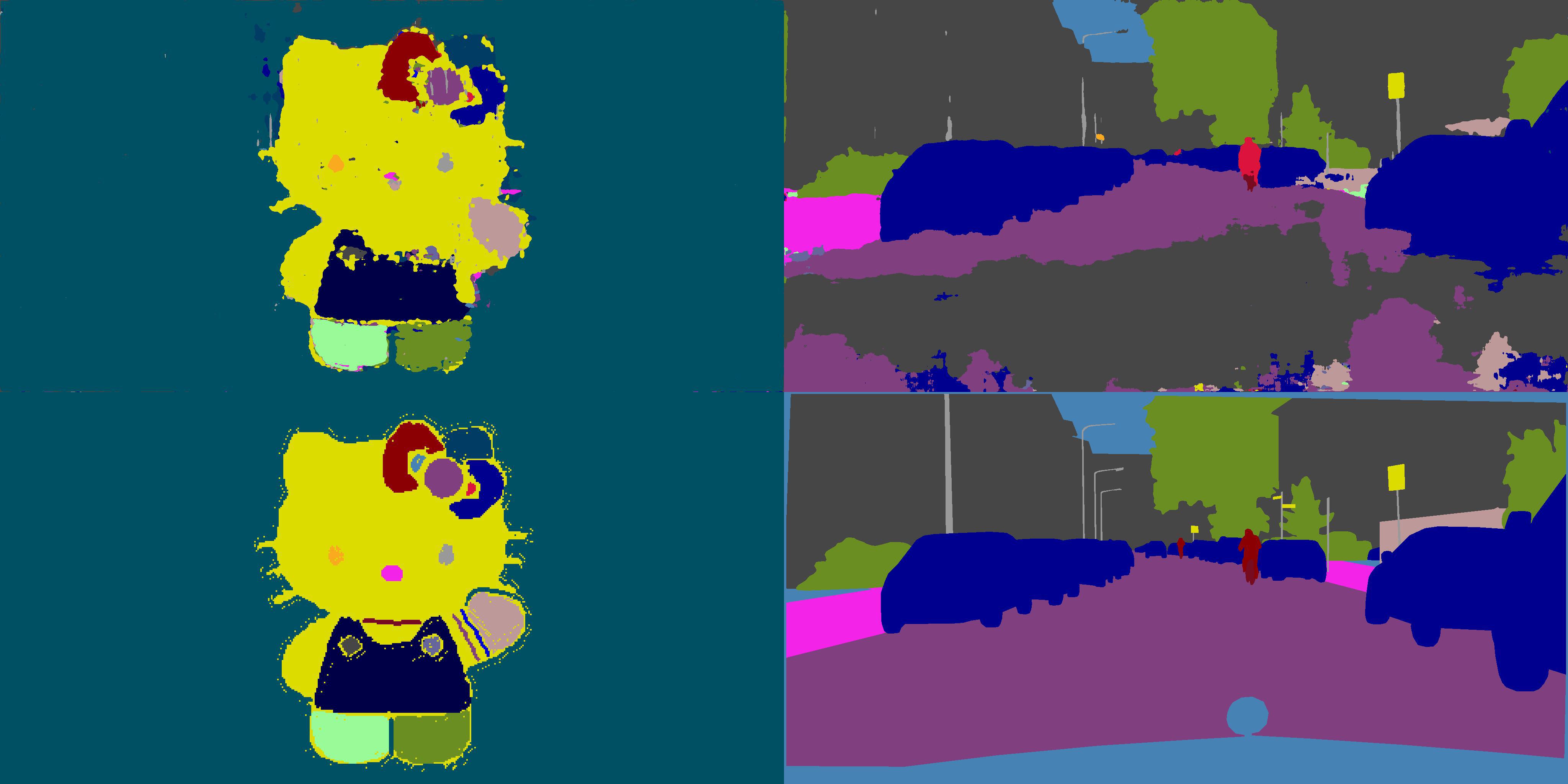}
     \caption{DAG $|$ Hello Kitty $|$DLA34UP}
     \label{fig:attention-a}
     \end{subfigure}
    \end{minipage}
    \begin{minipage}{0.45\textwidth}
     \begin{subfigure}{\textwidth}
     \centering
     \includegraphics[width=\textwidth]{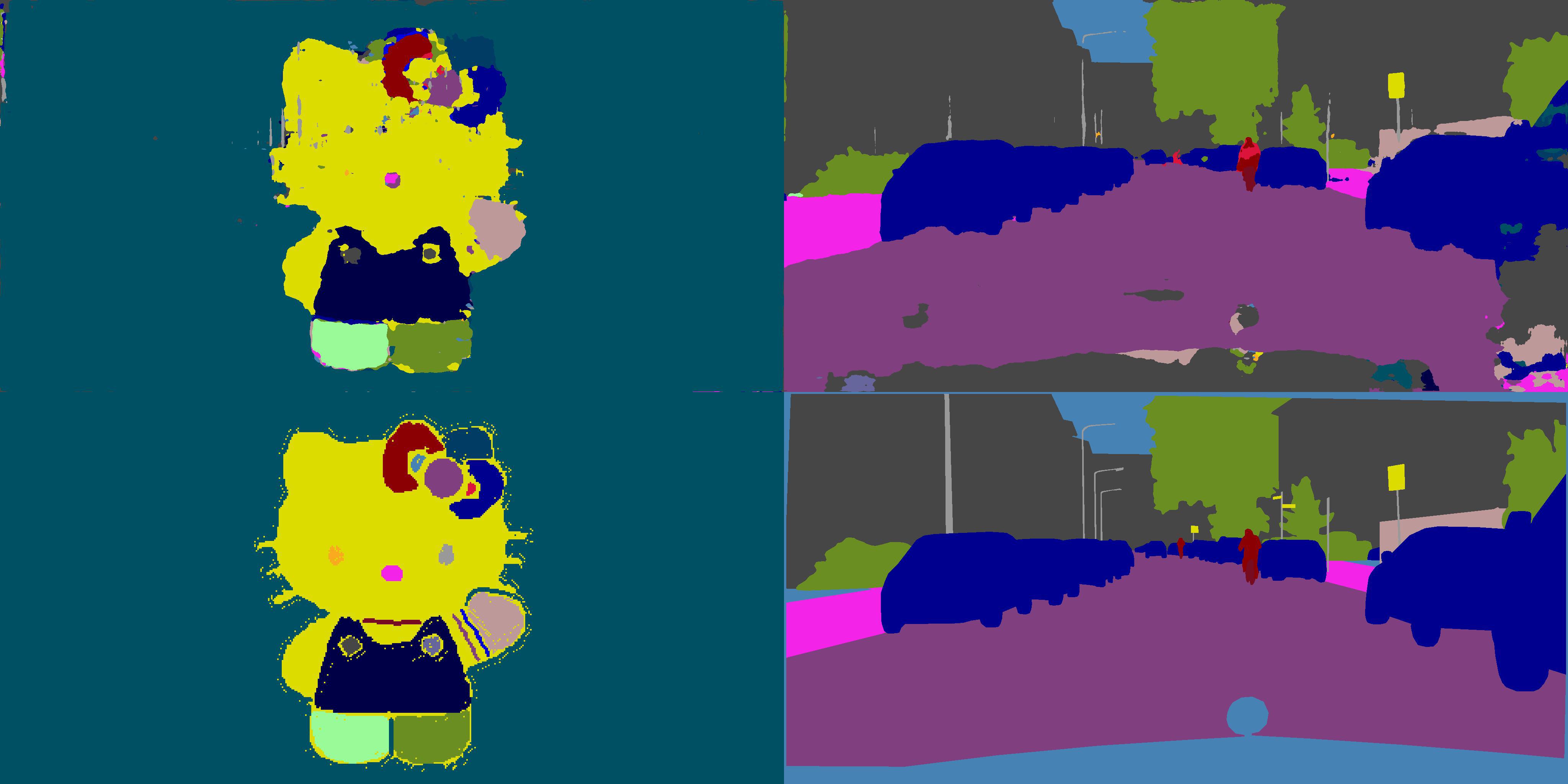}
     \caption{Houdini $|$ Hello Kitty $|$DLA34UP}
     \label{fig:attention-a}
     \end{subfigure}
    \end{minipage}
    \caption{Transferability visualization on CityScapes dataset. In each sub-figure, the first row presents
    the segmentation results of adversarial example on model A (targeted model) and model B. 
    The second row shows the adversarial target and the ground truth. We use format ``attack method $|$ attack target $|$ model '' to denote the caption of each sub-figure.}
    \label{fig:cityscapes-transfer-vis}
\end{figure}

\begin{figure}[h]
    \centering
    \begin{minipage}{0.45\textwidth}
     \begin{subfigure}{\textwidth}
     \centering
     \includegraphics[width=\textwidth]{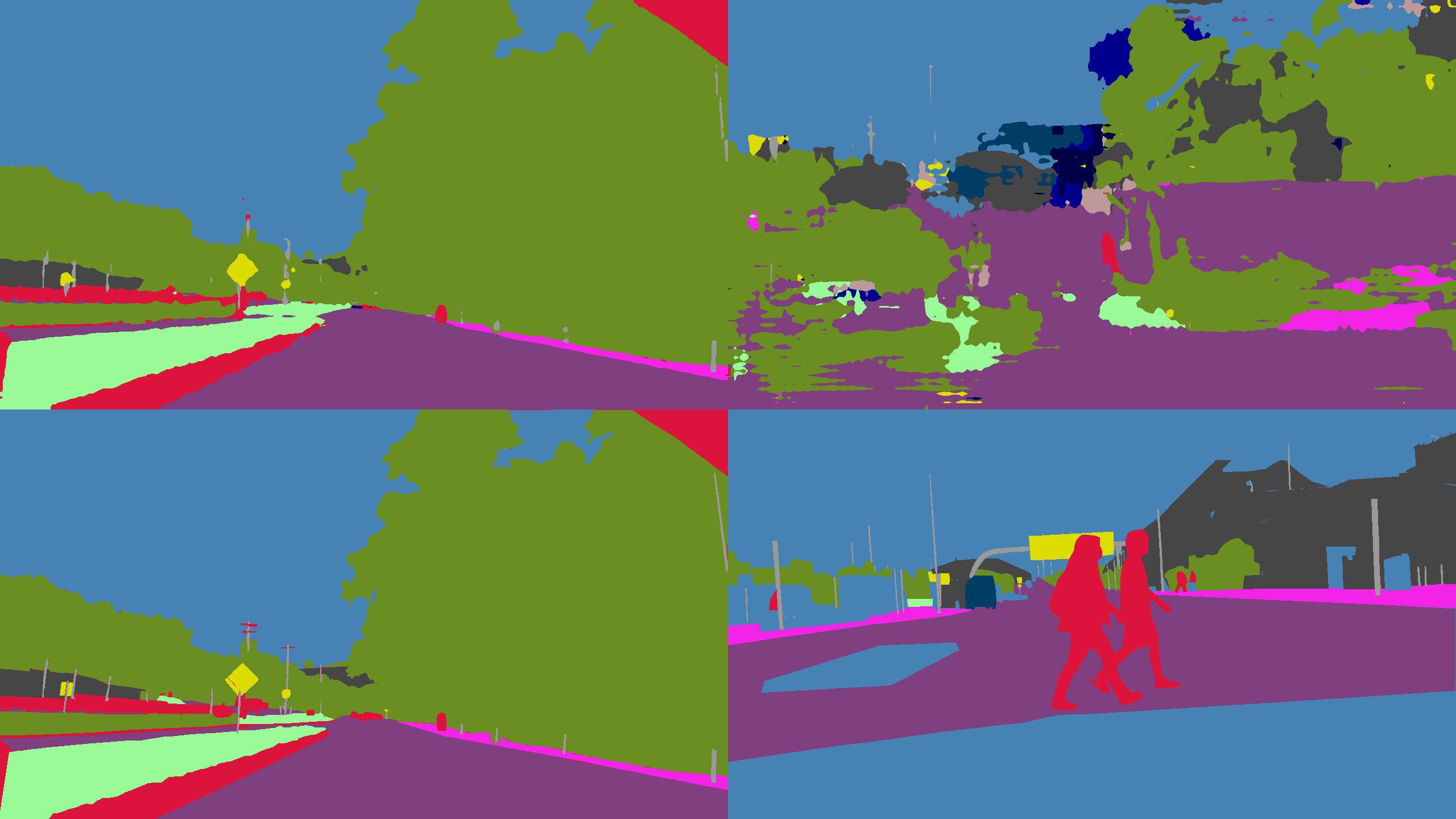}
     \caption{DAG $|$ Scene $|$ DRN-D-22}
     \end{subfigure}
    \end{minipage}
    \begin{minipage}{0.45\textwidth}
     \begin{subfigure}{\textwidth}
     \centering
     \includegraphics[width=\textwidth]{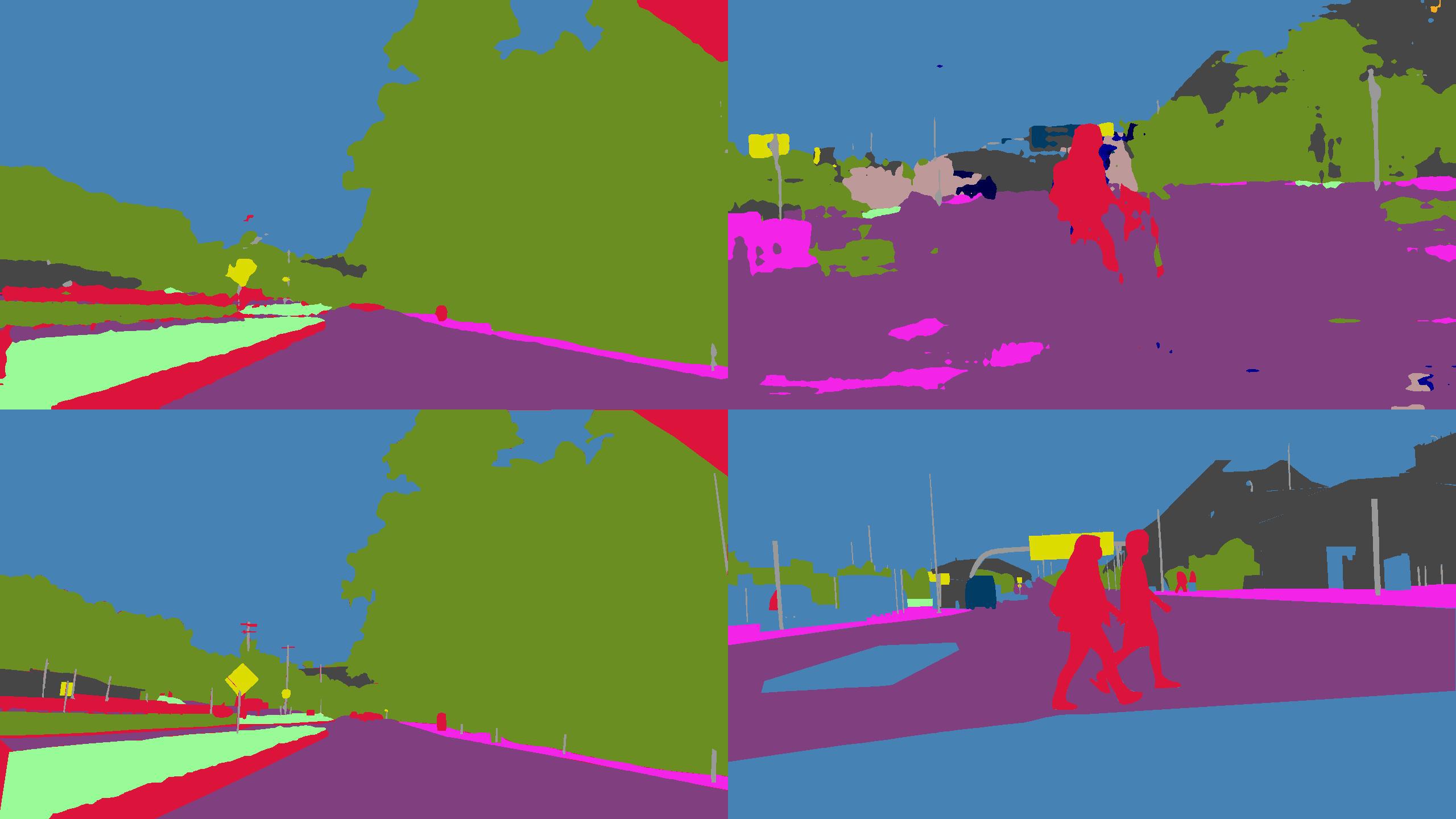}
     \caption{Houdini $|$ Scene $|$ DRN-D-22}
     \end{subfigure}
    \end{minipage}
    \begin{minipage}{0.45\textwidth}
     \begin{subfigure}{\textwidth}
     \centering
     \includegraphics[width=\textwidth]{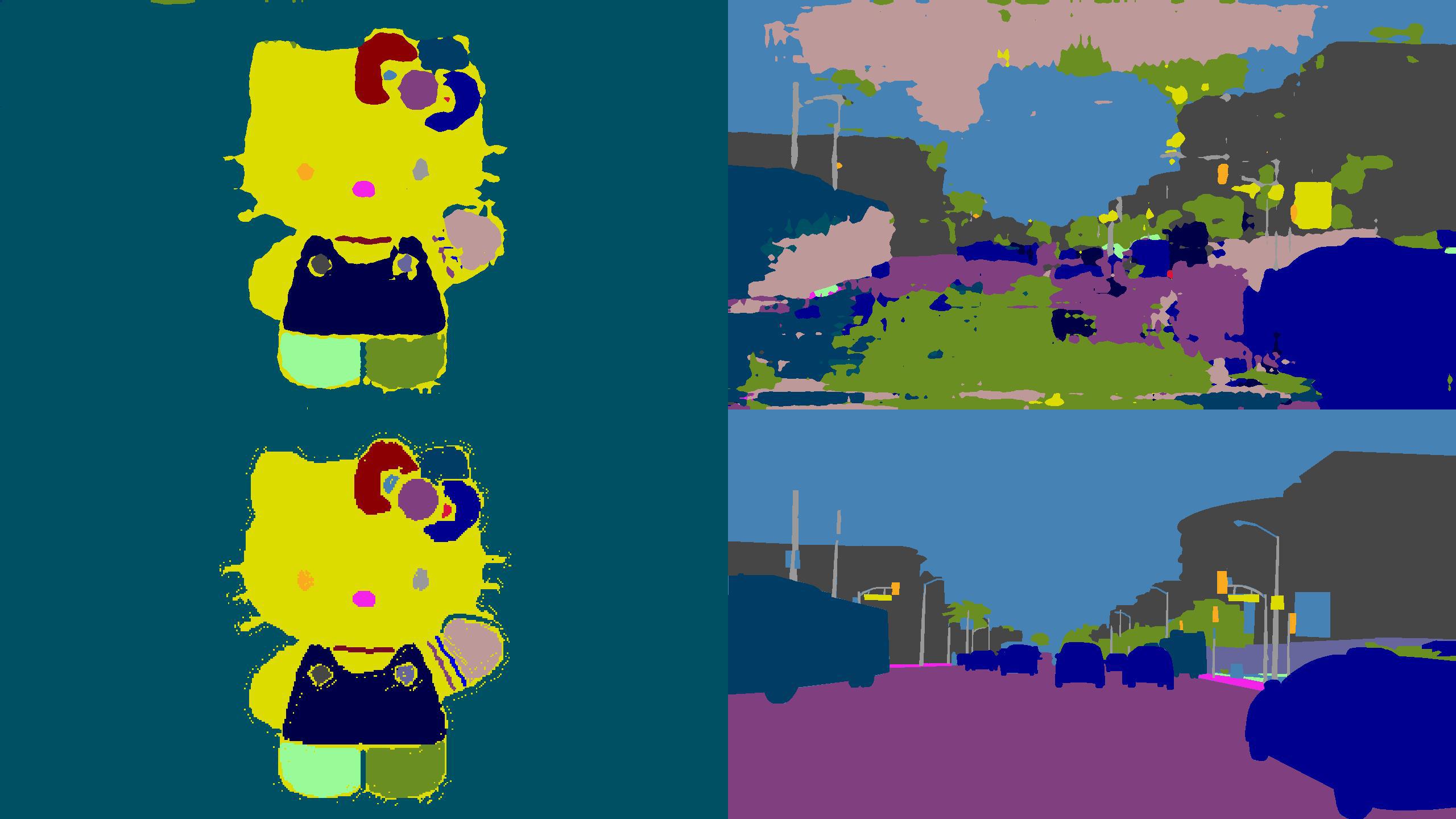}
     \caption{DAG $|$ Hello Kitty $|$ DRN-D-22}
     \end{subfigure}
    \end{minipage}
    \begin{minipage}{0.45\textwidth}
     \begin{subfigure}{\textwidth}
     \centering
     \includegraphics[width=\textwidth]{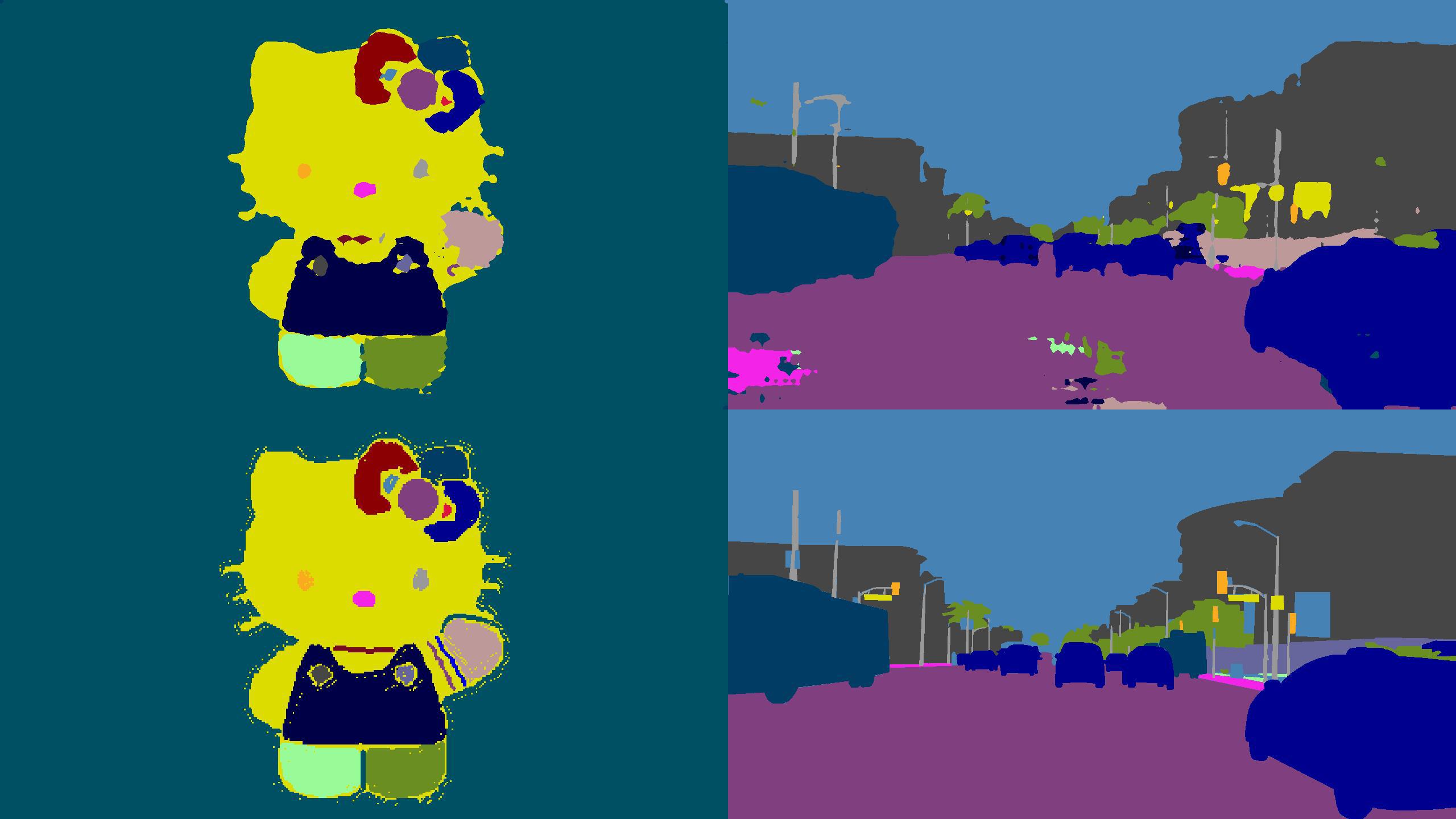}
     \caption{Houdini $|$ Hello Kitty $|$ DRN-D-22}
     \end{subfigure}
    \end{minipage}
    \begin{minipage}{0.45\textwidth}
     \begin{subfigure}{\textwidth}
     \centering
     \includegraphics[width=\textwidth]{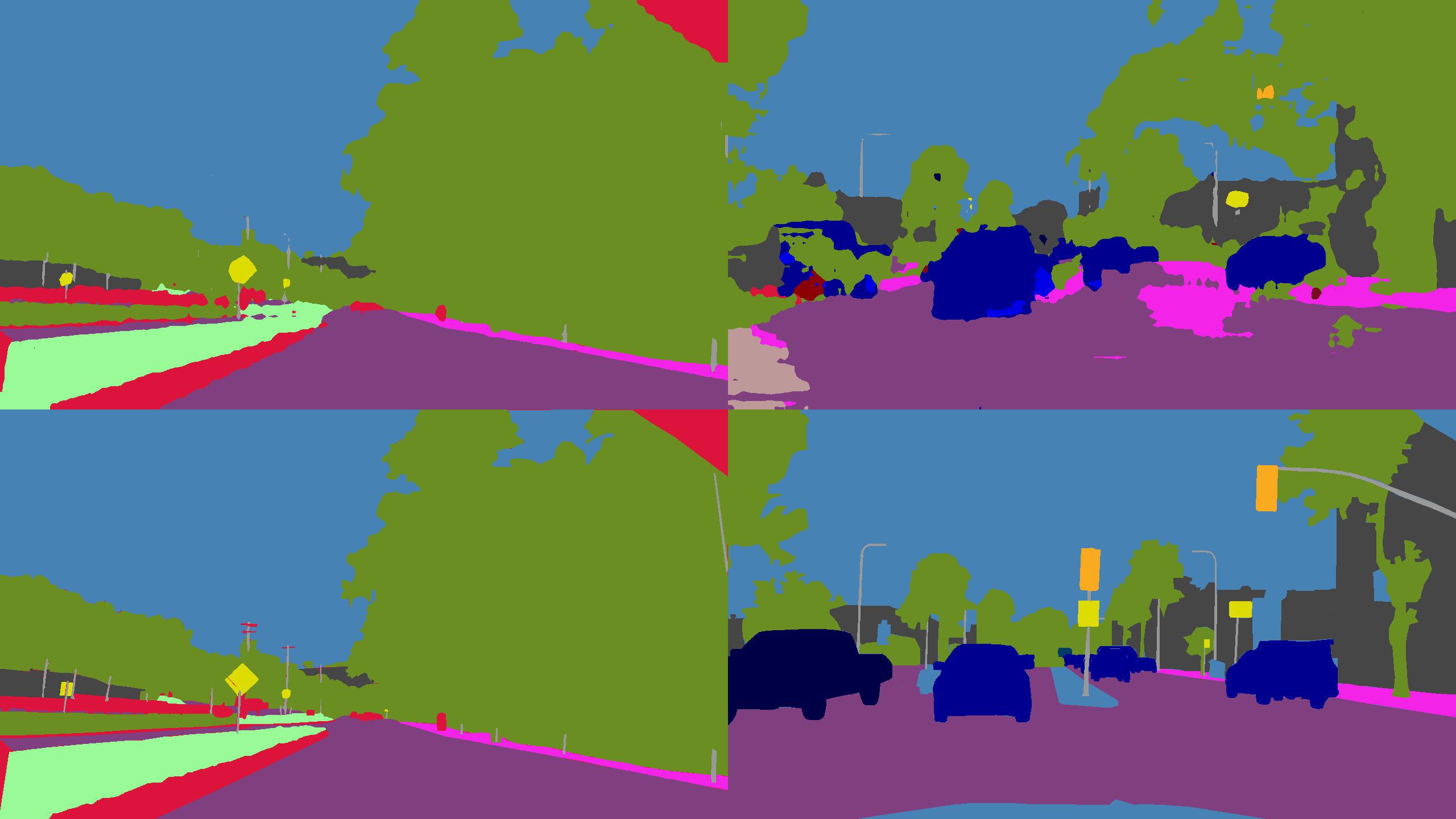}
     \caption{DAG $|$ Scene $|$ DRN-C-26}
     \end{subfigure}
    \end{minipage}
    \begin{minipage}{0.45\textwidth}
     \begin{subfigure}{\textwidth}
     \centering
     \includegraphics[width=\textwidth]{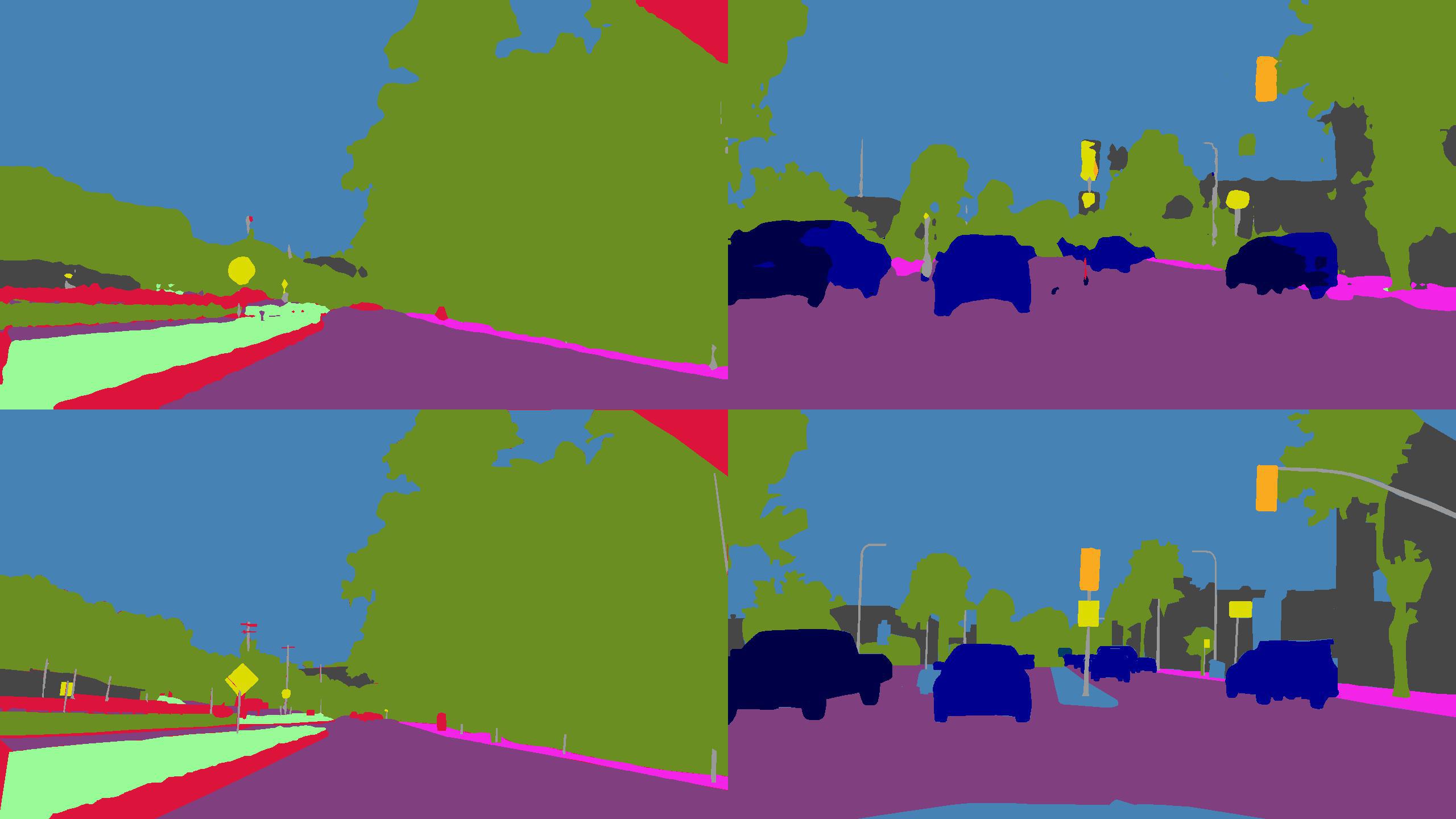}
     \caption{Houdini $|$ Scene $|$ DRN-C-26}
     \end{subfigure}
    \end{minipage}
\end{figure}

\begin{figure}[h]
    \ContinuedFloat
    \centering
    \begin{minipage}{0.45\textwidth}
     \begin{subfigure}{\textwidth}
     \centering
     \includegraphics[width=\textwidth]{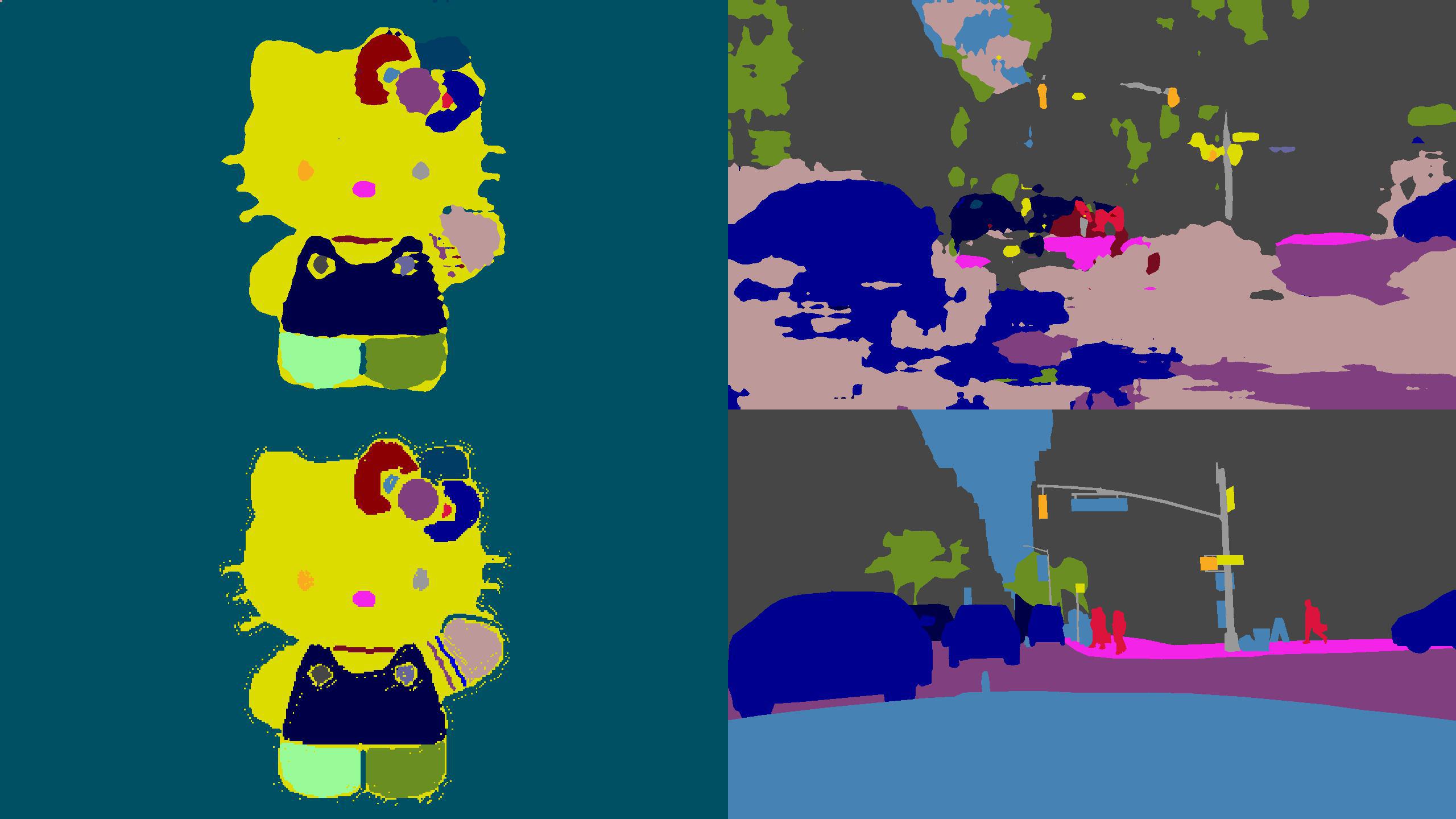}
     \caption{DAG $|$ Hello Kitty $|$ DRN-C-26}
     \end{subfigure}
    \end{minipage}
    \begin{minipage}{0.45\textwidth}
     \begin{subfigure}{\textwidth}
     \centering
     \includegraphics[width=\textwidth]{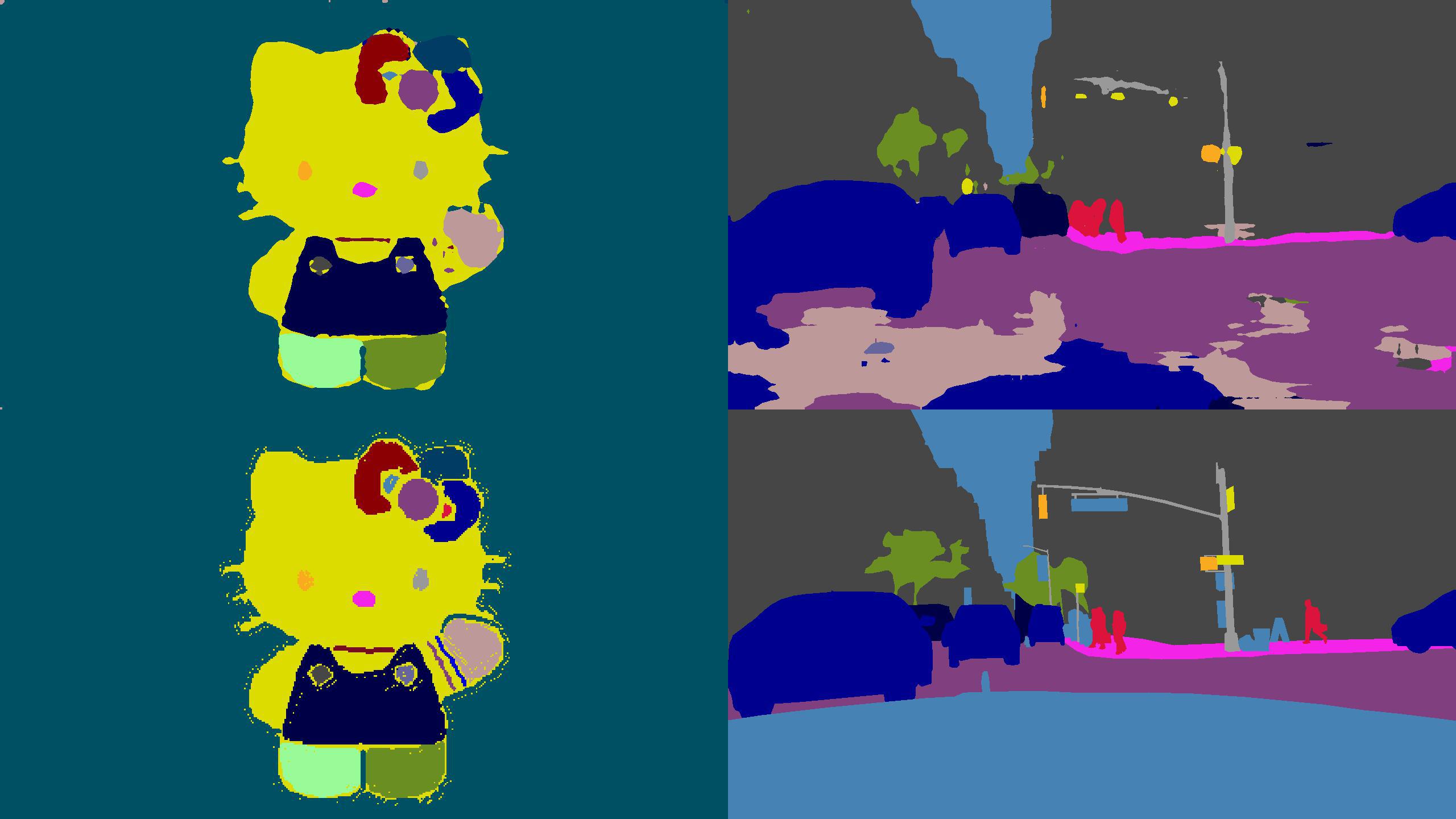}
     \caption{Houdini $|$ Hello Kitty $|$ DRN-C-26}
     \end{subfigure}
    \end{minipage}
    \begin{minipage}{0.45\textwidth}
     \begin{subfigure}{\textwidth}
     \centering
     \includegraphics[width=\textwidth]{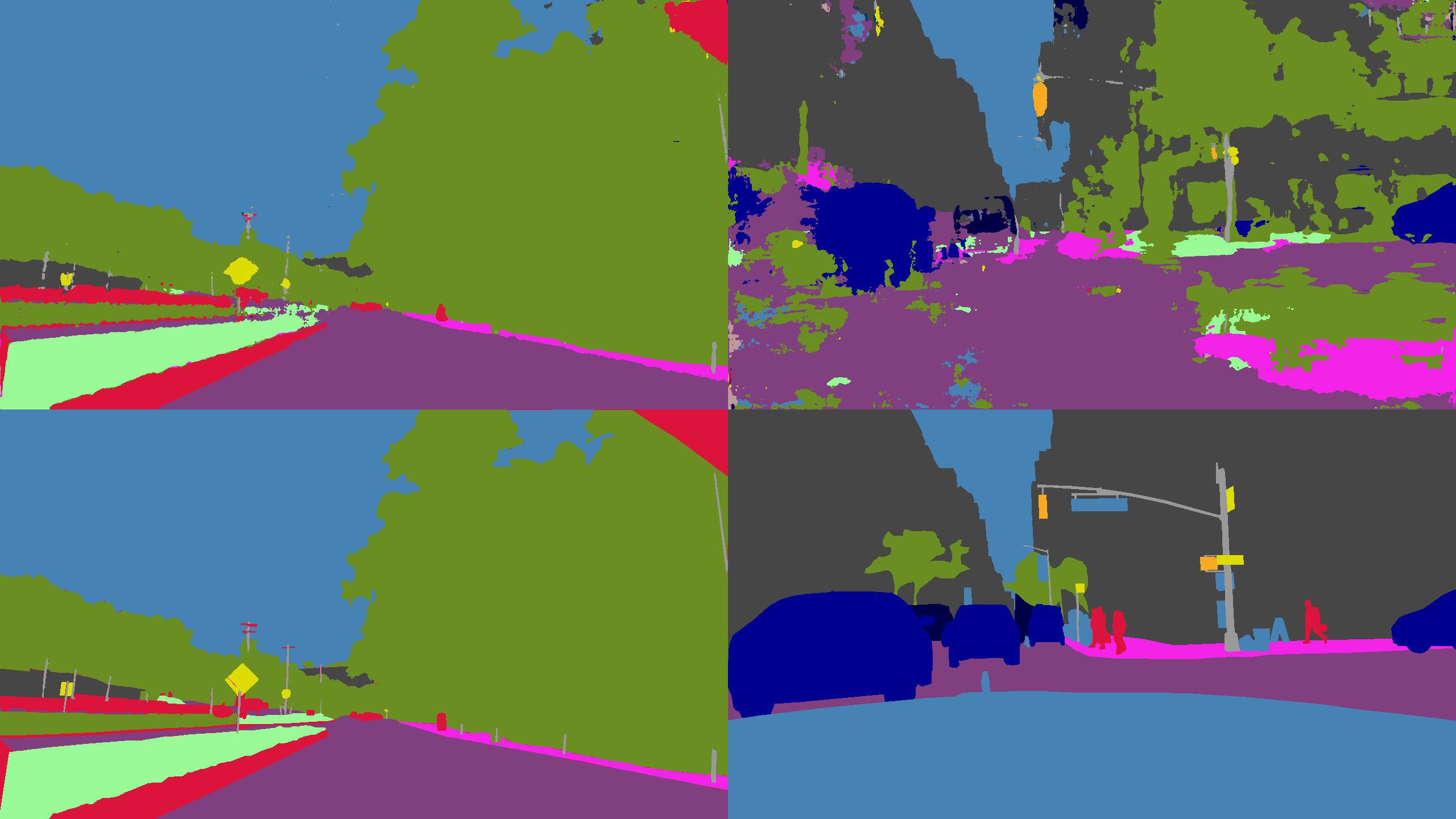}
     \caption{DAG $|$ Scene $|$ DLA34UP}
     \label{fig:attention-a}
     \end{subfigure}
    \end{minipage}
    \begin{minipage}{0.45\textwidth}
     \begin{subfigure}{\textwidth}
     \centering
     \includegraphics[width=\textwidth]{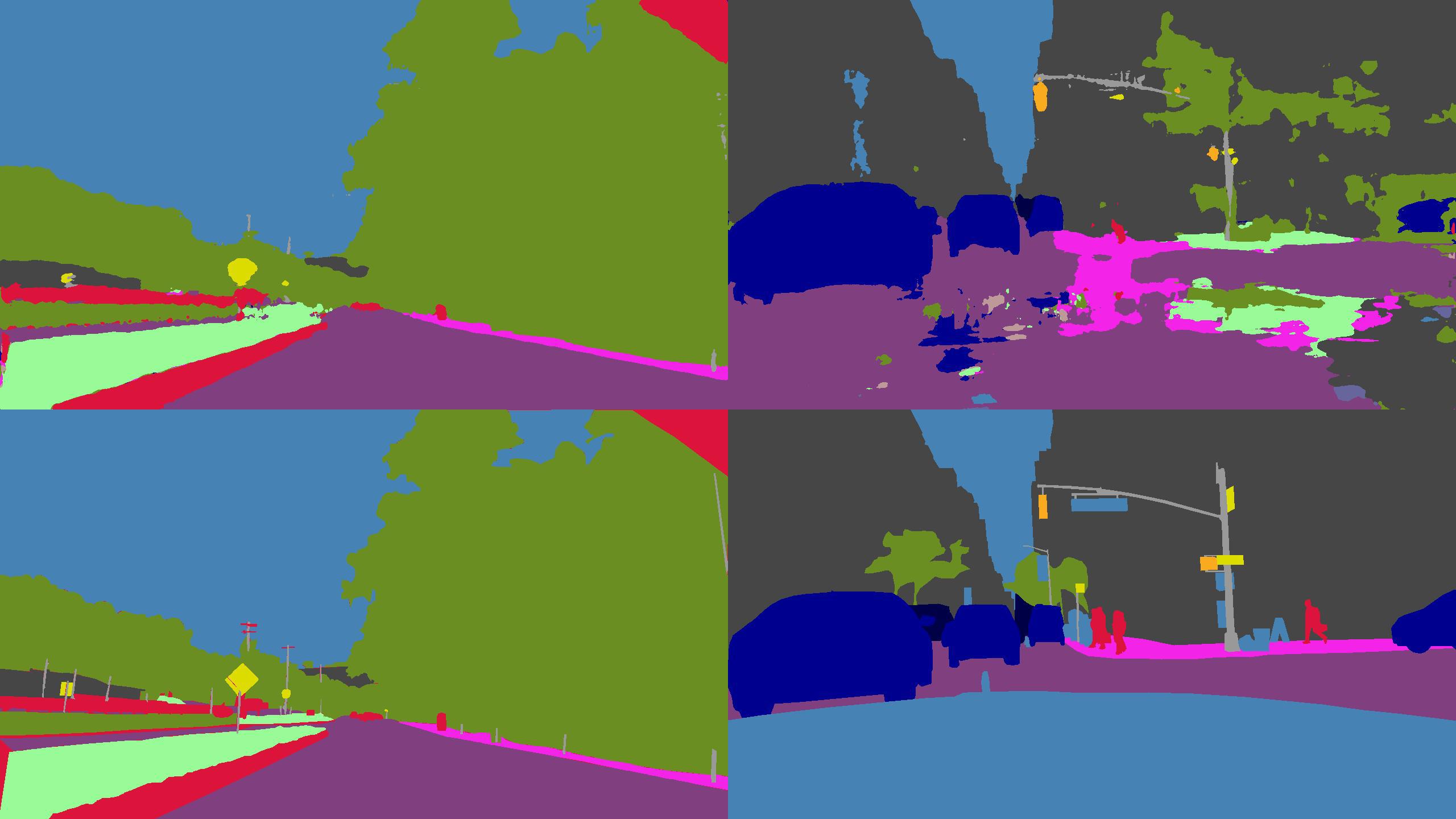}
     \caption{Houdini $|$ Scene $|$ DLA34UP}
     \label{fig:attention-a}
     \end{subfigure}
    \end{minipage}
    \begin{minipage}{0.45\textwidth}
     \begin{subfigure}{\textwidth}
     \centering
     \includegraphics[width=\textwidth]{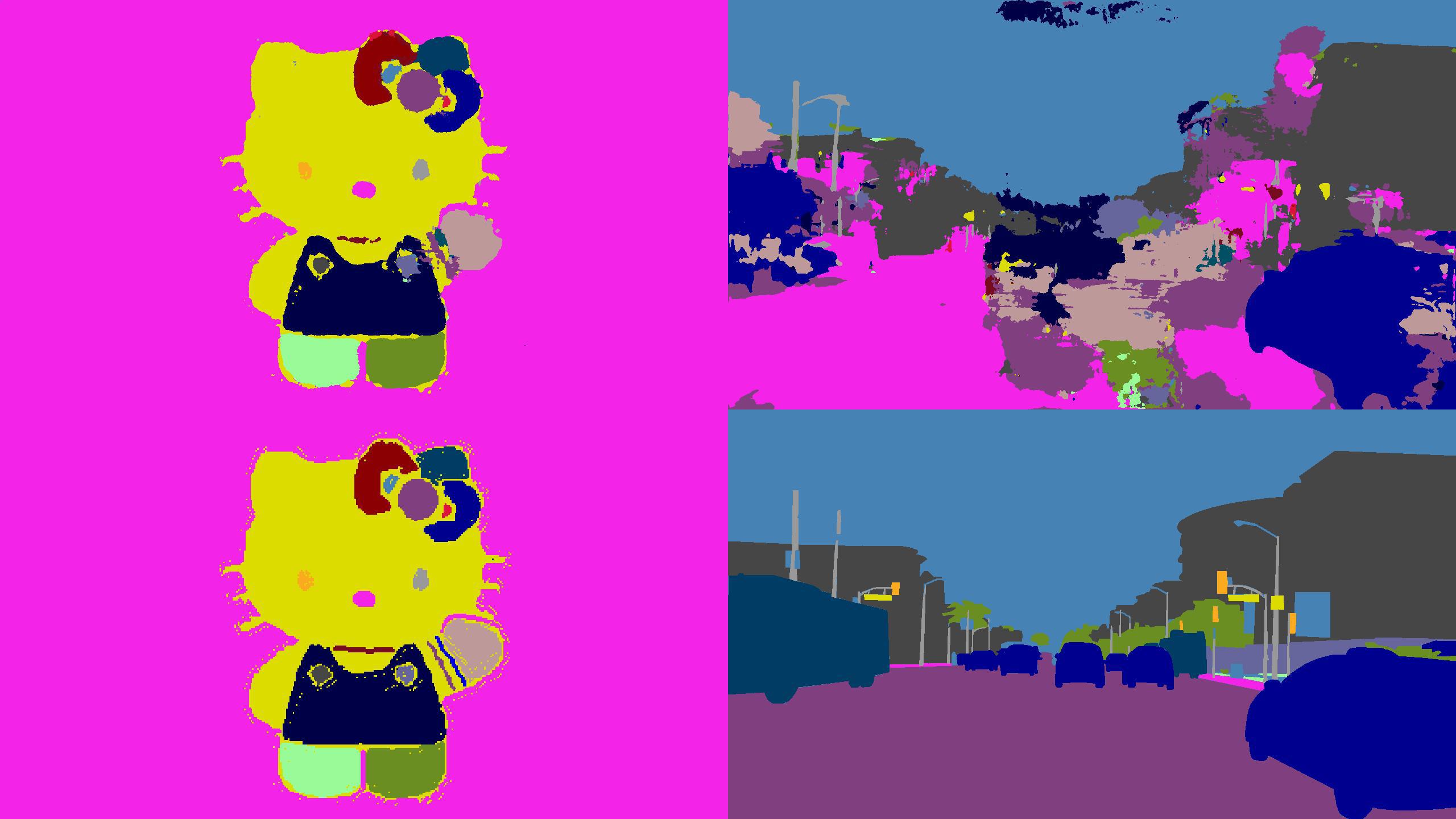}
     \caption{DAG $|$ Hello Kitty $|$ DLA34UP}
     \label{fig:attention-a}
     \end{subfigure}
    \end{minipage}
    \begin{minipage}{0.45\textwidth}
     \begin{subfigure}{\textwidth}
     \centering
     \includegraphics[width=\textwidth]{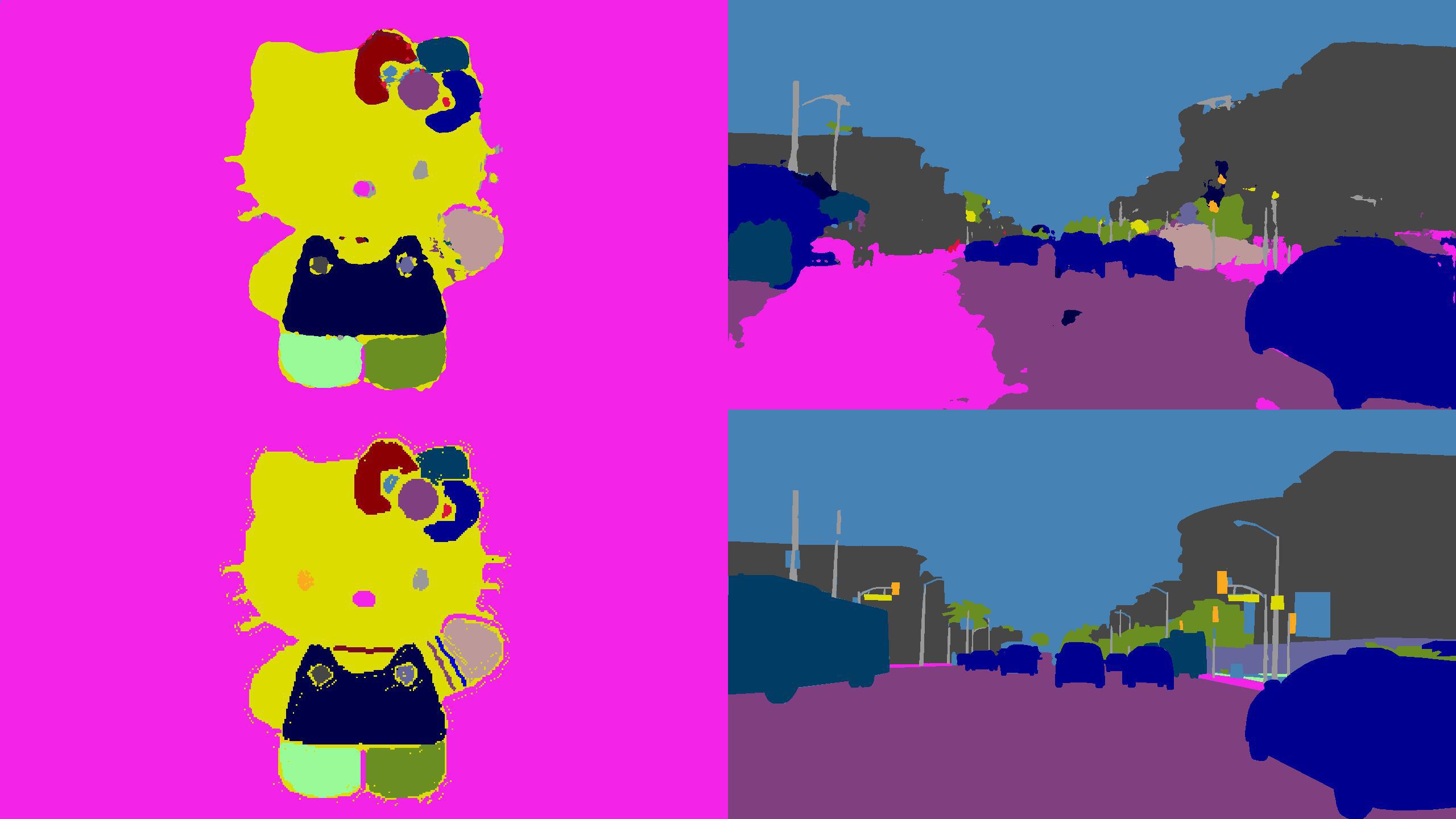}
     \caption{Houdini $|$ Hello Kitty $|$ DLA34UP}
     \label{fig:attention-a}
     \end{subfigure}
    \end{minipage}
    \caption{Transferability visualization on BDD dataset.}
    \label{fig:bdd-transfer-vis}
\end{figure}

